\documentclass[final,times,12pt,aps,prd]{elsarticle}
\usepackage{graphicx, bm, color, amsmath, amssymb, xfrac}
\usepackage[usenames,dvipsnames,table]{xcolor}

\usepackage{scrextend}
\usepackage{hyperref}

\usepackage[explicit]{titlesec}
\titleformat{\paragraph}[runin] {\normalfont\normalsize}{}{}{{\it #1\!\!\!}}
\titleformat{\section}{\normalfont\large\bfseries}{\thesection.}{0.4em}{#1}
\titleformat{\subsection}{\normalfont\bfseries}{\thesubsection.}{0.4em}{#1}

\usepackage[total={6.7in, 8.2in}, top=3cm, bottom=3cm]{geometry}

\definecolor{lightgray}{gray}{0.96}

\journal{Physics of the Dark Universe}

\newcommand{\prd}{Phys. Rev. D}
\newcommand{\prl}{Phys. Rev. Lett.}
\newcommand{\mnras}{MNRAS}
\newcommand{\jcap}{JCAP}
\newcommand{\apj}{ApJ}
\newcommand{\apjs}{ApJS}
\newcommand{\apjl}{ApJL}
\newcommand{\aj}{AJ}
\newcommand{\aap}{Astron. Astrophys.}

\newcommand{\nat}{Nature}

\newcommand{\lcdm}{$\Lambda$CDM}

\newcommand{\be}{\begin{equation}}
\newcommand{\ee}{\end{equation}}

\hypersetup{
    colorlinks = true,
    citecolor = {MidnightBlue},
    linkcolor = {BrickRed},
    urlcolor = {BrickRed},
}

\begin{document}

\begin{frontmatter}



\title{Beyond \lcdm{}: Problems, solutions, and the road ahead}

\author[Caltech,JPL,Oslo]{Philip Bull\footnote{\label{note1}Editors and corresponding authors --- \href{mailto:Philip.J.Bull@jpl.nasa.gov}{Philip.J.Bull@jpl.nasa.gov},~\href{mailto:Yashar.Akrami@thphys.uni-heidelberg.de}{Yashar.Akrami@thphys.uni-heidelberg.de}}}
\address[Caltech]{California Institute of Technology, Pasadena, CA 91125, USA}
\address[JPL]{Jet Propulsion Laboratory, California Institute of Technology, 4800 Oak Grove Drive, Pasadena, California, USA}
\address[Oslo]{Institute of Theoretical Astrophysics, University of Oslo, PO\ Box 1029 Blindern, 0315 Oslo, Norway}

\author[Heidelberg,Oslo]{Yashar Akrami\footref{note1}}
\address[Heidelberg]{Institut f\"ur Theoretische Physik, Ruprecht-Karls-Universit\"at Heidelberg,\\ Philosophenweg 16, 69120 Heidelberg, Germany}


\author[GEN]{\\ Julian Adamek}
\address[GEN]{D\'epartement de Physique Th\'eorique \& Center for Astroparticle Physics, Universit\'e de Gen\`eve,\\ 24 Quai E.\ Ansermet, 1211 Gen\`eve 4, Switzerland}

\author[Oxford]{Tessa Baker}
\address[Oxford]{Dept. of Astrophysics, University of Oxford, Denys Wilkinson Building, Keble Road, Oxford, OX1 3RH, UK}

\author[EB0]{Emilio Bellini}
\address[EB0]{ICCUB, University of Barcelona (IEEC-UB), Mart{\'\i} i Franqu{\`e}s 1, E08028 Barcelona, Spain}

\author[JB1]{Jose Beltr\'an Jim\'enez}
\address[JB1]{CPT, Aix Marseille Universit\'e, UMR 7332, 13288 Marseille, France}

\author[EB1,EB2]{Eloisa Bentivegna}
\address[EB1]{Dipartimento di Fisica e Astronomia, Universit{\`a} degli Studi di Catania, Via S. Sofia 64, 95123 Catania, Italy}
\address[EB2]{INFN, Sezione di Catania, Via S. Sofia 64, 95123 Catania, Italy}

\author[SC0]{Stefano Camera}
\address[SC0]{Jodrell Bank Centre for Astrophysics, The University of Manchester,\\ Alan Turing Building, Oxford Road, Manchester M13 9PL, UK}

\author[SC2]{S\'ebastien Clesse}
\address[SC2]{Institute for Theoretical Particle Physics and Cosmology (TTK), RWTH Aachen University, D-52056 Aachen, Germany}

\author[JHD1]{Jonathan H. Davis}
\address[JHD1]{Department of Physics, Kings College London, London WC2R 2LS, UK}

\author[Trieste,FVN2]{Enea Di Dio}
\address[Trieste]{Osservatorio Astronomico di Trieste, Universit{\`a} degli Studi di Trieste, Via Tiepolo 11, 34143 Trieste, Italy}
\address[FVN2]{INFN, Sezione di Trieste, Via Valerio 2, I-34127 Trieste, Italy}

\author[OKC,SU]{Jonas Enander}
\address[OKC]{Oskar Klein Centre, Stockholm University, Albanova University Center, 106 91 Stockholm, Sweden}
\address[SU]{Department of Physics, Stockholm University AlbaNova University Center, 106 91 Stockholm, Sweden}


\author[AH1]{Alan Heavens}
\address[AH1]{Imperial Centre for Inference and Cosmology, Department of Physics, Imperial College London, Blackett Laboratory, Prince Consort Road, London SW7 2AZ, UK}

\author[LH1]{Lavinia Heisenberg}
\address[LH1]{Institute for Theoretical Studies, ETH Zurich, Clausiusstrasse 47, 8092 Zurich, Switzerland}

\author[BH1]{Bin Hu}
\address[BH1]{Institute Lorentz, Leiden University, PO Box 9506, Leiden 2300 RA, The Netherlands}

\author[CL1,Oslo]{Claudio Llinares}
\address[CL1]{Institute for Computational Cosmology, Department of Physics,\\ University of Durham, South Road, Durham, DH1 3LE, UK}

\author[RM1,ICG]{Roy Maartens}
\address[RM1]{Department of Physics, University of the Western Cape, Cape Town 7535, South Africa}
\address[ICG]{Institute of Cosmology \& Gravitation, University of Portsmouth, Portsmouth PO1 3FX, UK}

\author[OKC,SU]{Edvard M{\"o}rtsell}

\author[ICG]{Seshadri Nadathur}

\author[Oxford]{Johannes Noller}

\author[RP1]{Roman Pasechnik}
\address[RP1]{Department of Astronomy and Theoretical Physics, Lund University, SE 223-62 Lund, Sweden}

\author[MP1]{Marcel S. Pawlowski}
\address[MP1]{Department of Astronomy, Case Western Reserve University,
10900 Euclid Avenue, Cleveland, OH, 44106, USA}

\author[TP1]{Thiago S. Pereira}
\address[TP1]{Departamento de F\'isica, Universidade Estadual de Londrina,\\
Rodovia Celso Garcia Cid, km 380, 86051-990, Londrina -- PR, Brazil}

\author[MQ1]{Miguel Quartin}
\address[MQ1]{Instituto de Fisica, Universidade Federal do Rio de Janeiro, 21941-972, Rio de Janeiro, RJ, Brazil}

\author[AR1]{Angelo Ricciardone}
\address[AR1]{Faculty of Science and Technology, University of Stavanger, 4036, Stavanger, Norway}

\author[Oslo]{Signe Riemer-S\o{}rensen}

\author[MR1,MR2]{Massimiliano Rinaldi}
\address[MR1]{Department of Physics, University of Trento, Via Sommarive 14, 38123 Trento, Italy}
\address[MR2]{INFN -- TIFPA, Via Sommarive 14, 38123 Trento, Italy}

\author[ICG]{Jeremy Sakstein}

\author[IS1]{Ippocratis D. Saltas}
\address[IS1]{Instituto de Astrof\'isica e Ci\^{e}ncias do Espa\c{c}o, Faculdade de Ci\^{e}ncias, Campo Grande, PT1749-016 Lisboa, Portugal}

\author[VS1]{Vincenzo Salzano}
\address[VS1]{Institute of Physics, University of Szczecin, Wielkopolska 15, 70-451 Szczecin, Poland}

\author[GEN]{Ignacy Sawicki}

\author[ARS1,Heidelberg,ARS2]{Adam R. Solomon}
\address[ARS1]{DAMTP, Centre for Mathematical Sciences, University of Cambridge, Wilberforce Rd., Cambridge CB3 0WA, UK}
\address[ARS2]{Center for Particle Cosmology, Department of Physics and Astronomy,\\ University of Pennsylvania, Philadelphia, PA 19104, USA}

\author[OKC,SU]{Douglas Spolyar}

\author[GS]{Glenn D. Starkman}
\address[GS]{CERCA/Department of Physics/ISO, Case Western Reserve University, Cleveland, OH 44106-7079, USA}

\author[DS1]{Dani\`ele Steer}
\address[DS1]{APC, Universit\'e Paris VII, B\^atiment Condorcet, 75205 Paris Cedex 13, France}

\author[IT1,IT2]{Ismael Tereno}
\address[IT1]{Instituto de Astrofisica e Ciencias do Espaco, Tapada da Ajuda, 1349-018 Lisboa, Portugal}
\address[IT2]{Department of Physics, Faculty of Sciencies, University of Lisbon,  Campo Grande, 1749-016 Lisbon, Portugal}

\author[LV1,EB0,LV3,Oslo]{Licia Verde}
\address[LV1]{ICREA (Instituci\'o catalana de recerca i estudis avan\c{c}ats)}
\address[LV3]{Radcliffe Institute for Advanced Study, Harvard University, MA 02138, USA}

\author[Trieste,FVN2]{Francisco Villaescusa-Navarro}

\author[MvS1]{Mikael von Strauss}
\address[MvS1]{UPMC-CNRS, UMR7095, Institut d'Astrophysique de Paris, GReCO, \\
98bis boulevard Arago, F-75014 Paris, France}

\author[Oxford]{Hans A. Winther}


\begin{abstract}
Despite its continued observational successes, there is a persistent (and growing) interest in extending cosmology beyond the standard model, \lcdm{}. This is motivated by a range of apparently serious theoretical issues, involving such questions as the cosmological constant problem, the particle nature of dark matter, the validity of general relativity on large scales, the existence of anomalies in the CMB and on small scales, and the predictivity and testability of the inflationary paradigm. In this paper, we summarize the current status of \lcdm{} as a physical theory, and review investigations into possible alternatives along a number of different lines, with a particular focus on highlighting the most promising directions. While the fundamental problems are proving reluctant to yield, the study of alternative cosmologies has led to considerable progress, with much more to come if hopes about forthcoming high-precision observations and new theoretical ideas are fulfilled.
\end{abstract}


\end{frontmatter}

\newpage
Cosmology has been both blessed and cursed by the establishment of a standard model: \lcdm{}. On the one hand, the model has turned out to be extremely predictive, explanatory, and observationally robust, providing us with a substantial understanding of the formation of large-scale structure, the state of the early Universe, and the cosmic abundance of different types of matter and energy. It has also survived an impressive battery of precision observational tests -- anomalies are few and far between, and their significance is contentious where they do arise -- and its predictions are continually being vindicated through the discovery of new effects (B-mode polarization \citep{2013PhRvL.111n1301H} and lensing \citep{2007PhRvD..76d3510S, 2011PhRvL.107b1301D} of the cosmic microwave background (CMB), and the kinetic Sunyaev-Zel'dovich effect \citep{2012PhRvL.109d1101H} being some recent examples). These are the hallmarks of a good and valuable physical theory.

On the other hand, the model suffers from profound theoretical difficulties. The two largest contributions to the energy content at late times -- cold dark matter (CDM) and the cosmological constant ($\Lambda$) -- have entirely mysterious physical origins. CDM has so far evaded direct detection by laboratory experiments, and so the particle field responsible for it -- presumably a manifestation of ``beyond the standard model'' particle physics -- is unknown. Curious discrepancies also appear to exist between the predicted clustering properties of CDM on small scales and observations. The cosmological constant is even more puzzling, giving rise to quite simply the biggest problem in all of fundamental physics: the question of why $\Lambda$ appears to take such an unnatural value \citep{2012CRPhy..13..566M, Burgess:2013ara,Padilla:2015aaa}. Inflation, the theory of the very early Universe, has also been criticized for being fine-tuned and under-predictive \citep{2011SciAm.304d..36S}, and appears to leave many problems either unsolved or fundamentally unresolvable. These problems are indicative of a crisis.

From January 14th--17th 2015, we held a conference in Oslo, Norway to survey the successes and shortcomings of the \lcdm{} model. The aim was to decide which theoretical and observational directions {\it beyond \lcdm{}} will be the most fruitful over the coming decade -- if looking beyond the model is indeed the right thing to do. To structure thinking around this topic, our discussion was divided into three broad themes, corresponding to ``pillars'' of the standard model:
\begin{itemize}
 \item {\bf Laws of physics:} The question of whether the basic theoretical frameworks that we rely on for building cosmological models are correct/fit for purpose. These frameworks include general relativity (GR) as the theory of gravitation; quantum field theory and the standard model of particle physics as the fundamental description of the Universe's matter content; and inflation as the theory of the early Universe.
 \item {\bf Foundational assumptions:} The question of whether the fundamental assumptions underlying the \lcdm{} model are correct, and whether they can be tested. Such assumptions include the statistical homogeneity and isotropy of space and matter, the Gaussianity of initial fluctuations in energy density, and that the Big Bang/inflationary picture accurately describes the origin of the cosmos.
 \item {\bf Constituents of the Universe:} The question of the nature and physical origins of the various contributions to the energy density of the Universe, including what dark energy really is and whether it evolves; what dark matter is and how it clusters; what the inflaton is; and the mass of neutrinos and existence of other relativistic degrees of freedom.
\end{itemize}
Each of these themes was explored through plenary presentations and guided discussion sessions to determine which issues are real and relevant; whether proposed theoretical solutions and observational campaigns are on the right track to resolving them; and which directions should be prioritized as most promising (or likely to result in a breakthrough in understanding) in the coming decade.

This paper is a summary of these efforts.\footnote{Slides are available for some of the talks: \url{http://www.mn.uio.no/astro/english/research/news-and-events/events/conferences/beyond-lcdm/slides/}} It is divided into seven broad topic areas, with each section consisting of summaries from the plenary talks to outline the state-of-the-art in each topic,\footnote{Summaries of plenary talks that are marked with `*' were not prepared by the plenary speaker; errors and omissions are our own.} and setting the stage for further discussion. In the final section, we also report on the results of a poll to canvas the opinions of the cosmological community on current problems and possible directions beyond \lcdm{}.

\setcounter{tocdepth}{2}
\tableofcontents

\section{\lcdm{}: the road ahead}

\lcdm{} is not the standard model of cosmology without good reason -- a significant and impressive body of observational evidence supports its theoretical predictions. In this section we begin by reviewing some of the evidence that has led to its establishment as the standard model, and its status in light of recent observational programs like the Planck CMB mission~\cite{Adam:2015rua}.

Looking to the future, there is a tremendous imperative to understand the physics of cosmic acceleration. We review some of the observational efforts in this direction, mostly from the perspective of large-scale structure surveys, and ask what we can expect to learn about dark energy from them -- and whether we are asking the right questions in the first place.

\subsection{\lcdm{}: successes and status}
\label{sec:george}
\noindent{\it Plenary speaker: G. Efstathiou*}


The \lcdm{} model has survived more than a decade of -- increasingly stringent -- precision tests, which culminated recently with the announcement of the latest results from the Planck satellite~\cite{Adam:2015rua}. In many ways the Planck 2015 cosmological results \cite{2015arXiv150201589P} highlight the successes of the \lcdm{} model, and so we shall briefly review them here.

Planck's precision measurements of the CMB temperature and E-mode polarization auto- and cross-spectra (TT, TE, EE) are in remarkable agreement -- the best-fit 6-parameter \lcdm{} model to the temperature-only data predicts the measured TE and EE spectra with remarkable accuracy, for example. The cosmological parameters derived from these spectra are also highly consistent, regardless of which combination of spectra is used for the parameter estimation (although low-level systematics nevertheless remain in the polarization data). Expected secondary effects are detected with high significance; Planck's 2.5\% constraint on the CMB lensing power spectrum amplitude corresponds to a $40\sigma$ detection of this effect, for example. The Planck 2015 results are also consistent with flatness ($\Omega_K = 0 \pm 5\times10^{-3}$, 95\% CL), and constrain the effective number of relativistic species to its standard model value of $N_{\rm eff} = 3.046$ to within 10\%.

The best-fitting 2013 Planck parameters \cite{2014A&A...571A..16P} were consistent between the CMB power spectrum and CMB lensing, as well as with external baryon acoustic oscillations (BAO) experiments, but possible tensions were observed between various measurements of the $\sigma_8$ parameter (e.g. from weak lensing and cluster number counts) and $H_0$. A $2\sigma$ tension with Type Ia supernova data was also observed. In the 2015 results, a possible tension with the CFHTLenS weak lensing data was again observed in the $\Omega_M - \sigma_8$ plane (also see Ref.~\cite{Raveri:2015maa}). There is also an apparent tension between measurements of the growth rate, $f\sigma_8$, from redshift-space distortions and the value predicted by the Planck best-fit parameters (e.g. Ref.~\cite{2013PhRvL.111p1301M}). If taken at face value, these tensions can, in the most extreme case (Planck + CFHTLenS weak lensing), be translated into a significant deviation from a cosmological constant equation of state, $w=-1$, or general relativity ($\mu_0=\eta_0=1$) \cite{2015arXiv150201590P}. The cause of these tensions is not yet known; it does seem that the base Planck \lcdm{} model has a definite preference for a higher $\sigma_8$ than other experiments, however, and that this cannot be alleviated by changing assumptions about relativistic degrees of freedom such as massive neutrinos.

The constraints on inflationary parameters are very tight, with some potentials (e.g. $\phi^2$) effectively being ruled-out \cite{2015arXiv150202114P}. A reconstruction of the primordial power spectrum is also found to be strongly consistent with a `standard' pure adiabatic, featureless, tilted spectrum. The upper limit on the tensor-to-scalar ratio is $r < 0.11$ (95\% CL), as in the 2013 results, although it should be pointed out that this is a model-dependent constraint. Large parts of the $n_s - r$ plane remain viable if one is willing to relax other assumptions (e.g. by allowing $\Delta N_{\rm eff} \neq 0$). Observed large-angle anomalies are not strongly significant once the look-elsewhere effect is taken into account \cite{Ade:2015hxq} (see, however, Sect.~\ref{sec:cmbanomalies}).

In conclusion, \lcdm{} fits the Planck 2015 data, including polarization, very well, and there is as yet no convincing evidence for simple extensions to the model. There are some tensions with other data, however, particularly those that measure the amplitude of matter fluctuations at late times.

\subsection{Hunting dark energy with galaxy redshift surveys}
\label{sec:huntingde}
\noindent{\it Plenary speaker: B. Reid*}

While observations of CMB anisotropies have now reached fantastic levels of precision, they only give us a partial view of the Universe -- as it was on a thin spatial shell at early times. Much has happened since $z \approx 1100$ -- fluctuations in the matter distribution have grown by more than a factor of a thousand, for example -- and a large volume of the observable Universe remains to be explored. Large-scale structure (LSS) surveys such as SDSS/BOSS have been instrumental in filling in the picture at low and intermediate redshifts, through detections of galaxies ($z < 0.7$) and the Lyman-$\alpha$ forest ($z \approx 2.4$) respectively.

In this section, we discuss what we have learned from recent LSS datasets, what we can expect to learn in the future, and some of the practicalities of extracting cosmological information from them.

\paragraph{LSS regimes.} Large-scale structure can be roughly divided into three regimes, characterized by a range of characteristic distance scales:
\begin{itemize}
  \item Linear scales ($> 30 h^{-1}$ Mpc), where fluctuations are Gaussian, and almost fully characterized by the two-point function.
  \item Mildly non-linear scales ($2 - 30 h^{-1}$ Mpc).
  \item Highly non-linear scales ($< 1 h^{-1}$ Mpc), where galaxies are forming and galaxy cluster cosmology becomes important.
\end{itemize}
The linear regime is the easiest to work in from a theoretical standpoint, as linear cosmological perturbation theory is well understood and gives definite predictions. These scales have been used to test \lcdm{} in a number of ways, and it has passed each test so far -- albeit with some potentially-interesting $\sim 2 - 2.5\sigma$ anomalies cropping up along the way.

\paragraph{Consistency tests.}
Whether the LSS and CMB data are consistent is an important test of the model. Comparing the matter power spectrum inferred from the Planck data with the SDSS-II LRG \cite{2010MNRAS.404...60R} and SDSS-III BOSS DR11 \cite{2014MNRAS.441...24A} samples reveals good agreement. The $\chi^2$ with the LRG data is $40.4$ (40 dof) for the Planck-only best-fit spectrum, for example, while the LRG-only spectrum gives $\chi^2 = 40.0$ -- excellent agreement considering that the two datasets are independent.

Other consistency tests with \lcdm{} fare similarly well. The clustering signal on linear scales matches the expectation for cold dark matter very well, with alternatives like TeVeS and warm dark matter being (strongly) disfavored \cite{2011IJMPD..20.2749D}. The expansion history, as measured through observations of the BAO scale, is also consistent between CMB and LSS data -- the BAO scale, $D_V / r_s$, is independently constrained to around the 1\% level by both Planck and low-redshift BOSS data, and again the measurements agree \cite{2015arXiv150201589P}.

There is a tension of around $2.5\sigma$ between the Planck and BOSS Lyman-$\alpha$ BAO measurements, however \cite{2015A&A...574A..59D}. The best-fit Lyman-$\alpha$ angular diameter distance, $D_A$, and inverse Hubble scale, $D_H = c / H$, at $z=2.34$ are offset from the Planck \lcdm{} expectation by $-7\%$ and $+7\%$ respectively. The flat \lcdm{} model does however give an acceptable fit to the data, despite the tension \cite{2014arXiv1411.1074A}.

\paragraph{Redshift-space distortions.}
The anisotropy of the clustering pattern in redshift space (redshift-space distortions, RSDs) gives a measure of peculiar velocities and thus the growth history of the Universe \cite{1987MNRAS.227....1K}. In particular, the RSD signature can be used to measure the linear growth rate, $f\sigma_8$, which can be used to place stringent constraints on alternative models of gravity (e.g. Ref.~\cite{Dodelson:2013sma}).

If a \lcdm{} {\it expansion} history with Planck best-fit parameters is assumed, recent RSD measurements at $z < 0.8$ prefer a growth rate that is $\sim 1.9\sigma$ lower than the Planck expectation \cite{2013MNRAS.429.1514S, 2013PhRvL.111p1301M, 2014MNRAS.444..476R}. This should be compared with tensions between Planck and several other LSS probes that measure $\sigma_8$ (see Sect. \ref{sec:george}).

\paragraph{Mildly non-linear regime.} Information from sub-linear scales is often ignored in cosmological analyses due to difficulties in accurately modeling the matter distribution in this regime -- perturbative analyses are typically no longer sufficient, and simulations must be used. Nevertheless, a considerable amount of useful information is available here; the matter distribution is non-Gaussian, so higher-order statistics can be considered, for example. This is also an interesting regime for testing modified gravity, where screening effects and other non-linear phenomena begin to become important (e.g. Refs.~\cite{2006NJPh....8..323C, 2006PhRvD..74h4007S, 2008PhRvD..77b4048L, 2012PhRvD..86d4015B, 2013MNRAS.428..743L}). A good understanding of the connection between the galaxy field and the underlying dark matter field is needed to make robust inferences, however.

As an example, consider the analysis in Ref.~\cite{2014MNRAS.444..476R}, which uses the non-linear regime to constrain the growth rate. The anisotropic clustering signal depends on two main effects -- coherent infall due to the mutual gravitational attraction of galaxies, and incoherent small-scale velocity dispersion. The analysis relies on simulations to predict theoretical quantities, and bias parameters must be marginalized over to reflect the uncertainty in the mapping between the galaxy and underlying matter field. Several different simulation methods are compared, and there is consistency between their results. The resulting constraint on $f\sigma_8$ is some $2.5\times$ tighter than the linear-only analysis, demonstrating the value of attempting to extend analyses further into the non-linear regime.

There are some limitations to this method, however. The observations must be compared with N-body simulations of modified gravity theories before those theories can be definitively tested/ruled-out by this approach, and producing the requisite suite of simulations can be very computationally demanding. The relative accuracy of the N-body codes must also be considered, especially in the presence of baryonic effects, radiative processes and the like (see Sect.~\ref{sec:mgdev}).

Other promising options include combining galaxy-galaxy lensing information with galaxy clustering data \cite{2013MNRAS.432.1544M}, which offers an alternative way of measuring $f \sigma_8$; and the use of Bayesian inference codes to reconstruct the non-Gaussian, non-linear 3D density field jointly with the galaxy power spectrum and nuisance parameters such as galaxy bias \cite{2014IAUS..306....1L, 2014IAUS..306..258A}.

\subsection{Observing dark energy in the era of large surveys}
\label{sec:lahav}
\noindent{\it Plenary speaker: O. Lahav*}

The immediate future of the \lcdm{} model will probably be determined by a new crop of large-scale structure surveys that seek to make precision measurements of the properties of dark energy. To understand this trajectory, some historical context is needed to place us on the ``development curve'' of \lcdm{} as a theory, so we begin by commenting on the checkered history of $\Lambda$, and how the paradigm shifted in the 1990's from ``Standard CDM'' towards ``\lcdm{}'' \citep{2008A&G....49a..13C}.

Before the famous Type Ia supernovae results that ``clinched'' the cosmic acceleration result, there was roughly a decade of hints from various sources about the existence of a non-zero $\Lambda$ contribution. As is often the case, Peebles \cite{1984ApJ...284..439P} was one of the first along this track; in 1984, he recognized that the matter density of the Universe was coming out low from observations, so that if the flatness prediction of inflation was taken seriously, a significant $\Lambda$ contribution would be required to fit the data. Results from the APM experiment in 1990 \citep{1990Natur.348..705E} lent significant weight to this picture, but alternative explanations were also supported \citep{1991MNRAS.251..128L} -- a strong belief in inflation's prediction of flatness remained necessary to conclude that $\Lambda$ was non-zero. Evidence continued to build for a low matter density, however \citep{1993Natur.366..429W}, and a combination of other data promoted the \lcdm{} model to a more or less favored status by 1995 \citep{1995Natur.377..600O}. The supernova data finally ruled-out a significantly open Universe in 1998, therefore making definitive the discovery of dark energy: $\Omega_\Lambda > 0$ \citep{1998AJ....116.1009R, 1999ApJ...517..565P}.

After the initial excitement died down, it became clear that an extensive observational program would be needed ``to determine the dark energy properties as well as possible'' \citep{2006astro.ph..9591A}. A variety of current and future imaging and spectroscopic galaxy surveys (e.g. DES, DESI, Euclid, LSST, SKA, WFIRST) constitute that program, and represent our best hope for exploring physics beyond the ``vanilla \lcdm{}'' model in the near term. The current roadmap, extending out to 2030, describes an impressive array of more than 10 large survey experiments -- imaging and spectroscopic, space- and ground-based -- that will detect billions of galaxies at a cost of around \$1 per galaxy.

The Dark Energy Survey, DES, is arguably the first of the ``Stage III'' imaging experiments. It saw first-light in 2012, and completed a second season of observations in early 2015. DES employs a multi-probe approach, enabling cluster counts, weak lensing, large-scale structure, and supernova detection from a single instrument. Current highlights from the initial DES dataset include hundreds of candidate SN Ia detections, quasars at $z=6$ and new detections of high-redshift galaxy clusters, as well as weak lensing mass maps \citep{2015PhRvL.115e1301C}, a detection of galaxy-galaxy lensing \citep{2015arXiv150705552T}, and various cross-correlations of DES galaxies/clusters with other probes (like CMB lensing \citep{2015arXiv150705551G}).

Later surveys will significantly improve upon DES. The Large Synoptic Survey Telescope, LSST, another imaging survey, will collect some 30 TB of data per night (compared to DES's 1 TB) for example, resulting in a catalog of some 20 billion objects. The Dark Energy Spectroscopic Instrument, DESI, will be a spectroscopic experiment around 10 times the size of BOSS, yielding around 18 million emission-line galaxy (ELG) spectra at $z \sim 1$, 4 million luminous red galaxy (LRG) spectra at low redshift, and around 1 million Lyman-$\alpha$ quasars at $z \gtrsim 2$. This will allow it to measure the distance scale (using the baryon acoustic oscillation feature) to around 0.4\%.

Optimal use of this massive influx of data requires a multi-probe approach, however. The combination of imaging and spectroscopic information can significantly improve constraints on dark energy and modified gravity parameters -- DES-like weak lensing plus DESI-like spectroscopic data (including their cross-correlation) can improve the dark energy figure of merit by a factor of 20-30 over what an individual experiment would be capable of, for example \citep{Kirk:2013gfc}.

As a final point, note that while the current cosmological model may only need 6 parameters to fit a wide range of data, those data in fact require hundreds of nuisance parameters! A particular example is that of bias models, for which there is something of a gap between the rich theoretical literature and actual implementations of bias models in surveys. Using an incorrect bias model can lead to incorrect cosmological inferences, so solutions such as the simple parameterization of Ref.~\citep{2015MNRAS.448.1389C},
\be
b(z) = C + (b_0 - C) / D^\alpha(z),
\ee
which incorporates many other models, may be needed to avoid this.

\subsection{Stress-testing \lcdm{}}
\label{sec:stresstest}
\noindent{\it Discussion session chairs: P. Bull, S. Nadathur \& L. Verde}

In the current era of large cosmological surveys and their associated observational advances, {\it stress-testing} a theory typically means pushing it to its limits and finding the point where it is observed to ``crack''.  Ideally one should devise tests for which {(a)} the expected result is ``pass'' or ``fail'', and {(b)} in the case of a failure, there is enough information to give hints as to what could be a better theory.

For many researchers, \lcdm{} is not strictly considered to be a fundamental physical theory. At a classical level, or at high redshift, \lcdm{} {\it can} be considered a physical theory since $\Lambda$ is either irrelevant or, classically-speaking, not a problem. Indeed, if we define a physical theory as being one for which we can write down a pathology-free Lagrangian description, \lcdm{} certainly qualifies. 
Most would agree that \lcdm{} is an effective model in most regimes though -- an approximation to which many different physical theories might reduce. In its current state, the model encompasses many possible physical theories (e.g. of dark matter and inflation, and even different solutions to the cosmological constant problem), and these are indistinguishable given current observational uncertainties. As such, ruling out (e.g.) a given model of inflation is not a problem for \lcdm{}, as long as other viable inflation models still remain under the \lcdm{} ``umbrella''.

Stress-testing might take on a different meaning in this context. A model may show theoretical inconsistencies (``cracks''), but this is only to be expected for an effective model, as it is not supposed to describe the whole truth. 
Stress-testing may therefore be better understood as finding the limits of where this overarching effective model can provide a good description of the data, including newer data. In performing such stress-tests, we should regard ``\lcdm{}'' as shorthand for the full package of the standard cosmological model -- including assumptions of statistical isotropy and homogeneity, an inflationary origin and Gaussianity of fluctuations, Friedmann-Lema\^{i}tre-Robertson-Walker (FLRW) background, general relativity etc. -- not just $\Lambda$ and cold dark matter.

As such, two different aims may be pursued:
 \begin{enumerate}
 \item Devise tests to distinguish between all of the different theoretical models contained under the \lcdm{} umbrella (e.g. rule in or out specific inflationary models, or specific dark matter models).
 \item Devise tests that may show that the \lcdm{} framework does not work at all (e.g. the Universe is not FLRW, GR is not a good description of gravity on cosmological scales and so on).
 \end{enumerate}

\paragraph{How do we stress-test \lcdm{} in practice?}
The great value of \lcdm{} is that it offers a simple framework to access many possible physical theories observationally. The base \lcdm{} model is powerfully predictive, with little room for maneuver and only a few free parameters, so it is therefore easier to test than more baroque models. Theories beyond \lcdm{} are often less predictive, for example due to the introduction of arbitrary functional degrees of freedom, but \lcdm{} offers a benchmark, a starting point around which one can look for deviations.

Some of the options for practical stress-tests include:
\begin{enumerate}
\item Coming up with alternative theories and comparing them to \lcdm{}. A simple way of doing this is to insert extra parameters (e.g. changing the effective number of neutrino species, the neutrino mass, or the dark energy equation of state) and then to check if the \lcdm{} reference value of the parameter is recovered. This is closely related to model comparison and model selection analyses, which have become popular in cosmology \cite{2008ConPh..49...71T}.
\item Predicting future (as-yet unobserved) data and checking \lcdm{} against the measurements once the new data have been gathered. For example, baryon acoustic oscillations in the galaxy distribution were predicted, and then later observed to match the \lcdm{} prediction. Similarly, \lcdm{} makes definite predictions for what relativistic effects on extremely large scales should look like (see Sect.~\ref{sec:maartens}), and these could be observed in the future.
\item Checking for consistency across different types of observation. For example, consider two or more different datasets and check if they are fitted by a \lcdm{} model with the same parameter values (see Ref.~\cite{Raveri:2015maa} for a recent example).
\item Considering the significance of anomalies/tensions, and whether they can be explained more reasonably by alternative models.
\end{enumerate}

In effect, \lcdm{} has been stress-tested in many different ways for the last 20 years, and has successfully accounted for new data of vastly increased precision -- in other words, precision cosmology is a stress-test, and the \lcdm{} model has survived it so far. But while this has shown impressive self-consistency, it is still dependent on fundamental assumptions such as statistical isotropy and homogeneity, and GR, each of which we would like to test individually. This may not always be possible -- for example, we cannot test the isotropy of CMB fluctuations without assuming Gaussianity and vice versa \cite{Ade:2015hxq}.
 
The ``anomalies'' issue deserves particular attention. Anomalies have also been studied extensively, but when should they be taken seriously as an indication of a failure of the model? It is still an open issue as to how one should decide whether an anomaly indicates new physics or a probable systematic effect (see Sects.~\ref{sec:peiris} and \ref{sec:cmbanomalies}). When using anomalies for stress-testing, the clearest, most useful situation is to have viable alternative models to compare with. For instance, if some anomaly in the data is predicted (or even post-dicted) by an alternative model, that anomaly will be taken more seriously, whereas in the absence of such a model it is more likely to be regarded as a statistical fluke. In the words of Eddington, ``never trust an observation until it is confirmed by theory''.

As an example consider the tension between the CMB-predicted value of $H_0$ within a \lcdm{} model, and the local measurements from supernovae \cite{2014A&A...571A..16P}. In this case, an alternative model such as a non-local gravity model \cite{2015JCAP...04..044D} is of interest because it appears to fit most data as well as \lcdm{}, but also improves the fit for the local $H_0$ measurement. One does need to guard against over-interpreting ``designer'' models that are constructed specifically to solve individual anomalies while reducing to \lcdm{} in all other situations, however.
 
It is also important to protect analyses from confirmation bias. Modern cosmological datasets are extremely complex, often relying on lengthy calibration processes and the subtraction of poorly-understood systematic effects. The vast majority of anomalies are caused by unexpected issues with the data, so if one is observed, it is tempting to first ascribe it to some previously unknown systematic and to continue trying new ways of filtering the data until the anomaly disappears. On the other hand, if the initial data analysis comes out consistent with \lcdm{} expectations, researchers are less likely to be motivated to continue digging for problems that may still be present. This is clearly detrimental to effective stress-testing -- real anomalies risk being spuriously ``corrected'', or missed because of compensating systematic effects.

It is hard to quantify how pervasive confirmation bias is in the cosmological community, but the parameters measured by a variety of experiments have been observed to cluster around the best-fit values determined by certain ``high-profile'' experiments like WMAP \cite{2011arXiv1112.3108C}. This suggests some tendency to work towards an expected result. Still, unexpected results are published reasonably frequently (e.g. Ref.~\cite{2014PhRvL.112x1101A}), and so can at least be studied in more detail (even if many are ultimately found to be spurious). One possible solution is to perform {\it blind analyses}, where the final result of a data analysis pipeline is obscured until the last possible moment to prevent comparison with other results. This technique is widespread in the particle physics community, and has been used in cosmology in the past (e.g. Ref.~\cite{2011ApJ...741..111Q}), but it is (as yet) far from the norm.

The assumption of \lcdm{} is often deeply ingrained in cosmological analyses too. 
For example, SNe Ia data reduction often assumes that \lcdm{} is the correct model in order to marginalize over nuisance parameters, and error bars for BAO measurements are usually determined using \lcdm{} simulations. Angles and redshifts are also often transformed into three dimensional distances assuming a \lcdm{} distance-redshift relation etc. While in practice these assumptions may not introduce quantitatively important biases yet, they might for future data.
 
Finally, one should ask whether we are looking in the right place. In the \lcdm{} model, some observations carry no useful information and are overlooked, while in alternative models they may be of vital importance. An example is the slip relation: at late times, the metric potentials $\Phi$ and $\Psi$ are equivalent in \lcdm{}, so can be used interchangeably. They often differ in modified gravity scenarios however, so observing such a difference would be a smoking gun for a breakdown of \lcdm{}. Redundancy -- observing the same thing in different ways -- is often invoked (and sold to the funding agencies) as a way to control systematics, but can also be used to perform effective null tests of \lcdm{} predictions. A number of tests of fundamental assumptions such as the Copernican Principle and statistical homogeneity that use redundant distance and spatial curvature measurements have been proposed, for example \cite{2008PhRvL.101a1301C, 2008PhRvL.101r1301Z, 2012PhRvD..85d3506C, 2014PhRvD..90b3012S}.

\paragraph{What are the best places to look for problems?}
If stress-testing means looking for problems with \lcdm{}, where are the most promising places to look for such problems? And when, by invoking (unknown) systematic effects, may we actually be papering over the cracks?
 
Consider small-scale structures as an example. We have several indications that the model is not a good fit to the data on these scales (see Sect.~\ref{sec:smallscale}). On the other hand, small scales are most strongly affected by systematic effects, notably the so-called baryonic effects. It is still an open issue as to whether these effects can be understood sufficiently well that small scales can be used in precision cosmology, and whether the observed anomalies are problems of the model or problems of our understanding.

To conclude, we list some promising possibilities for future tests of \lcdm{}:
\begin{itemize}
\item 
 Further testing of current anomalies -- although given that most of them are in the CMB, it will be difficult for future data to shed much more light on them unless they are also present in other regimes (see Sect.~\ref{sec:symm}).
 \item 
 A more realistic short-term prospect is the tension between CMB lensing (Planck) and galaxy lensing (CFHTLenS) measurements \cite{2015PhRvD..91f2001H, 2015PhRvD..92f3517L, 2015arXiv150201589P}. It is potentially possible to quantify systematic effects and assess whether the tension can be fully attributed to them, or if an indication of new physics is more likely.
  \item The bispectrum from Planck CMB observations could be a good test of primordial physics and recombination physics, as well as GR (via the ISW-lensing-induced $f_{\rm NL}$, where ISW stands for Integrated Sachs-Wolfe).
  \item Spectral distortions of the CMB provide what is probably the cleanest way to probe primordial small scales away from messy, hard-to-model astrophysics effects \cite{2012ApJ...758...76C}.
  \item Various consistency tests of FLRW can be performed using only distances and geometry (see Sect.~\ref{sec:maartens}) -- although if deviations from FLRW are small, the data will not be good enough to constrain this for some time.
\item Large-scale relativistic effects on clustering, as a way of testing GR (see Sects.~\ref{sec:mgde} and \ref{sec:maartens}).
\end{itemize}

\subsection{Model selection vs. parameterizations: what do we expect to learn?}
\label{sec:modelsvsparams}
\noindent{\it Discussion session chairs: E. M{\"o}rtsell, V. Salzano \& I. Tereno}

In this section, we discuss the relative merits of two approaches to testing dark energy theories: model selection versus parameterizations.
Models (of course, containing parameters) are defined as being the more theoretically-fundamental entities (e.g. derivable from a Lagrangian), while parameterizations
are more phenomenological in nature. An example of the former might be
a bimetric gravity model \cite{Hassan:2011zd}, while the latter could be
to parameterize the dark energy equation of state as \cite{Chevallier:2000qy}
\begin{equation}\label{eq:CP}
w(a)=w_0 + (1-a)w_a.
\end{equation}
In this parameterization, $a$ denotes the scale factor of the Universe, $w_0$ the equation of state today, and $w_0 + w_a$ the equation of state in the far past.

We agree that parameterizations are valuable working tools, e.g. in the design phase of instruments, surveys etc. A potential drawback is that the output from such considerations may depend on the parameterization used, however. The parameterization of Eq.~\ref{eq:CP} has been used extensively in the literature, but since it is not ``God-given,'' one should be careful about how it affects the outcome of (e.g.) survey optimizations. Alternatives do exist, both in terms of parameterizations \cite{Hannestad:2004cb} and other approaches \cite{Huterer:2002hy}, and it is
desirable that a multitude of methods should be considered and compared.

If we are not content to only provide a compact description of observational phenomena, but aim to also {\it explain} the data after they are taken, we need to make the effort
to investigate and test theoretically-justified models. This can be illustrated by the following example. Imagine that, using some dataset, we fit the value for a constant dark energy equation of state and find a best-fit value of $w=-0.5$. What would we make of such a result? There is no fundamental {\it physical} model (known to us) that has a constant dark energy equation of state different from $w=-1$.

\paragraph{Information criteria and Bayesian evidence.} The viability of different models compared to each other can be quantified
using various information criteria, many of which are variations on a theme -- i.e. the $\chi^2$-value of the best-fit plus some terms of a different nature, depending on the number of data points, parameters of the model, and so on \cite{schwarz1978,1100705}. Usually such criteria penalize models with a large number of free parameters.

An approach generally considered more reliable is to compare models using the Bayesian evidence. This method also has potential pitfalls and/or open questions, however \cite{2013JCAP...08..036N}. How can we understand a higher evidence in terms of ``this model is really better than others''?  What is the weight of the prior probabilities? The importance of the latter question can be illustrated like so. Imagine that it has been shown that {\it any} constant dark energy equation of state, $w$, can constitute a viable model (although in an extremely contrived manner). Using some dataset, we then investigate the likelihood for different constant $w$, and find a best-fit value of $w=-0.5$, with the \lcdm{} value ($w = -1$) being $5\sigma$ off. Whether this result will make you renounce \lcdm{} or not will then depend on the prior probabilities you assign to different values of constant $w$. One could try to get around this by assigning equal probabilities to all values of $w$ and claiming to ``let the data decide'', but when integrating the likelihood over the prior probabilities, the result will still depend on whether you assign equal prior probabilities to equally large intervals of $w$, or equally large intervals of the logarithm of $w$, or $w^2$, or $\sin (w)$, or any other function. Some prior must always be chosen, and the choice may be unavoidably subjective.

\begin{figure}
\centering
\includegraphics[width=100mm]{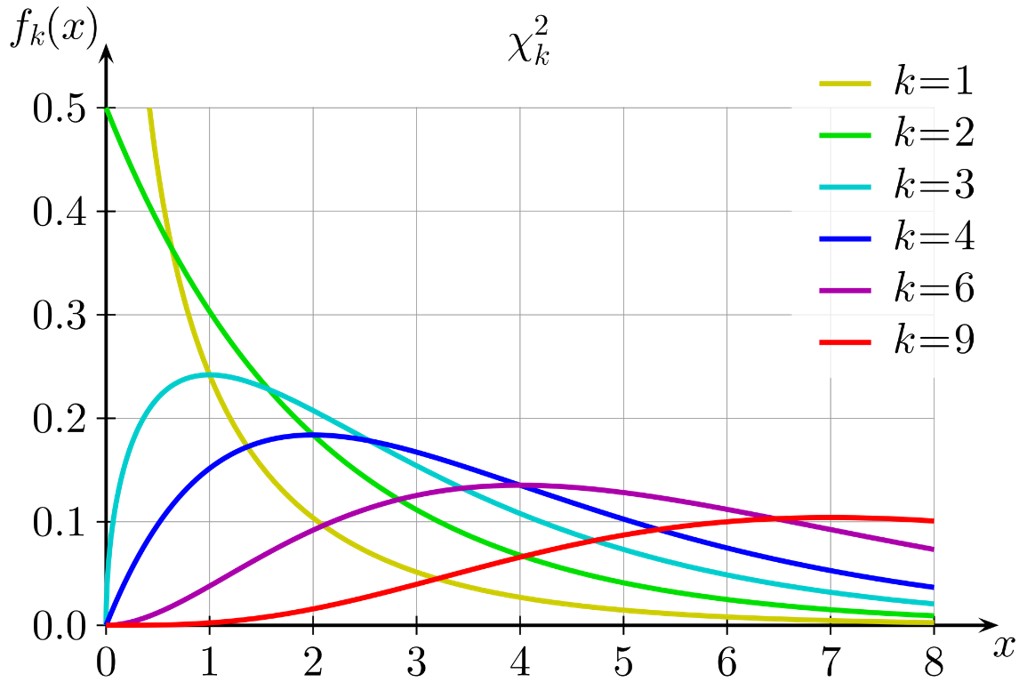}
\caption{The $\chi^2$-distribution [$f_k(x)$] as a function of best-fit $\chi^2$ value (denoted $x$), with $k$ denoting the number of degrees of freedom. (Wikimedia Commons)}\label{fig:chi2dist}
\end{figure}

\paragraph{When is a model ruled out?} It is one thing to try to rank different models based on how they compare in terms of their fit to data, numbers of parameters etc., but quite another to rule out a model in absolute terms. Specifically, one could argue that in the previous case discussed, if
we again find a best-fit value of $w=-0.5$, and the \lcdm{} value
($w = -1$) is $10\sigma$ off, this should necessarily mean that \lcdm{} is ruled out, simply because it should be a bad fit to data. This may not be the case, however.
For Gaussian random variables, the best-fit $\chi^2$ values will be distributed
according to the $\chi^2$-distribution, $f_k(\chi^2)$ (see Fig.~\ref{fig:chi2dist}). Here, $k$ denotes the number of degrees of freedom -- basically the number of data points minus the number of free parameters in the fit. As expected, the peak of the $\chi^2$-distribution occurs when the best-fit $\chi^2$ is roughly similar to the number of degrees of freedom. It is also evident that the distributions get flatter as $k$ grows however (which of course has to be the case, since the integrated probability should be unity for each value of $k$). This flattening has the interesting effect of making values of $\chi^2$ at large distances from the maximum of $f_k(\chi^2)$ less and less unlikely as $k$ grows -- that is, as our dataset grows.
For 1,000 data points, we will get a value that is $10\sigma$
off in 1.5\% of the realizations. For 10,000 data points, this grows to
25\%. In other words, while we may have ruled out $w=-1$ with $10\sigma$ confidence compared to the case of $w=-0.5$ using 1,000 data points, the cosmological constant still provided a decent fit to the data in itself.

Only if you are a Bayesian, assigning a flat linear probability in $w$, you can safely rule out \lcdm{} in this case. If you instead choose to renounce all models with constant $w \neq -1$ on the grounds of them being too contrived, you would interpret the observational results as showing a fairly plausible ($98.5\%$-level) deviation from the expected $w=-1$ model. If we ever come to a similar situation with real data, it will be interesting to see which of the two people prefer to give up on: their favorite model, or their favorite statistical interpretation.

It will therefore not be easy for any test to disprove \lcdm{}, at least at the level of the background expansion (and without going into questions on the complexities of systematic errors and so on). This forces us to study perturbations around the background, which often requires the use of specific physical models.

\paragraph{When to stop testing.} While the \lcdm{} model currently fits the available data quite well, we still feel the need for additional confirmation or surprises. Imagine that we are in the same situation in 20 years; the model works well, perhaps with a few things sticking out of the picture. When do we stop asking for more confirmation? There is a human tendency to believe that ``we are almost there,'' but the future might show something completely different. At least this is something to hope for, unless we are content with simply confirming a standard model to higher and higher precision.

\begin{figure}[h]
{\centering
\includegraphics[width=\columnwidth]{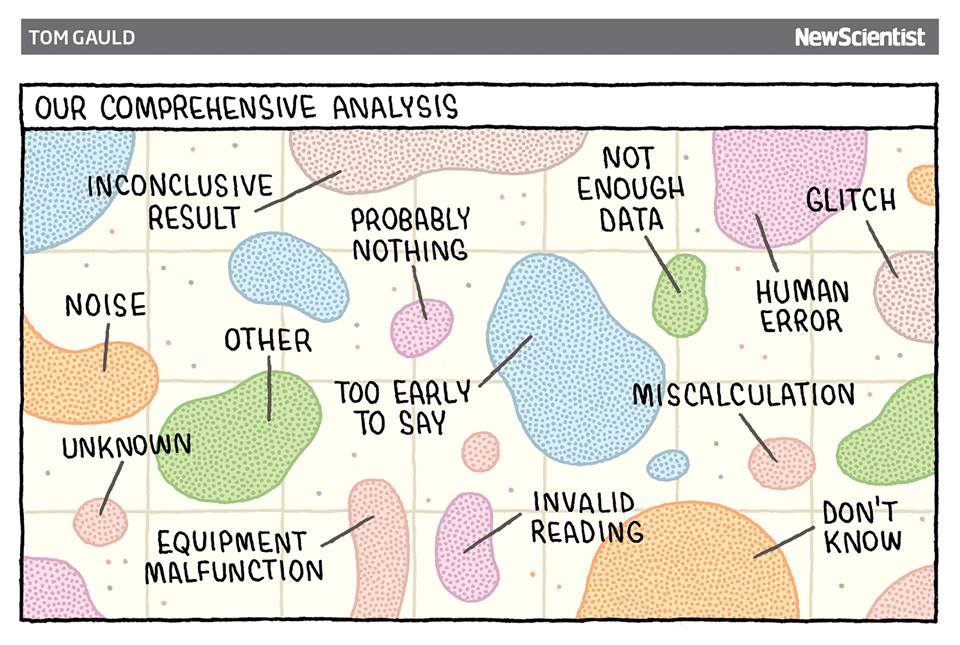}
\caption{The space of beyond-\lcdm{} models. (Tom Gauld, reproduced with permission.)}\label{fig:companalysis}}
\end{figure}

As for which future data that has the power to do this, we honestly don't know, but again, we can at least hope for something new. Remember that baryon acoustic oscillations, now perhaps the most powerful way to map the expansion history of the Universe, have only been seriously discussed as an important cosmological tool in
the new millennium. Thus, we summarize with Fig.~\ref{fig:companalysis}$\ldots$

\section{Cosmological constant: the $\Lambda$ in \lcdm{}}
\label{sec:cc}

The {\it cosmological constant problem} is frequently described as the biggest problem in fundamental physics; a fine-tuning of the highest degree; an embarrassment to the otherwise substantial progress in understanding the basic laws of nature. The observational necessity of introducing a $\Lambda$-like term into the standard cosmological model has inspired a tremendous variety of creative solutions to both the {\it old} and {\it new} cosmological constant problems, ranging from complex extensions and restructurings of cherished theories such as GR, to an almost philosophical shrugging of shoulders in the guise of anthropic explanations. The common thread is that no firm proposal has yet inspired even a glimmer of acceptance (or even particular excitement) in the cosmology and high-energy physics communities, and so the feeling is very much that we are still stuck at square one as far as finding explanations is concerned.

Is it too pessimistic to call this a crisis in physics? In this section, we examine whether we are justified in feeling so vexed by the cosmological constant problems, and ask whether there are any promising leads that might help in understanding them. We then take stock of the theoretical landscape, asking whether alternatives to the {\it ad hoc} addition of a $\Lambda$ term are necessary and, if so, whether the industries that produce them are on the right track.

\subsection{Cosmology and the cosmological constant problem}
\noindent{\it Plenary speaker: C. Burgess*}
\label{sec:burgess}

We live at a time of contradictory messages about how successfully we understand gravity. General relativity seems to work well in the Earth's immediate neighborhood, but arguments abound that it needs modification at very small and/or very large distances. Here we try to put this discussion into the broader context of similar situations in other areas of physics, and summarize some of the lessons that our good understanding of gravity in the Solar System has for proponents for its modification over very long and very short distances. The main message is mixed: On one hand, cosmology seems to like features (like light scalars and small vacuum energies) that are not generic to the long-wavelength limit of fundamental theories. This is a crucial clue that should not be ignored. On the other hand, although some mechanisms to contrive light scalars are known, so far there are none that everyone agrees could incorporate small vacuum energies (even in principle), making this a clue that is very difficult to use -- at least until a convincing example is found.

\paragraph{Naturalness: a central clue.} The era of modern cosmology can be characterized by the fact that we now have an enormous amount of precision data provided by various types of cosmological probes. One important task of cosmologists is therefore to find a model that describes the data, but many such models exist. The immediate question is therefore how to choose amongst the models. Arguably, it is now widely accepted in the cosmological community that the models should ideally satisfy two criteria: {\it simplicity}, and {\it naturalness}. With naturalness, we demand that our description arises as the low-energy limit of the theory that describes the rest of physics. Naturalness is a story of hope and change as, for example, light scalar fields that seem to provide good descriptions of various phenomena in cosmology are unnatural \cite{1998PhRvD..58b3503F}. This has provided opportunities for developing interesting ideas however, such as natural inflation \cite{1983veu..conf..251S, 1990PhRvL..65.3233F} and exponential potentials \cite{1987PhLB..185..341H}. Small vacuum energies also seem to be favored by observations, but they are also unnatural, while similarly providing interesting opportunities for cosmology.

\paragraph{Against gravitational exceptionalism: quantifying quantum effects.} Quantum field theory is a precision science; we can compute theoretical values of some quantities in quantum electrodynamics (QED) with $\sim10$ digits of precision, for example, and measure the actual values through experiments with similar degrees of precision. In addition to the experimental errors on the values, the renormalizability of QED, as an important part of its calculability, underpins the {\it theoretical} errors. General relativity is also a precision science, as one can compute various quantities for Solar System tests and binary pulsars theoretically to very high precision, and compare them to experimental measurements.

This comparison is meaningless if the size of quantum effects is unknown in the theory, however. Inferences about inflation from observations also rely on quantifying theoretical errors. But can we quantify the theoretical errors, coming from quantum effects, for general relativity? This is an important question, and has proved to be difficult to tackle. One can compute the amplitude for the graviton-graviton scattering about a static background in GR, but as soon as we try to include quantum loops, the loop integrals diverge, and higher-order loops diverge more and more \cite{1979grec.conf..790W}. Contrary to QED, all divergences cannot be absorbed into the parameters of the theory, which is Newton's constant for GR. The divergences can be absorbed only if GR is just the first term in a derivative expansion that includes all possible local interactions allowed by symmetries~\cite{Burgess:2003jk}. The question now, in an effective field theory (EFT) framework, would be how to interpret the non-GR terms in the action. One could therefore consider GR as a low-energy EFT that would be obtained if we integrated out a collection of particles with masses higher than a particular energy scale. If this EFT interpretation is valid, the low- and high-energy regimes should be treated as being decoupled.

\paragraph{Naturalness problems: the electroweak hierarchy vs. the cosmological constant problem.} Naturalness and hierarchy problems are not specific to the cosmological constant (CC). Such problems already exist -- although not as severely as for the CC -- in the electroweak (EW) sector of the Standard Model (SM) of particle physics. The SM is the most general renormalizable theory that is possible given its particular particle content. Ideas for what lies beyond the SM are largely driven by {\it technical naturalness} arguments \cite{2013PDU.....2....1B}, and are motivated by the belief that the SM is an effective field theory~\cite{Burgess:2007zi}.

An effective theory can be defined at many scales, however. Let us assume that the SM is valid up to a scale $M\sim 10^{12}\,{\rm GeV}$, for example. As the masses of the particles in the QED sector of the SM are set through their interactions with the Higgs field, a mass of $m^2\sim 10^{2}\,{\rm GeV}$ is required for the Higgs to obtain masses of a similar order for the other QED particles, as measured experimentally. The measured mass $m$ is a combination of a ``bare" mass $m_0$ and quantum corrections set by the scale $M$: $m^2\approx m_0^2+M^2$. Therefore, one will need a cancellation to $\sim20$ decimal places. Although this enormous tuning is not a theoretical inconsistency, it signals the existence of new physics beyond the SM, and at high energies. This latter possibility has led to various proposals for solving the tuning puzzle of the hierarchy problem, including the theories of composite Higgs, electroweak supersymmetry, and extra dimensional extensions of the SM (see e.g. Ref.~\cite{Burgess:2007zi}).

In addition to the Higgs mass, the SM has another unnatural parameter; that brings us to the problem of the cosmological constant, which is considerably more unnatural than the EW hierarchy. The measured value of the CC is quite small ($\sim3\times 10^{-3}\,{\rm GeV}$), while its theoretically natural value from the SM (in an EFT framework) is several orders of magnitude larger. In our modern picture of particle physics, there is no unique {\it classical} theory; there are instead many {\it effective} theories. EFT calculations of the SM show that the CC receives contributions from all massive particles in the SM, proportional to the fourth power of their masses~\cite{2012CRPhy..13..566M}. Ignoring neutrinos, the smallest contribution comes from electrons, from which the CC should receive a contribution of $m_\mathrm{e}^4$, which is $32$ orders larger than the observed value. That is, even if we ignore all the particles heavier than the electron, we seem to need a cancellation to at least $32$ decimal places already. This increases to $55$ orders of magnitude if we include all the SM particles. A pertinent question now is: can we change the properties of {\it low-energy} particles, like the electron, so that their zero-point energy does not gravitate, {\it even though quantum effects do gravitate in atoms}? One would then need only to change gravity to explain the value of the CC, and not any of the other well-tested properties of the SM particles.

\paragraph{What must a solution do?} The CC problems seem to be severe, and there is as yet no consensus in the theoretical physics community on what the solution would be. But there are at least three properties that one would expect from a solution:
\begin{enumerate}
 \item Going beyond the classical approximation seems unavoidable, as it is hard to beat a cosmological constant at the classical level.
 \item The solution must be applied at energies larger than the CC scale, as quantum effects are not a problem until particles with masses heavier than the vacuum energy are included.
 \item The solution should be ``harmless,'' i.e. it must not affect well-tested properties of particle physics or cosmology.
\end{enumerate}

\paragraph{Popular proposals: roads well traveled.} The literature is full of proposals for how to tackle the CC problem, although none has been accepted by the majority of the particle physics and cosmology communities. Listing all the proposed ways out is beyond the scope of this article (see e.g. Ref.~\cite{Joyce:2014kja}, instead), although some of them will be discussed briefly in some of the following sections. Such proposals cover a vast spectrum of possibilities, ranging from: {\it denial}, with the hope that the problem is fully harmless in all areas of interest in particle physics and cosmology; to {\it anthropic arguments}, where multiverse explanations are employed for why naturalness might not be needed; and {\it modified gravity} proposals, where one tries to, for example, screen the CC in a {\it degravitation} framework using (e.g.) the graviton mass or non-localities (i.e. to solve the {\it old} CC problem) or, instead, tries to generate the cosmic acceleration {\it if} the CC is not present (i.e. to solve the {\it new} CC problem) -- see Sect.~\ref{sec:ccdisaster} for the definition of the old and new CC problems. Some of these proposals, especially in the context of modified theories of gravity, are quite promising, but all still suffer from various difficulties that need more investigation.

\paragraph{A way forward.} We end our discussion of existing proposals for solving the CC problem with a closer look into one interesting class of solutions within the framework of extra dimensions. The CC problem comes from the fact that the Einstein equations make a Lorentz-invariant vacuum energy (which is generally large) an obstruction to a close-to-flat spacetime (which we see around us). There is however a loophole in this statement: it need not be true if there are more than four dimensions~\cite{ArkaniHamed:2000eg, Giddings:2001yu,Carroll:2003db,Aghababaie:2003wz}. The reason is that 1) extra dimensions need not be Lorentz invariant, and 2) vacuum energy might curve extra dimensions, rather than the ones we see in cosmology.

Now let us imagine a {\it brane-world} picture: we are trapped on a four-dimensional brane in a bulk with six (or more) dimensions. Notice that in this picture, particle physics remains four-dimensional, and only gravitational response sees the extra dimensions. We should now re-ask the CC question in this context: What choices give near-flat geometries? Sufficient conditions for that are to assume supersymmetry in the extra dimensions, and the absence of coupling of the brane to a specific bulk field~\cite{Tolley:2005nu}; the latter is currently the biggest worry. Supersymmetry in the bulk helps because it forbids a higher-dimensional cosmological constant. More generally, we assume that at least one supersymmetry is unbroken {\it locally} everywhere in the extra dimensions, but breaks {\it globally} once all branes are viewed together (assuming that there are more than one brane in the bulk). In this case, the shortest wavelength that knows that supersymmetry is broken is the size of the extra dimensions, i.e. $\lambda \sim 1/r^4$. This does not mean that particle physics should look supersymmetric; only gravity need be.

In order for this proposal to work, it is required that:
\begin{itemize}
 \item The radius, $r$, is as large as microns (this is currently in agreement with experimental tests if the extra dimensions are smaller than 45$\mu$m, and particles are stuck on the branes~\cite{Hoyle:2000cv});
 \item At most two dimensions can be this large, otherwise the high-dimensional Planck scale must be too low to get the four-dimensional Planck scale right. Remarkably, this is the same size as needed by the extra-dimensional solutions to the EW hierarchy problem~\cite{ArkaniHamed:1998rs};
 \item One must include the {\it back-reaction} of the branes on the extra dimensions. The extra-dimensional curvature cancels the brane tension against the four-dimensional vacuum energy~\cite{Aghababaie:2003wz}; this is hard to do, and is why the mechanism had not been found earlier.
\end{itemize} 
We conclude here that a solution to the cosmological constant problem using supersymmetric large extra dimensions (SLED) may yet exist, though work is ongoing to see if that is the case.

\subsection{Theoretical foundations of the cosmological constant problem}
\label{sec:ccdisaster}
\noindent\textit{Discussion session chairs: Y. Akrami, R. Pasechnik \& M. von Strauss}

While observers may feel comfortable adding a CC as an extra phenomenological parameter, the small and uniformly positive CC raises a separate series of theoretical problems and conceptual dissatisfactions within the quantum field theory (QFT) paradigm that are very difficult to address. Explaining them remains one of the major unsolved problems of theoretical physics \cite{Weinberg:1988cp}. In this section, we discuss a number of theoretical perspectives on the origin and significance of the cosmological constant problem (or problems), and how it relates to problems in QFT.

\paragraph{Fixing a value for the CC.} Starting from a renormalizable quantum field theory, one can simply fix a value for the CC at some arbitrary scale. The major problem, however, is to have control over various vacuum condensate contributions to this, as well as the renormalization group running (with an account of intermediate thresholds), without having a real high energy theory of gravity. The real issue therefore lies in obtaining a quantum theory of gravity, which would potentially provide a natural answer to why we observe a small positive CC, as well as unifying all four of the fundamental interactions in nature. A majority of scientists believe that the CC problem is real, but that it is not clear how to resolve it until we better understand quantum gravity. Certain aspects of the CC problem can be addressed and potentially resolved even before such a theory has been constructed, however.

One often considers two versions of the CC problem: the strong, or ``old'', CC problem (why $\Lambda$ is small and positive) and the weak, or ``new'', CC problem (why $\Lambda$ is non-zero and exists at all). A consistent theoretical approach should address both of these problems at the same time of course, although if we can understand why $\Lambda$ is small then we will likely also understand (or at least not question) its value. In other words, if we have a mechanism to make the observable $\Lambda$ small, then its precise value will probably not be surprising, but will either come out with a natural relation to other fundamental parameters, or simply as a measurement to be accepted.

\paragraph{UV/IR completions and naturalness.} A useful theoretical tool for discriminating between various theories is the ``naturalness'' argument (see Sect.~\ref{sec:burgess}), favoring theories that are technically natural, \`a la 't Hooft \cite{tHooft:1979bh}. Naturalness arguments can also be misleading however, so one should not be too locked in to that perspective when searching for resolutions.\footnote{c.f. the remarkable similarity in the angular size of the Moon and Sun as seen from Earth, which allows total solar eclipses to occur. This apparently ``fine-tuned'' situation is generally seen as a mere coincidence, rather than as an unnatural occurrence in need of explanation.} In any case, if we take naturalness seriously, we have to identify whether the CC problem is really a problem with unknown infrared (IR) or ultraviolet (UV) dynamics, or perhaps even a combined IR/UV problem, and to clearly define what IR and UV scales actually mean in this context.

The UV completion of the standard model (SM) of particle physics still appears to be far from complete, as it does not account for gravity and dark matter. Additionally, accelerator experiments (e.g. at the Large Hadron Collider, LHC) have only probed physics over a relatively small window of energies, very far from the Planck scale. IR dynamics at large distances (such as effects of graviton condensation at the horizon scale, or the existence of cosmological Yang-Mills or scalar fields with special properties) may not be fully understood either. Intuitively, an IR or long-range problem arises due to the usual locality, factorization, or decoupling principles that separate dynamics at very distinct scales. AdS/CFT~\cite{Maldacena:1997re} analogies also suggest an intricate interplay/duality between UV and IR physics however, indicating that a full resolution to the CC problem may involve a modification at both ends.

Deep UV or IR limits may also be important in themselves due to non-perturbative vacuum effects that are not fully accounted for in the standard approaches to the CC problem. At the moment it is not entirely clear if a consistent resolution lies within the gravity sector (e.g. through a modified gravity approach such as degravitation~\cite{ArkaniHamed:2002fu} or partially massless gravity~\cite{Schmidt-May:2015vnx}), the particle sector, or an intricate mixture of both. All possible solutions are welcome, but any good theory must make sense of physics at all (IR and UV) scales, or at least within its effective range of validity. At the same time, consistent theories should make verifiable predictions so that observers can test them, with the basic aim of confirming or excluding the models without having to care too much about consistency issues at the fundamental level.

\paragraph{Coincidence problem.} Another related problem is the coincidence, or ``why now'' problem, which concerns the fact that the CC, though rather small in absolute value, strongly dominates the evolution of the Universe today. Its value is also believed to be important for the structure formation epoch. 
The coincidence problem is more serious if the time evolution of $\Lambda$ becomes important. In this case it could potentially be addressed by treating the CC as an attractor solution, or by invoking the anthropic principle. If $\Lambda$ is really constant at all times then this is still a problem, but not as urgent or important as the other CC problems. For a truly non-dynamical $\Lambda$, we can probably only resort to an anthropic answer to this problem at the moment, which may not be satisfactory for theorists looking for reasonable resolutions within the conventional QFT framework. 

\paragraph{Vacuum catastrophe.} What are some promising resolutions of the CC problem that do not invoke extra exotic degrees of freedom or phenomena beyond traditional QFT? One direction is towards a better understanding of quantum dynamics of the ground state of the Universe, its evolution in time, and its possible relation to both the early (inflation) and late-time (CC) acceleration. There are no robust predictions for the CC value within the standard QFT paradigm that account for all existing vacuum contributions from quantum field dynamics (i.e. condensates) at various scales -- ranging from the quantum qravity scale, $M_{\rm Pl}\simeq 1.2\cdot 10^{19}$ GeV, to the quantum chromodynamics (QCD) confinement scale, $M_{\rm QCD}\simeq 0.1$ GeV. The well-studied quark-gluon and Higgs condensates alone (responsible for chiral and gauge symmetry-breaking in the SM respectively) have contributions to the ground state energy of the Universe that far exceed the observed absolute CC value today \cite{2012CRPhy..13..566M}. Regardless of how the observed CC is explained, these huge quantum vacuum contributions must be eliminated. Any consistent solution of this problem, known as the ``vacuum catastrophe'', must rely on a compensation of short-distance vacuum fluctuations by the ground state density of the Universe to many tens of decimal digits \cite{Polchinski:2006gy}. A dynamical mechanism for such gross cancellations (without a major fine-tuning) is not known, and should be regarded as a new physical phenomenon anyway \cite{Copeland:2006wr,Dolgov:2003fw, Dolgov:2003at,Pasechnik:2013poa}.

One hope is to realize a consistent cancellation mechanism of weakly-coupled (perturbative) vacua in supersymmetry or supergravity theories. A cancellation of the strongly-coupled non-perturbative quark-gluon contribution would require a dynamical understanding of the QCD confinement mechanism \cite{Pasechnik:2013poa}, or the existence of extra cosmological Yang-Mills fields \cite{Pasechnik:2013sga} however. A deeper understanding of these mechanisms could potentially also address the nature of dark energy within the standard QFT paradigm.

There is still no real consensus in the community on what the resolution to the CC problem is or should be. This is quite an unusual situation in physics, where traditionally there has tended to be a consensus on at least a general direction to look in. Any possible resolution here really needs other testable predictions in order to convince a majority of physicists, in addition to addressing the CC problem.

\subsection{Can we accept $\Lambda$ at face value?}
\label{sec:justlambda}
\noindent{\it Discussion session chairs: J. Enander, L. Heisenberg \& I. D. Saltas}

While \lcdm{} is highly successful as an {\it effective} cosmological model, it suffers from both theoretical and observational drawbacks. Theoretical, since the observed value of the cosmological constant is hard to reconcile with technical naturalness arguments in quantum field theory (see also Sect.~\ref{sec:ccdisaster}). Observational, since dark energy has so far only been inferred from its gravitational effects, and because it is hard to distinguish between different dark energy components (although the cosmological constant is favorable due to its simplicity). In this section we confront the $\Lambda$ in \lcdm{} from both perspectives to show ways in which its ad hoc nature is considered unsatisfactory, and how this motivates the desire for a deeper physical understanding of the CC.

\paragraph{Technical naturalness.} The Einstein-Hilbert action is invariant under general coordinate transformations, and a cosmological term can be included in this action without breaking this symmetry. In fact, it {\it must} be included from an effective field theory point of view. This cosmological term is the bare cosmological constant. At the classical level, it can just be treated as a free parameter that can be arbitrarily fixed to any value, and in particular to the value required by cosmological observations. If one decides to stick to the usual interpretation of the cosmological constant as the term representing the energy of the vacuum coming from all different matter fields present in the theory however, one also has to consider the quantum corrections on top of its bare value. This is the point at which the problem of technical naturalness becomes alive \cite{Burgess:2013ara, 2012CRPhy..13..566M,Padilla:2015aaa} (see also Sects.~\ref{sec:burgess} and~\ref{sec:ccdisaster}).

Probably the best way of understanding the issue of technical naturalness is the Wilsonian approach to calculating quantum corrections. What Wilson teaches us is that quantum corrections should be successively calculated as one continuously moves the cut-off scale of the theory from lower to higher energies. This gives rise to the concept of the Wilsonian 
effective action. In this context, one can immediately see the problem by looking at the corrections to the bare cosmological constant coming from the lightest of the Standard Model particles, like the electron; these are already enough to render the renormalization procedure for $\Lambda$ unstable. In this sense, the observed value of the cosmological constant is a problem of low energy physics, which cannot be cured by a high energy completion. This is therefore an indication that the gravitational properties of the vacuum are not clearly understood, even in the regime where other interactions have been mapped out in great detail.

\paragraph{Symmetry breaking and the history of the vacuum.} Another important point is that the vacuum has a history \cite{2015arXiv150204702B} (see also Sect.~\ref{sec:ccdisaster}). The Universe has undergone a series of symmetry breakings throughout its history, each of which changes the ground state. All of these transitions contribute to the vacuum energy today. There is therefore a whole chain of contributions at different energy levels that one has to take into account. From this perspective, the vacuum energy today is indeed an infrared problem, but taking the entire evolution of the Universe into account, it concerns all energies. In this sense, explaining $\Lambda$ requires an approach that incorporates both low- and high-energy physics. A possible mixing of UV and IR physics might be the ultimate road for tackling the problems associated to the vacuum energy. Due to its nature, it definitely concerns both particle physics and gravity.

Note that while the prediction of the vacuum energy is performed mostly on flat spacetimes, there are indications that the result does not change on curved backgrounds \cite{2012CRPhy..13..566M}. However, it might be the case that our tools for computing the vacuum energy on curved spacetimes are inadequate. 

\paragraph{Classical problems with $\Lambda$.} Besides the quantum problems of $\Lambda$, there are still classical conundrums. What does it mean geometrically? How does one relate the value of Newton's constant and $\Lambda$? Are they intertwined somehow? It is worth mentioning at this point that the various modified gravity models suggested in the literature cannot provide any convincing solution, since the fundamental problem of vacuum energy remains. On the other hand, alternative suggestions for screening out the cosmological constant from the classical equations, such as unimodular gravity (see Sect.~\ref{sec:gravcosmo}), do not appear to provide an explanation so far either \cite{Padilla:2014yea}. 

From an observational perspective, we only really have constraints on $\Lambda$ at low redshifts so far. To argue that dark energy must stem from a cosmological constant is thus a rather weak statement, since higher-redshift constraints are still too weak to reveal its possible time evolution. Furthermore, while there is an intense effort towards a direct detection of dark matter (see Sect.~\ref{sec:dmdetection}), there are as yet no compelling proposals for laboratory measurements of $\Lambda$. It will therefore only be probed on galaxy cluster scales and beyond for the foreseeable future.

\paragraph{\lcdm{} as an effective description.} The consensus of this discussion session was towards accepting the \lcdm{} model only as an effective description of some unknown underlying model that can originate from a fundamentally different gravitational theory or, equally, from general relativity but with a better understanding of the vacuum energy and its computation. Without a deep understanding of all the above-mentioned problems, it is unhelpful to insist that $\Lambda$ represents the vacuum energy as calculated using the standard techniques of quantum field theory. 

Data currently allows for a dynamical dark energy component while also being consistent with the cosmological constant scenario. Since the latter is the simplest model, alternative cosmological models -- such as those presented in modified gravity scenarios -- that do not solve the problem of the reconciliation of gravity and the quantum properties of the vacuum are hard to motivate.

A final judgment on the meaning of $\Lambda$ must incorporate both UV and IR physics, together with an understanding of how the vacuum has changed during different periods of the Universe's history. Without these, $\Lambda$ will continue to present one of the most intriguing challenges in cosmology.

\subsection{Alternatives to $\Lambda$: what works and what doesn't}
\label{sec:lcdmalt}
\noindent\textit{Discussion session chairs: J. Adamek, J. Noller \& A. R. Solomon}

As discussed above, the cosmological constant problem is the elephant in the room in cosmology, defying our understanding of low-energy physics in a way that cannot be rectified purely by modifying high-energy physics. Radical departures from long-held principles may therefore be required to solve it. Motivated by the CC problem (and at any rate, in the spirit of good scientific practice), it is therefore interesting to consider possible alternatives to \lcdm{}. In this section we consider what types of departure offer the most promise, as identified by the participants of the discussion session. Our focus is on significant departures from the standard cosmological model, as opposed to small modifications that can approach arbitrarily close to it.

\paragraph{Assessment of alternative models.} Building a successful theory is difficult, taking a great deal of time and effort. While the need to replicate the robust observational successes of \lcdm{} is a daunting obstacle, we suggest that alternatives should be considered even if they do not perform better than \lcdm{} phenomenologically. The assessment of an alternative model can also be obstructed by difficulties in making robust predictions, for example when one has to deal with a strongly-coupled theory. We should remain maximally open-minded when faced with such problems, discarding alternatives only if they perform considerably worse in explaining the observed data.

\paragraph{Fundamental principles.} Several fundamental principles underpin the \lcdm{} model. We will examine the significance of each of them in turn:
\begin{itemize}
\item \textit{Locality}
is a feature of the standard model of particle physics as well as most theories of gravity (including general relativity). The study of non-local theories (see Sect.~\ref{sec:gravcosmo}) is a worthwhile
exercise however, as it is not immediately clear that such theories are necessarily inconsistent or pathological.
\item \textit{Lorentz invariance} is again a property of the standard model of particle
physics, but it is (spontaneously) broken in cosmology where the cosmic microwave background defines a preferred frame. It therefore seems that this principle can quite comfortably be relaxed in the gravitational sector, especially since invariance under local Lorentz transformations can always be restored in a theory without changing the dynamics.
\item The existence of \textit{ghosts}, on the other hand, is considered unacceptable. Stability is therefore a minimal requirement for any viable alternative to \lcdm{}. In fact, this is one of the most useful selection criteria when constructing modified gravity theories (or theories in general).
\item Finally, the \textit{action principle} lies at the heart of almost all physical theories.
While giving up on the action principle certainly calls into question whether a robust microphysical description of the theory in terms of path integrals exists, theories that are constructed in a different way (e.g. directly at the level of equations of motion) are still considered interesting to explore, at least as far as phenomenological models are concerned.
\end{itemize}

Of these options, the front-runners (principles that were deemed dispensable by a significant fraction of participants in the discussion) that emerged were locality and Lorentz invariance, as well as the action principle.

\section{Gravity on the largest scales}
\label{sec:gravity}

While there is almost universal acceptance that we do not fully understand gravity, it remains unclear exactly where (and how) our understanding breaks down. GR is in startling agreement with experiment and observation on Solar System and pulsar binary scales, and so far appears to work flawlessly even in the vastly different arena of cosmology -- if we are content to add a couple of somewhat mysterious ``dark'' components to the stress-energy tensor. The question of where to look for problems with GR, and what those problems might look like, has preoccupied many theoretical cosmologists in recent years, not least because of possible connections with dark energy.

In this section we give an overview of what cosmological observations can tell us about the validity of general relativity, and examine how this relates to efforts to understand cosmic acceleration. We also attempt to predict how recent developments in modified gravity theory will pan out in the near future, and ask what motivates their continued development -- are they really prime candidates to explain the big problems in cosmology, or just elaborate foils for increasingly complex tests of GR?

\subsection{Gravity in cosmology}
\noindent{\it Plenary speaker: P.~G. Ferreira*}
\label{sec:gravcosmo}

Gravity is central to our picture of the Universe. Without Einstein's theory of general relativity at hand, we would arguably not have come this far in understanding the make-up and evolution of the cosmos. The non-linear but highly constrained structure of GR, its wide regime of validity and applicability, and its remarkable predictive power, have enabled us to construct a detailed model of the Universe, based on a very few parameters, which very accurately describes its evolution from early after its birth until the present time. On the other hand, cosmology has provided an enormous laboratory to test our theory of gravity on various scales and in a wide range of environments. This symbiosis between the cosmological model and gravitational theory has proved to be highly fruitful, and seems likely to strengthen in coming years. In this section we focus on two aspects of this beneficial relationship, and ask: 1) can gravity solve the most difficult problem in cosmology, i.e. the problem with $\Lambda$? (see Sect.~\ref{sec:cc}), and 2) can we use cosmology to (further) constrain gravity?

\paragraph{Modifying gravity.} There are two main ways to systematically arrive at GR. The {\it traditional} description of GR is based on a geometrical picture: GR is a theory of spacetime and its metric. Assuming gravity is described solely by one metric tensor, one arrives uniquely at GR after adding a few assumptions about the structure of the equations of motion for the metric (see the discussion about the Lovelock's theorem in Sect.~\ref{sec:mgdev}). Our {\it modern} description of GR is, on the other hand, free from geometrical concepts, and is instead based on the framework of theories of classical and quantum fields. In this picture, GR is the {\it unique} theory of massless, spin-2 particles, if one includes their self-interactions (Feynman/Weinberg theorem).

Like any other scientific theory, GR contains several assumptions, in both the geometrical and field-theoretical pictures. Although there are good theoretical and observational reasons to believe that these assumptions hold in a broad range of regimes and environments, there might still be room for modifying some of the assumptions in particular situations. Although GR has been tested observationally to a great extent, no {\it direct} observational tests have been made on very large scales (larger than our Solar System) or very small scales (smaller than a fraction of a millimeter). These are exactly the scales at which some of our most fundamental questions arise, such as the problems with the cosmological constant, or the construction of a consistent theory of quantum gravity. One might therefore ask whether it could be possible to resolve such problems by assuming that our standard theory of gravity, i.e. GR, is modified on very large and/or very small scales.

\paragraph{The $\Lambda$ problems.} The cosmological constant problem was discussed in Sect.~\ref{sec:cc}, where it was pointed out that various attempts have been made to use modified gravity theories to address at least one of the three problems with $\Lambda$: the {\it old CC} (or naturalness/hierarchy) problem, the {\it new CC} (or dark energy) problem, and the {\it coincidence} (or why-now) problem. These mostly include modifications on extremely large scales, as the problem with $\Lambda$ seems to be an infrared one. The number of proposed IR-modified theories of gravity is rather large, each with interesting motivations, and we cannot discuss all of them here (see Fig.~\ref{fig:mg} for a diagrammatic representation of the theories that have been studied in the literature). Here we only list and briefly comment on three of the most popular, but (so far) failed, modified gravity attempts to address the CC problems (see Ref.~\cite{Clifton:2012ke} for more examples):

\begin{figure}[t]
\centering
\includegraphics[width=\hsize]{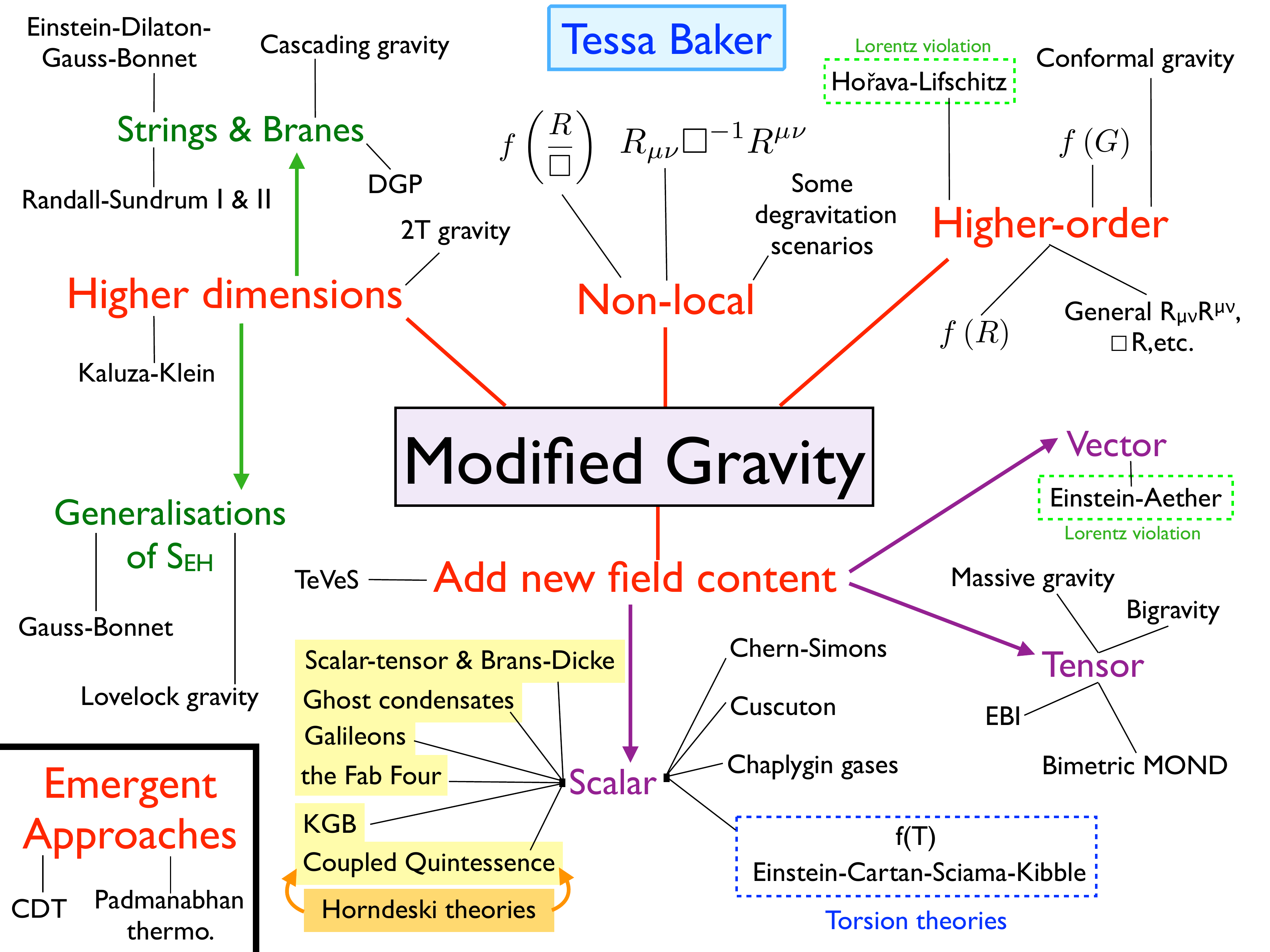}
\caption{Tree diagram of modified theories of gravity. (Tessa Baker, reproduced with permission.)}\label{fig:mg}
\end{figure}

\paragraph{Massive, bi-metric and multi-metric gravity.} Apart from theoretical importance of knowing whether gravitons can consistently receive a nonzero mass, and whether several interacting spin-2 fields could consistently exist, arguably the main motivation for massive gravity has been the initial hope that these theories can be used to both {\it degravitate} $\Lambda$ and provide self-accelerating solutions. This possibility arises as the graviton mass may suppress the effects of long wavelength sources, and hence solve the old CC problem (see Sect.~\ref{sec:cc} for some discussions on the idea of degravitation). This idea was however shown to fail as soon as a consistent, non-linear theory of massive gravity was constructed \cite{2010PhRvD..82d4020D,2011PhRvL.106w1101D,2012PhRvL.108d1101H,2012JHEP...04..123H, Hassan:2011zd} after about seven decades of theoretical work (see Refs.~\cite{2012RvMP...84..671H,2014LRR....17....7D,Schmidt-May:2015vnx,Solomon:2015hja} for reviews of recent developments).

Massive gravity, and its cousin theory of bimetric gravity, could still provide a solution to the new CC problem, however. A small mass (or interaction) parameter $m$ in these theories -- if able to generate a late-time cosmic acceleration -- would be protected against quantum corrections and therefore remains small. This is because $m \rightarrow 0$ restores a symmetry (diffeomorphism invariance or general covariance). Although the simplest massive gravity theory was shown not to admit a flat FLRW universe, its bimetric generalization was indeed able to provide self-accelerating solutions~\cite{Volkov:2011an, vonStrauss:2011mq,Comelli:2011zm}, consistent with all existing observational data at the background level~\cite{2013JHEP...03..099A,Akrami:2013pna}. Since then, an extensive amount of work has been done to study the viability of the theories through metric perturbation theory and structure formation studies \cite[e.g.][]{Berg:2012kn,Fasiello:2013woa,2014JCAP...10..066S,2014PhRvD..90l4014K, 2015PhRvD..91h4046E,2014JCAP...12..026L, 2015JCAP...05..030C,2015JCAP...05..052A}. Unfortunately, although the simplest bigravity models are able to provide viable self-accelerating background expansions, all such models suffer from ghost and/or gradient instabilities~\cite{2012JHEP...06..085C,2014PhRvD..90l4014K, 2014JCAP...12..026L,2015PhRvD..91j4019K}. While it is possible to push these instabilities back to unobservably early times, beyond the regime of validity of the theory, without losing self-acceleration and obtaining a technically natural acceleration parameter~\cite{2015PhLB..748...37A}, the theory becomes observationally indistinguishable from $\Lambda$CDM in this case. While this may render the theory less favorable from an Occam's razor perspective, the fact that a small mass is protected by the symmetry of diffeomorphisms makes the theory more favorable than $\Lambda$CDM from the perspective of naturalness. It is then mainly a matter of subjective taste and further observational tests to decide which theory, if any, is actually realized in nature. As a different resolution of the problem, it is conjectured that the instabilities will be cured if non-linear effects in structure formation are taken into account~\cite{Mortsell:2015exa}.

Various attempts are ongoing in order to modify or generalize both massive and bimetric gravity in such a way that they provide observational signatures that distinguish them from GR while providing healthy cosmological models (see e.g. Refs~\cite{Akrami:2013ffa,Tamanini:2013xia,Akrami:2014lja,Yamashita:2014fga,deRham:2014naa, deRham:2014fha,Noller:2014sta,Enander:2014xga,Solomon:2014iwa,Schmidt-May:2014xla,Heisenberg:2014rka,Gumrukcuoglu:2014xba,Noller:2014ioa,Gumrukcuoglu:2015nua, Comelli:2015pua,Hinterbichler:2015yaa,Lagos:2015sya} for generalized couplings to matter, Ref.~\cite{deRham:2014gla} with mass parameters promoted to Lorentz-invariant functions of the St\"uckelberg fields, and Refs.~\cite{DeFelice:2015hla,DeFelice:2015moy} for the minimal theory of massive gravity with two physical degrees of freedom). Both massive gravity and bigravity are considered as effective theories, while their cutoff scales are yet to be found. Generalizations of the theories therefore seem to be essential in order to construct high-energy completions; such complete theories are expected to behave better cosmologically.  

\paragraph{Unimodular gravity.} This class of modified gravity theories is interesting because they contain fewer degrees of freedom than GR, and are completely insensitive to vacuum fluctuations, hence potentially providing a solution to the old CC problem. Here, the Einstein equations are recovered with $\Lambda$ as an integration constant~\cite{Ellis:2010uc}. These theories can also be made massive with a technically natural mass, as with GR~\cite{Bonifacio:2015rea}. It is not clear whether this integration constant is protected from radiative corrections however \cite{Smolin:2010iq,Padilla:2014yea}, and the massive versions have the same problems as the ordinary massive gravity~\cite{Bonifacio:2015rea}. Unimodular gravity seems therefore to be essentially equivalent to GR.

\paragraph{Non-local gravity.} Non-local effects are generic in the effective action of gravity with massless (gauge) fields, and naturally arise in effective field theories of gravity where contributions from high-energy gauge and matter fields are integrated out (e.g. to construct a one-loop effective action for gravity~\cite{Barvinsky:1994cg,Barvinsky:2002uf}). Such non-local actions are in general immune from naturalness and why-now problems. It is also possible to modify gravity in this framework without adding new degrees of freedom to the gravity sector~\cite{2007PhRvL..99k1301D}, and this class of theories can admit (super-)renormalizability and suggest possible resolutions of singularities in black holes and cosmology. The idea of degravitation or screening the vacuum energy may also be possible to implement within theories of non-local gravity, potentially resolving the old CC problem~\cite{ArkaniHamed:2002fu}. Although no consistent theory of non-local gravity that solves the old CC problem has been constructed so far, there are classes that can provide non-$\Lambda$ self-accelerating solutions, with interesting phenomenologies~\cite{2007PhRvL..99k1301D,Deser:2013uya, Maggiore:2014sia, 2014JCAP...06..033D} that can be observationally distinguishable from GR. There seem to be at least two obstacles that need to be resolved in order to have further progress with these theories, however: 1) there is no systematic way of constructing a model in an EFT framework, and 2) the generic theory seems to suffer from instabilities~\cite{Ferreira:2013tqn}. Work is in progress to address these problems and develop the theories further (see e.g. Refs.~\cite{Conroy:2014eja,Maggiore:2015rma}).

\paragraph{Seeking inspiration from observations.} The answer to the question ``can gravity solve the $\Lambda$ problem in cosmology?'' therefore seems to be ``not yet.'' Although modified gravity model building is still a very active field of research, one can alternatively try to test the standard theory, GR, independently of any particular modification framework. This could be of vital importance, as knowing where the standard picture break downs would make it much easier to know what types of modifications are actually needed. This will also tell us more about how well the standard gravity theory is able to describe various phenomena, and what the range of validity of the theory is. After all, GR, like any other scientific theory, needs to be tested observationally.

Predictions of GR have been extremely well tested using local observations, such as weak-field tests in the Solar System, or strong-field tests using binary pulsars. These tests have placed strong constraints on deviations from GR on Solar System scales, such that any modified theory is constrained to essentially reduce to GR on such scales, phenomenologically. GR has not yet been tested very well on cosmological scales however, and almost all of the tests so far have been dependent on several extra assumptions in cosmology~\cite{Baker:2014zba}.

Various proposals have been put forward for large-scale tests of gravity, which are expected to be possible using future cosmological observations. These are discussed in more detail in Sect.~\ref{sec:mgde}, but we summarize some of them here.

Most of the proposed tests are for intermediate cosmological scales, where the evolution of large-scale structure can be treated {\it linearly} or {\it quasi-linearly}~\cite{Ferreira:2010sz,Baker:2012zs,Caldwell:2007cw, DiPorto:2007ym,Bertschinger:2006aw, Pogosian:2010tj,Bean:2010zq,Baker:2013hia,2011PhLB..706..123D,Baker:2012zs,Silvestri:2013ne}. Many of these tests additionally rely on the so-called {\it quasi-static} approximation, which makes the analyses considerably easier if the relevant scales are all well inside the horizon. Such an approximation is valid for GR, as well as a large number of modified gravity theories, but depending on what precision we aim for in our cosmological analyses, we may need to include effects beyond the quasi-static regime.

Another interesting place to test gravity cosmologically is with small-scale structures, where non-linear effects become important. Many modified gravity theories have distinct signatures in the non-linear regime (see Sects.~\ref{sec:mgdev} and \ref{sec:sim}). Also, one has more statistical power on these scales, as more Fourier modes are available (see Sect.~\ref{sec:huntingde}). The problem is that non-linear structures cannot (for the most part) be treated analytically, and one therefore needs N-body simulations in order to extract the non-linear, small-scale implications of gravity and compare them to observations. This is often very computationally expensive.

Finally one can look for gravitational signatures on ultra-large scales~\cite{Bonvin:2011bg,Challinor:2011bk, 2013PhRvD..87j4019L, 2015arXiv150507596A,2015ApJ...811..116B} (see Sect.~\ref{sec:mgde}). These scales are of particular interest, as they directly probe the scales that IR modifications of gravity must affect (i.e. the Hubble scale at late times). Observations of ultra-large scales can provide access to more and different information, and in that respect are unique. The problem with such scales is however that uncertainties could be large, as the amount of information for each type of probe is highly limited due to the large cosmic variance effect. By combining several probes on such scales, however, there is hope to beat the cosmic variance and gain invaluable knowledge about the properties of gravity on the largest scales.

\paragraph{Summary.} So far there is no convincing gravitational solution to the $\Lambda$ problem, although it probably remains one of the most promising routes to tackling it. Fundamental assumptions are currently being explored in order to construct further consistent theories of modified gravity, which may address important questions in cosmology such as the CC problem. On the phenomenological side, cosmological tests of gravity are being performed. Such tests are completely understood in the quasi-static regime, although it might be important to go beyond that regime for particular classes of theories. Non-linear tests on small cosmological scales have proved to be very interesting, but highly complicated and computationally expensive. Some studies of ultra-large scales exist, but there is still much to be done, as such scales provide unique, new, and direct probes of the gravitational effects on the largest scales, i.e. exactly where alternative theories of gravity are supposed to make gravity behave differently.

\subsection{Theoretical directions, motivations, and the role of simulations in developing modified gravity theories}
\label{sec:mgdev}
\noindent\textit{Discussion session chairs: T. Baker, J. Beltr{\'a}n Jim{\'e}nez \& C. Llinares}

General relativity has received remarkable support from observations over a wide range of scales. This success has not precluded intense research activity on alternative theories of gravity, however, for a variety of reasons. In this section we examine the motivations for modifying GR, and survey current theoretical developments. The future role of simulations in developing these theories is given special attention.

\paragraph{Theoretical and observational motivations for modifying GR.} On the theoretical side, GR is known to possess some uncomfortable features such as the cosmological constant problem, the existence of singularities, and its non-renormalizability. The CC problem was discussed extensively in Sect.~\ref{sec:cc}, so we will simply mention that some attempts to solve (or at least alleviate) it have resorted to modifying GR, e.g. IR modifications exhibiting degravitating solutions ~\citep{Dvali:2002fz, ArkaniHamed:2002fu, Dvali:2007kt, deRham:2010tw}, or the presence of a Weyl symmetry \citep{Barcelo:2014qva, Carballo-Rubio:2015kaa, Alvarez:2015pla}. The existence of black hole or big bang singularities indicate a failure of GR in the high-curvature regime. Likewise, its non-renormalizability also requires a successful UV completion. Both problems might be related and, in fact, singularities are commonly expected to be regularized by quantum effects, although modifications of the high-curvature regime are also used to avoid singularities \citep{Novello:2008ra,Banados:2010ix}. 

From a more observational standpoint, the standard cosmological model, \lcdm{}, requires the presence of three unknown components -- namely, dark matter, dark energy, and the inflaton field. It is commonly thought that dark matter corresponds to an as-yet unobserved particle, although modifications to gravity have also been invoked to remove the need for dark matter \citep{Famaey:2011kh, Khoury:2014tka}. In addition, given the current lack of a dark matter particle detection (both from astrophysical observations and accelerators), it seems worthwhile to further explore modified gravity alternatives. It is worth mentioning that, even if dark matter is a new particle, it might give us some hints about modifications to gravity that could explain cosmic acceleration \citep{Cembranos:2003mr, Cembranos:2008gj}. 

An explanation of dark energy (or cosmic acceleration) is the driving motivation behind many modifications of gravity. This problem is actually related to the CC, but in most cases one hopes for some unknown mechanism to tame the CC (see the discussions of degravitation, above), whilst the current acceleration of the Universe is driven by some other component, usually a scalar degree of freedom arising from a modification of gravity. Also, in standard inflation, a (quasi-) de Sitter phase is assumed to be driven by a scalar degree of freedom. Again, this field could be associated to the gravitational sector, and some proposals use modified gravity to generate inflation. Moreover, some alternatives to inflation (e.g. bouncing universe scenarios) also require modifications of gravity in the high-curvature regime.

It is worth stressing that even if alternative theories of gravity do not fully achieve resolutions of the theoretical problems for which they were first designed, they have at least provided substantial advancement in our understanding of GR. They have also led to the realization of new theoretical phenomena, e.g. screening mechanisms, massive gravity and bigravity, etc. It is crucial to have theoretical models against which observations can be tested; in this respect, our current modified gravity theories are useful toy models for deviations from GR, even if they are ultimately not realized in nature. For a historical comparison, we mention that scalar-tensor and vector-tensor theories were used as counterfoils to GR in the parameterized post-Newtonian formalism (PPN), which is used to constrain deviations from GR in Solar System experiments. In an analogous manner, the discovery of cosmic acceleration has triggered the development of general formalisms to parameterize dark energy models and/or modified linear cosmological perturbations, as well as motivating the use of new cosmological observables \citep{Baker:2012zs, Gubitosi:2012hu}.

\paragraph{Theoretical directions.}
It is helpful to attempt some classification of the alternative gravity theories that have proliferated during the past decade. A commonly-used scheme is to classify theories according to the manner in which they evade the consequences of Lovelock's theorem. Lovelock's theorem can be stated as follows (in vacuum or with conserved sources) \citep{Lovelock1,Lovelock2,Clifton:2012ke}:
\begin{quote}
\textit{The only local, second-order gravitational field equations that can be derived from a four-dimensional action that is constructed solely from the metric tensor, and satisfy the conditions of being symmetric and divergence-free (i.e. admitting Bianchi identities), are those of the Einstein field equations with a cosmological constant.}
\end{quote}
To build a gravity theory that differs from GR, then, one must implement one of the following changes to the gravitational action: (i) the introduction of fields other than the metric tensor; (ii) more than four spacetime dimensions; (iii) the introduction of non-local terms; or (iv) acceptance of higher-than-two-temporal-derivative field equations (in this instance, care is required to ensure that the theory is energetically stable). 

All of these options have been explored in recent literature, with the first option -- invoking new fields to participate in gravitational interactions -- proving the most popular. Scalar-tensor field theories tend to allow the most freedom in their construction; the requirement of spatial isotropy restricts vector-tensor actions \citep{2008PhRvD..77h4010Z,2009PhRvD..80f3512B, Zuntz:2010hc}, and stability of perturbations has been found to place quite strict constraints on permitted bigravity (tensor-tensor) actions \citep{2012JHEP...06..085C,2014PhRvD..90l4014K}. However, with the (re-)discovery of Horndeski gravity \citep{Horndeski1974,PhysRevD.84.064039} -- the most general scalar-tensor theory resulting in second-order field equations -- one might well ask whether this particular avenue of model-building (i.e. adding new fields) has nearly been exhausted. 

It is not surprising then that some attention has recently been given to more novel methods of modifying GR (see also Sect.~\ref{sec:lcdmalt}). For example, a series of papers \citep{Maggiore:2013mea,Maggiore:2014sia, 2014JCAP...06..033D, 2014IJMPA..2950116F, 2015JCAP...04..044D} has explored the viability -- both theoretical and observational -- of introducing non-local terms to the gravitational field equations, e.g. $\sim m^2\left(g_{\mu\nu}\square^{-1}{\rm R}\right)^{\rm T}$, where $\square^{-1}$ is the inverse d'Alembertian, and $R$ is the Ricci scalar (see also Refs.~\cite{2010JCAP...11..008B, 2007PhRvL..99k1301D} for earlier works). It is possible to construct an action that gives rise to these kinds of terms in the field equations, but the precursor action must be supplemented by additional information to ensure that causality is respected (e.g. by specifying that only the retarded Green's function for the non-local operator $\square^{-1}$ is used), although it has been argued, e.g. in Ref.~\cite{Foffa:2013sma}, that the causality condition is automatically satisfied if one interprets the non-local action as a quantum effective one.

Perhaps an even more radical option for modifying GR is to relinquish the concept of deriving gravitational field equations from a fundamental action altogether. The central tenet of this approach, sometimes termed `emergent gravity', is that gravitational forces may be a macroscopic phenomenon that emerges from some more fundamental microphysical theory of spacetime. In this picture, the Einstein field equations are given the same status as, say, the equations of classical fluid dynamics \citep{Jacobson:1995ez,2014GReGr..46.1673P}. In this analogy, quantization of GR is like attempting to describe the properties of fluids on very short distance scales without knowing about the existence of atoms. That is to say, our present failure to develop a consistent theory of quantum gravity is unsurprising. If the natural arena for the underlying microphysical theory is the Planck scale, then we will only be able to study the properties of the coarse-grained, emergent gravitational phenomena.

Finally, we note that there is another way to alter the gravitational field equations that is not subject to Lovelock's theorem. Lovelock's theorem enumerates the possibilities for modifying the gravitational action, but says nothing about the coupling between the energy-momentum tensor of matter and the metric. A celebrated argument by Weinberg \citep{PhysRev.135.B1049, 9781139644167} restricts the low-energy limit of the matter coupling to be that dictated by the equivalence principle; however, it is possible that the coupling could be subject to high-energy corrections. A concrete example of this mechanism is found in Palatini theories, in which integrating out the connection field results in modified couplings~\cite[e.g.][]{Delsate:2012ky}.

It cannot be denied that the field of modified gravity is subject to trends that wax and wane, with recent `hot topics' including DGP gravity, Galileons, Horndeski gravity and bigravity \citep{2000PhLB..485..208D,2011CQGra..28t4003T,2011PhLB..706..123D, Schmidt-May:2015vnx}. We choose to defend this culture, however -- theories that experience a rapid rise in popularity generally do so because they have feature(s) of particular appeal. They may incorporate a natural screening mechanism or be able to fit the expansion history of the Universe with the same number of free parameters as \lcdm{} \citep{2014JCAP...06..033D}, for example. The fact that they are subsequently found to be non-viable is testament to the energy of this research community, and the machinery we have in place to rapidly take theories from their initial mathematical development to observational constraints.

\paragraph{Role of simulations.}
The presence of screening in some gravity theories means that cosmological simulations have become a central tool for calculating testable predictions. Indeed, neglecting screening effects -- as  done in a linear analysis -- can lead to incorrect predictions of the power spectrum of density perturbations over a range of scales that will soon be observable with forthcoming surveys such as Euclid \citep{2014JCAP...07..058F}.  Furthermore, screening must be taken into account when making predictions for discrete objects such as galaxies or galaxy clusters. Recent years have seen increased interest in studying the effects of modified gravity on these objects, and the community already owns several independent and highly developed codes for this purpose \citep[e.g.][]{2012JCAP...01..051L, 2013MNRAS.436..348P, 2014A&A...562A..78L}.

At present, the key challenge to modified gravity simulations is as follows: the scales at which deviations from GR are expected to become significant coincide with the scales at which baryonic processes and the effects of neutrinos can play an important role. It is already known that there is a degeneracy between modified gravity effects and these processes \citep{2014MNRAS.440...75B, 2014arXiv1404.3639S}. Finding ways to break this degeneracy -- for instance, by using alternative observables to the matter power spectrum -- is a topic of active research. Input from future surveys such as Hetdex may assist in improving the simulation prescriptions for baryonic and neutrino effects, thereby alleviating some of the degeneracy.

Simulation techniques are still maturing (e.g. see the recent code comparison project in Ref.~\cite{2015arXiv150606384W}), with codes continuing to use some approximations whose validity is still under debate. One of these is the `quasi-static' approximation, which permits time derivatives in the equations of motion that correspond to new degrees of freedom to be neglected. The quasi-static approximation has been tested for some specific models in the non-linear regime \citep{2013PhRvL.110p1101L, 2014PhRvD..89h4023L,2015PhRvD..92f4005W}, and work has also been done in the linear regime \citep{2014PhRvD..89b3521N}. The quasi-static approximation remains to be tested for models in which the effective sound speed of the extra degree of freedom is less than the speed of light, however; different fields could then have different horizon sizes, possibly resulting in new observable signatures \citep{2015arXiv150306831S}.

\subsection{Distinguishing dark energy and modified gravity theories}
\label{sec:mgde}
\noindent\textit{Discussion session chairs: S. Camera \& J. Sakstein}

Attempts to explain cosmic acceleration have commonly focused on models that introduce new degrees of freedom (see Sect.~\ref{sec:mgdev}, as well as Ref.~\citep{Clifton:2012ke} for a compendium), or theories that postulate a modification of the Einstein equations such as non-local models and gravitational aethers \cite{Afshordi:2008xu,Woodard:2014iga, Maggiore:2014sia}. Whether individual models should be classed as dark energy or modified gravity models is often subjective. In this section, we ask the more natural question of whether or not it is possible to distinguish different theories from one another.

\paragraph{parameterizations vs. theories.} One approach is to attempt to constrain the parameters and functions appearing in parameterized cosmological frameworks 
\cite{Baker:2012zs,Battye:2013aaa, Gleyzes:2013ooa,Silvestri:2013ne}, while another is to study specific models that are theoretically well-motivated, or exhibit screening mechanisms \cite{Joyce:2014kja, Koyama:2015vza}.

The first has the advantage that it allows one to constrain many different models with one set of functions, and can easily be applied to new models. The drawback is that it does not distinguish between different theories, and typically does not apply to non-linear scales where many novel features can be exhibited. In particular, theories with screening mechanisms can exhibit drastically different behavior on small scales. The second approach requires one to compute observables on a model-by-model basis, but the non-linear features found can serve as smoking guns for particular theories. We will review the second approach here, discussing the merits and drawbacks of various probes on different scales.

\paragraph{Large and ultra-large scales.}
Access to increasingly large cosmological volumes allows for a major advance in tackling the fundamental question of whether GR holds on the largest scales. Tests of GR on cosmological scales are based on observations of the large-scale structure (see also Sect.~\ref{sec:gravcosmo}). Current constraints are weak, but forthcoming cosmological surveys, with their huge survey volumes and multiple probes, will take the lead in the next generation of tests (see e.g. Refs.~\cite{2011arXiv1110.3193L,2013LRR....16....6A, Dewdney:2009, Camera:2015fsa,2015ApJ...811..116B, 2015ApJ...812L..22F, Bull:2015lja} and Sect.~\ref{sec:lahav}). We can also tighten the current constraints on dark energy and modified gravity models by including much larger scales, thus increasing the statistical power of the observations and improving sensitivity to any scale-dependent deviations from GR.

In contrast to tests on `small' scales (see the next section), tests on horizon scales are still rather weak. It is important to test GR on these scales however, as the accelerated expansion of the Universe is a late-time (Hubble scale) phenomenon, making it very natural to look for hints concerning its physical nature in this regime.

As already discussed, phenomenological parameterizations can in principle describe the full range of deviations from GR relevant for cosmology. We could therefore exploit the information contained in the data in full by effectively modeling the metric perturbations (see Ref.~\cite{Kunz:2012aw} and references therein). Such phenomenological models are mostly useful to capture the evolution of linear perturbations, however, which effectively limits them only to large scales.

Many modified gravity theories also modify the expansion history of the Universe. This often happens by construction, as they are intended as alternatives to dark energy. Unfortunately, it has been shown that dark energy models are able to produce (almost) arbitrary equations of state, $w(z)$, and that modified gravity models can, in turn, be tuned to mimic the cosmological constant value of $w=-1$. The expansion history therefore fails as a `smoking gun' for modified gravity. On the other hand, the linear growth rate is a sensitive probe of the growth history. It can be accurately measured using redshift-space distortions from galaxy redshift surveys and intensity mapping experiments for example (see e.g. Ref.~\cite{Hall:2012wd}).

Another important quantity in this context is the scalar anisotropic stress. This is generally non-zero in modified gravity theories, especially on very large scales \citep{Saltas:2010tt}. Measuring the anisotropic stress is a promising test for deviations from \lcdm{}, as it can be measured without any assumptions on the primordial matter power spectrum or the galaxy bias by combining weak-lensing measurements with peculiar velocities \citep{Motta:2013cwa}. The presence of the anisotropic stress also appears to be connected to a modification of the propagation of gravitational waves, which can be used as a way to define what ``modified gravity'' means \citep{Saltas:2014dha,Camera:2013xfa}.

Summarizing, going to large scales is valuable for many reasons. As we shall see, on small scales, where perturbations are highly non-linear, modified gravity theories generically need to be screened to avoid violating strong Solar System constraints. If we assume that only linear and mildly non-linear scales can exhibit deviations from GR as a result, we then need to measure as many modes as we can in this regime to see the potentially small effects. This pushes us to look at larger volumes. The scale dependence of deviations from GR is also a crucial observable to distinguish between different models. We need a wide range of scales to be able to observe a scale dependence, and the horizon scale is a natural place to look for this effect.

\paragraph{Small-scale probes.}
Many theories predict deviations from GR in the Solar System, and this is the first test that any model should pass before it can be considered on cosmological scales. The parameters appearing in the parameterized post-Newtonian formalism (PPN) \cite{Will:2008dya} have been measured to extremely high precision, and many of the `classical' alternatives to general relativity are constrained to be cosmologically irrelevant as a result \cite{EspositoFarese:2004cc}. Any alternative theory where additional degrees of freedom couple to matter will typically make predictions that fit into the PPN form. A noteworthy exception is those theories where the inverse mass is $\lesssim\mathrm{AU}^{-1}$. These are best tested with table-top experiments, which probe Compton wavelengths down to a few $\mu$m \cite{Hoskins:1985tn,Adelberger:2005vu, Kapner:2006si,Lambrecht:2011qm}, and lunar laser ranging (LLR), which tests the inverse square law at a distance of 384,400 km, corresponding to the Earth-Moon separation \cite{Nordtvedt:2003pj}. LLR also provides a constraint on the time variation of Newton's constant, $\dot{G}/G=(2\pm 7) \times10^{-13}$yr$^{-1}$ and $\ddot{G}/G=(4\pm5)\times10^{-15}$ yr$^{-2}$ \cite{Merkowitz:2010kka}. In many cases, these measurements directly constrain the allowed cosmological parameters \cite{Sakstein:2014isa, Sakstein:2014aca}.

Theories that do not satisfy these tests must be discounted, and this has prompted a search for models that can evade them. Several models can accomplish this using a clever choice of parameter tunings (for example, Einstein-Aether theory \cite{Foster:2005dk}) or through a suitable choice of the free functions that can appear in the action (non-local models fall into this class \cite{Woodard:2014iga, Maggiore:2014sia}). In other cases, the PPN parameters constrain combinations of the cosmological quantities and fundamental parameters. In this case, the PPN parameters typically do not rule out the theories, but constrain the specific models \cite{Ip:2015qsa,Sakstein:2015jca}.

Finally, one can evade Solar System tests completely using screening mechanisms. These typically fall into two classes. {\it Chameleon} and chameleon-like theories \cite{Khoury:2003aq,Khoury:2003rn,Hinterbichler:2010es,Brax:2010gi} (which include $f(R)$ models \cite{Brax:2008hh}) screen by using non-linearities in the field equations to suppress the scalar charge. As a result, these theories do not satisfy the equivalence principle \cite{Hui:2009kc}, a fact that has been exploited by many proposed tests \cite{Davis:2011qf,Jain:2011ji, Jain:2012tn,Brax:2013uh, Sakstein:2013pda,Vikram:2013uba, Vikram:2014uza,Sakstein:2014nfa, Sakstein:2015oqa}.

The {\it Vainshtein mechanism} \cite{Vainshtein:1972sx} (see also the review~\cite{Babichev:2013usa}) suppresses the field gradients sourced by massive objects, and hence decouples the new degrees of freedom. It is exhibited in theories that typically involve higher-order derivatives such as Galileons \cite{Nicolis:2008in} and massive gravity \cite{2014LRR....17....7D}. Theories with this mechanism are harder to test due to negligible fifth-forces on small scales and because they satisfy the equivalence principle, although a few novel probes have been suggested \cite{Hui:2012jb,Hiramatsu:2012xj,Koyama:2015oma,Jimenez:2015bwa}. The best probes are currently linear perturbations however \cite{Barreira:2012kk,2012JHEP...06..085C}.

It is clear that there are several competitors to \lcdm{} that also behave identically to GR on Solar System scales. Any test to distinguish these theories from GR, and from one another, must therefore focus on reaching a consistency between all scales: ultra-large, linear, non-linear, and astrophysical.

To conclude, we report on some caveats that one has to keep in mind when testing alternative cosmological models.

\paragraph{Degeneracies with baryons and massive neutrinos.}
Accurate predictions on non-linear scales require dedicated $N$-body simulations for each theory -- as mentioned above, it is unclear how to use the unifying phenomenological frameworks on non-linear scales (although see Ref.~\cite{Gronke:2015ama}). Baryonic effects on small scales are also poorly understood (see Sect.~\ref{sec:cdm}), and add a systematic error that effectively renders these scales unusable for precision cosmology for the time being. So far, some specific dark energy and modified gravity theories have been tested in the non-linear regime with dedicated $N$-body simulations \citep[e.g.][]{Oyaizu:2008tb, Schmidt:2009sg,Khoury:2009tk, Brax:2012nk,Brax:2013mua, Baldi:2012ky,Llinares:2013jua}. These have allowed forecasts for future surveys to be performed \citep[e.g.][]{Beynon:2009yd,Camera:2011mg, Camera:2011ms,2013LRR....16....6A}.

As recently pointed out by \citet{2014MNRAS.440...75B} however, few attempts have so far been made to investigate the effects of baryons and massive neutrinos on the formation and evolution of linear and non-linear cosmic structures in the context of alternative dark energy and modified gravity models. For instance, it has been shown that suitable choices of parameters for a combination of massive neutrinos and modified gravity could in principle result in a large-scale matter distribution that is barely distinguishable from \lcdm{}\ \citep{2014MNRAS.440...75B}. Such results seem to indicate a theoretical limit to the effective discriminating power of cosmological observations, and clearly suggest the necessity of further investigations to properly account for possible degeneracies with small-scale effects. 

\paragraph{The dangers of circular logic.}
Circular reasoning is a logical fallacy where the reasoner begins with what she or he wants to end up with. While it may seem like a trivial error, it is not so easy to avoid when trying to test a theoretical alternative to a standard paradigm without full control over all the pieces of evidence used for this purpose. This is what can happen with dark energy, and modified gravity models in particular. We will illustrate this using a specific example by \citet{Diaferio:2011kc} that shows the dangers of testing alternative theories with what we only {\it suppose} are consistently-calibrated data.

In the late 90s, Beppo-SAX \citep{1997IAUC.6576....1C} provided the first sufficiently-accurate estimates of the celestial coordinates of gamma-ray bursts (GRBs), thus enabling the measurement of a host galaxy redshift and proving the extragalactic origin of GRBs. Since then, GRBs have been advocated as new standardizable candles, to be used much in the same way as Type Ia supernov\ae\ (SNe Ia), but at considerably higher redshifts (up to $z\simeq8$). The two-step procedure for using GRBs as distance indicators involves: (i) calibrating their temporal and spectral correlations; and then (ii) using the calibrated correlations to estimate the GRB luminosity distances of a given sample.

\citet{Diaferio:2011kc} adopted a fully Bayesian approach to infer both cosmological parameter values and the additional GRB calibration parameters. For the first time, they employed GRBs as cosmological probes without prior information from other data, thus neatly avoiding the inconsistencies of previous methods that were plagued by the so-called circularity problem. To test their method, they compared the \lcdm{}\ prediction with that of a conformal gravity model~\cite[e.g.][]{Mannheim:1989jh}, where the Universe's expansion has always accelerated. They found that, when properly analyzed, current data are consistent with distance moduli of GRBs and Type Ia supernovae that can be, in a variant of conformal gravity, $\sim15$ and $\sim3$ magnitudes fainter than in \lcdm{}, respectively. That is, their results show that currently available SN Ia and GRB samples can be accommodated by both theories, and do not exclude a continuous accelerated expansion. In other words, if we assume a specific theory when calibrating our observational data, we cannot possibly use that same calibration to constrain a competing theory without falling foul of the circular logic fallacy.

\section{Dark matter: the CDM in \lcdm{}}
\label{sec:cdm}

Cold dark matter (CDM) -- or something that closely mimics it -- must exist in substantial quantities if the consistency of recent precision cosmological tests is to be believed. A non-baryonic, pressureless, clustered fluid appears necessary to make sense of the well-measured baryon acoustic oscillation signature in the CMB and galaxy clustering, for example. The physical composition of this fluid remains unknown however, despite relentless searching for candidate dark matter particles by astrophysical and direct-detection experiments. In this section we consider the implications for cosmology if the physical cause of dark matter is never positively identified -- would it make sense to continue using CDM as a primary component of our model?

It has also become clear that CDM, though dominant in terms of its overall energy density, is only one ingredient in a complex, interacting system of matter components that are responsible for forming the structures that we see in the Universe. Driven by puzzling discrepancies between observations on small scales and increasingly sophisticated simulations, our picture of structure formation is becoming progressively more complicated; processes involving baryons and massive neutrinos can no longer be neglected, and may even dominate in some situations. We review the status of recent efforts to get simulations and observations to agree, and ask whether current simulation technology is fit for purpose as we enter the era of gigantic large-scale structure surveys and non-linear studies of modified gravity.

\subsection{Simulating the Universe}
\label{sec:sim}
\noindent{\it Plenary speaker: R. Teyssier*}

Forthcoming surveys are designed to improve the precision on cosmological parameters to beyond the 1\% level. Multiple large simulations will be required to make sense of the new data, calibrate error bars and so on, but there is a serious question as to whether existing simulation codes can provide information at the required level of accuracy -- Euclid will require simulations that are correct to at least 1\% up to highly non-linear scales, $k = 10 h/{\rm Mpc}$, for example.

\paragraph{Accuracy of simulations.}
Several `code comparison' projects are in progress, which attempt to validate simulation codes by quantifying their accuracy (or at least agreement between different methods). In the first instance, one can compare between collisionless (i.e. cold dark matter-only) simulations, without adding the complications of baryons and radiative transfer. In this case, recent works have shown that three popular simulation codes -- Ramses, Pkdgrav3, and Gadget3 -- agree to within 1\% at $k \le 1 h/{\rm Mpc}$, and 3\% at $k \le 10 h/{\rm Mpc}$ \cite{2015arXiv150305920S}.

One can also validate codes by running a suite of simulations with different resolution settings to check that they are self-consistent. This exercise was recently performed using the Dark Sky Simulations, where the halo mass function and mass power spectrum were found to be stable (i.e. consistent) at the 1\% level over 3 orders of magnitude in particle mass resolution \cite{2014arXiv1407.2600S}.

Both of these results show that current codes are falling short of the Euclid target (or marginally scraping through). There are some ideas for improving accuracy in the future -- e.g. by using Vlasov-Poisson solvers instead of N-body techniques \cite{2013MNRAS.434.1171H, 2015arXiv150101959H} -- but the path ahead is uncertain.


\paragraph{Modified gravity simulations.}
An important goal of many future surveys is to constrain deviations from GR. Modified gravity theories can alter structure formation in a range of complex and novel ways, typically leading to small (but measurable) deviations from the vanilla \lcdm{} model at the $\sim 1\%-10\%$ level. There is therefore an imperative to develop simulations of structure formation in a variety of modified gravity models, so that theoretical predictions can be made on the same footing as for \lcdm{}.

For a modified gravity simulation to be viable, one at least needs to be able to calculate the time evolution of the background scale factor, draw random initial conditions self-consistently within the theory, and to have a valid weak-field limit for the matter dynamics. A number of technical challenges can also arise in the presence of a different gravitational theory -- direct/Fourier convolution approaches may not be valid if the modified Poisson equation is non-linear, and non-linear solvers can be slow and struggle with convergence. (Non-linear multigrid and Newton-Raphson iteration techniques can help with this, however.)

Regardless, a number of modified gravity and Modified Newtonian Dynamics (MOND) simulation codes are now on the market, using a range of techniques, and accommodating a number of different theories, including $f(R)$ gravity \cite{2011PhRvD..83d4007Z, 2013MNRAS.428..743L, 2013MNRAS.436..348P} and MOND \cite{2015CaJPh..93..232L, 2015MNRAS.446.1060C}. Others are discussed in Sect.~\ref{sec:mgdev}, and a multi-theory code comparison was also recently performed \cite{2015arXiv150606384W}.

\paragraph{Baryonic physics.}
The difficulty of correctly modeling baryonic processes is a major limitation for cosmological simulations, and thus on precision cosmology itself. A number of baryonic effects are known to be important, and can have wide-ranging effects on physical quantities. For example, low-mass galaxies are thought to be dominated by stellar feedback \cite{1986ApJ...303...39D}, while higher-mass galaxies depend strongly on AGN feedback \cite{1998A&A...331L...1S}. The introduction of baryons can change the halo mass at the tens of percent level, with large differences in the size of the effect as a function of halo mass \cite{2014MNRAS.444.1518V, 2014MNRAS.445..175G}. A comparison of an analytic+simulation model with observations of galaxy clusters in the X-ray, and using the Sunyaev-Zel'dovich effect, found that baryons suppress modes on scales $k \gtrsim 0.5 h/{\rm Mpc}$, with a $\sim 10\%-25\%$ suppression observed at $k \approx 2 h/{\rm Mpc}$ \cite{2015arXiv151006034S}.

Baryonic effects are typically taken into account through hydrodynamic simulations, where radiative and collisional processes can be included (or at least modeled). There is considerable uncertainty in how to model many processes, however, resulting in a corresponding uncertainty in the simulation results. Take AGN feedback as an example -- super-massive black holes in high-mass elliptical galaxies can significantly affect galaxy formation (and structure formation more widely) through a combination of complex effects: thermal feedback \cite{2007MNRAS.380..877S, 2010MNRAS.405L...1B}, radiative feedback \cite{2013MNRAS.436.3031V, 2014MNRAS.442..440C}, jet feedback \cite{2007MNRAS.376.1547C, 2010MNRAS.409..985D, 2014MNRAS.442..440C}, cosmic rays \cite{2012ApJ...752...24P}, and bubbles \cite{2007MNRAS.380..877S}. Constructing adequate models for these effects, including them all in simulations, and correctly determining their relative contributions to the overall structure formation process, is a highly non-trivial process. Indeed, various code comparison projects over the years \cite{1999ApJ...525..554F, 2008MNRAS.390.1267T, 2012MNRAS.423.1726S, 2014MNRAS.442.1992H, 2014ApJS..210...14K} have shown disagreement between codes on the order of tens of percent, and comparison with observations~\cite[e.g.][]{2015arXiv151006034S} suggests that baryon-inclusive simulations are only good at the 10\% accuracy level so far.

Accuracy aside, hydrodynamic simulations are also computationally intensive, so it is useful to also have analytic models of baryonic physics. The halo model can be extended by providing analytic models for the gas and central galaxy in a halo, for example, in addition to the dark matter. The power spectrum can then be calculated by providing the mass of the central galaxy (through abundance-matching), its size, the total gas mass as a fraction of the total halo mass, and an adiabatic contraction for the cold dark matter. This model has been developed over a number of years \cite{2004APh....22..211W, 2004ApJ...616L..75Z, 2008ApJ...672...19R, 2010MNRAS.405..525G, 2011MNRAS.417.2020S, 2011MNRAS.415.3649V, 2015MNRAS.454.1958M}, and has been found to agree well with zooming hydrodynamical simulations \cite{2014MNRAS.440.2290M}. One can then use such models to marginalize over baryonic effects in surveys, e.g. for weak lensing.

Baryonic effects can be neglected in some circumstances, but only if one is willing to discard information from small scales -- which contain a great deal of useful cosmological information. For example, Mohammed et al. \cite{2014arXiv1410.6826M} found that, by neglecting baryonic effects, they introduced a negligible bias in cosmological parameters measured from weak-lensing 2-point statistics for $\ell_{\rm max} = 4000$, but found a $1\sigma$ bias for $\ell_{\rm max} = 5000$, and a $10\sigma$ bias for $\ell_{\rm max} = 10000$. Marginalizing over a model of the baryonic effects can reduce the bias while still allowing one to improve precision by going to higher $\ell_{\rm max}$, but biases can remain if the model is not good enough. Note that non-Gaussianity at small scales is not modeled in Ref.~\cite{2014arXiv1410.6826M}, and can also introduce biases if not taken into account.

\subsection{The small-scale problems of \lcdm{}}
\label{sec:smallscale}
\noindent{\it Discussion session chairs: M. S. Pawlowski, T. Sawala \& M. Viel}

Simulations predicting the properties of dark matter (sub-) halos and the (satellite) galaxies inhabiting them in the context of \lcdm{} have become increasingly detailed. Combined with improved observational constraints on these scales, an increasing number of ``small-scale'' problems have become apparent. Many observations on galaxy scales now seem to be in conflict with the \lcdm{} paradigm (or were at least not {\it a priori} expected within it). These include:
\begin{itemize}
\item \textit{Missing Satellites problem}: The over-abundance of the predicted number of halo substructures compared to the observed number of satellite galaxies \citep{Klypin1999,Moore1999}.
\item \textit{Too Big to Fail (TBTF) problem}: The discrepancy between the measured densities at the half-light radii of the brightest local dwarf galaxies and the (higher) densities of the most-massive subhalos in \lcdm{} simulations, i.e.~an over-abundance in simulations of massive substructures that are expected to host galaxies after reionization \citep{BoylanKolchin2012}.
\item \textit{Emptiness of Voids problem}: The discrepancy between the velocity function of galaxies observed in the ALFALFA survey compared to CDM simulations of an equivalent volume \citep{Zavala2009}.
\item \textit{Core/Cusp problem}: The difference between the \lcdm{} prediction that halos follow a universal, centrally peaked (cuspy) density profile, and observations of dwarf and low-surface-brightness galaxies indicating shallower (cored) profiles \citep{Dubinski1991,Walker2011}.
\item \textit{Void Phenomenon}: The observation that galaxies in voids have similar properties to galaxies in very different environments, even though \lcdm{} predicts their merger histories and gas accretion to be very different \citep{Peebles2001}.
\item \textit{Satellite Planes problem}: The alignment of satellite galaxies of both the Milky Way and Andromeda on relatively thin, rotating planes, which is an exceptionally rare phenomenon in \lcdm{} simulations \citep{Pawlowski2014}.
\item \textit{Baryonic Tully-Fisher Relation (BTFR)}: The tight correlation of the circular velocity of galaxies with their total baryonic mass ($M_{\mathrm{b}} \propto V_\mathrm{c}^4$), which has a different slope and less scatter than expected from \lcdm{} and hierarchical structure formation \citep{McGaugh2011}.
\item \textit{Mass-Discrepancy vs. Acceleration Relation (MDA)}: The tight correlation of the mass discrepancy, (baryonic mass)/(dynamically-required mass), in galaxies with local acceleration, which is not expected in \lcdm{} \citep{Famaey:2011kh}.
\end{itemize}

The list of proposed solutions to these problems is similarly long. It includes the effects of accurately modeling baryonic physics in \lcdm{} simulations; using different types of dark matter, such as warm dark matter (WDM), interacting dark matter (IDM), and self-interacting dark matter (SIDM); and modifications to the assumed laws of gravity, either introducing modified gravity (MG) on large scales in addition to dark matter (as discussed in the context of dark energy, see Sect.~\ref{sec:gravity}), or as an alternative to dark matter such as Modified Newtonian Dynamics (MOND).

\begin{table*}
 {\centering \small
 \caption{Small-scale problems in \lcdm{}, and whether various models can solve them.}
    \begin{tabular}{rcccccc}
\hline
    & Baryons & WDM & IDM & SIDM & MG & MOND \\
\hline
    Missing Satellites      & \checkmark & \checkmark & \checkmark & X & X & \checkmark \\
    Core/Cusp               & \checkmark? & X & X & \checkmark & X & \checkmark \\
    Void Phenomenon         & \checkmark? & ? & ? & ? & \checkmark? & ? \\
    Planes of Satellites    & X & X & X? & X & ? & ? \\
    BTFR                    & \checkmark? & ? & ? & ? & ? & \checkmark \\
    MDA                     & \checkmark? & ? & ? & ? & X & \checkmark \\
\hline
    \end{tabular}
 \\ \smallskip
 \label{tab:smallscale}}
 The participants of the discussion session voted on which alternatives or modifications of current models appear to be feasible solutions for which problems. The symbols indicate whether the effect is: (\checkmark) mostly thought to provide a solution; (X) unable to provide a solution; or (?) that it is currently unclear whether it addresses the problem.
\end{table*}

Table \ref{tab:smallscale} gives an indication of which solutions might address which problems according to the majority opinion of the $\sim20$\ participants of the small-scale problems discussion session. No proposed solution addresses every problem (there are at least question marks for some problems for every model).

While no consensus was reached, and uncertainties in the current hydrodynamic simulations remain large, many problems appear to be affected by baryonic physics, which is consequently considered to be a promising possibility to solve many of the problems \textit{within} \lcdm{}. Whether this will solve all the problems (and solve them simultaneously) is not clear, however. Furthermore, baryonic effects have only been extensively studied in the \lcdm{} context, while alternatives have largely been restricted to dark matter-only simulations. This clearly demonstrates the need to not only improve subgrid physics models and reduce the degeneracies of hydrodynamic simulations, but also to extend their scope \textit{beyond} \lcdm{} in order to make more direct predictions for alternative models.

MOND is also quite a promising solution according to Table \ref{tab:smallscale}. It naturally explains the BTFR and MDA relations, but not many tests for the other problems have been done yet. Its effect on the satellite planes problem might be indirect: in MOND they can consist of tidal dwarf galaxies (TDGs) instead of infalling primordial substructures \citep{Kroupa2015}. While TDGs are observed in the Universe, and are naturally distributed in co-orbiting planes, they are expected to be devoid of dark matter in \lcdm{}-like models. The missing satellites and TBTF problems do not apply to MOND because these are concerned with how galaxies populate dark matter sub-halos, and of course there are no such sub-halos in MOND.

Finally, one should note that the satellites planes problem is unique amongst the reported problems, as neither baryons nor any of the alternative dark matter models appear to offer a satisfactory solution -- particularly if such planes are a truly universal feature. Observations of satellite systems outside the Local Group promise to provide new insights.

\subsection{Are cosmological simulations fit for purpose?}
\label{sec:cosmosims}
\noindent{\it Discussion session chairs: E. Bentivegna, F. Villaescusa-Navarro \& H.~A. Winther}

In an almost homogeneous and isotropic Universe, cosmology is fairly simple: the evolution of perturbations is linear and consequently the different modes in the Fourier representation decouple from each other. This simplifies the analysis of perturbations significantly since we can study them one mode at a time. However, linear theory cannot take us too far, since as we go to smaller and smaller scales the evolution of the perturbations becomes more and more non-linear, and for matter perturbations this happens around $k\sim 0.1 h{\rm Mpc}^{-1}$ in the \lcdm{} model. Linear theory is also not able to encapsulate important baryonic effects, which take place in our Universe and affect observables such as the matter power spectrum. Cosmology on linear scales has been very successful over the last few decades with observations of the CMB and BAO giving us very precise constraints on the parameters of our cosmological models, but there is also a lot of interesting information hiding in the non-linear regime; this information will be accessed by future planned large-structure surveys. In order to use this information to further test our models we must carry out numerical simulations to really understand what our theories predict in this regime. There do exist semi-analytical approaches that can make predictions beyond linear theory, but to obtain really accurate predictions we need to perform numerical simulations. In addition to the small-scale physics, the question of large-scale relativistic effects, sourced by inhomogeneities, resurfaces regularly, and requires an even larger numerical effort as the gravitational interaction between bodies has to be extended and corrected.

Due to all these reasons, numerical simulations will continue to play a major role in modern cosmology and astrophysics (see Ref.~\cite{Kuhlen_2012} for a recent review on N-body simulations). Nowadays, supercomputers are capable of following the evolution of billions to trillions of particles in cosmological volumes; however, upcoming observational missions place higher and higher demands on the realism required by cosmological codes. This discussion session was devoted to the problem of identifying areas where the simulation infrastructure may fall short of the future observational requirements, and to the outline of possible strategies to improve it. 

\paragraph{What accuracy is needed for observations?}
Before discussing infrastructural requirements, we must establish what accuracy demands the observations are going to impose in the near future.
Galaxy surveys such as Euclid \cite{2011arXiv1110.3193L,2013LRR....16....6A,EuclidWeb} will measure the matter power spectrum on scales as small as $k=10~h/$Mpc. In order to extract the maximum cosmological information from those surveys we need theoretical predictions, which are accurate at the $1\%$ level. This represents a huge challenge for simulations: on one hand, the matter power spectrum from different N-body codes agree at the $1\%$ level at $k=1~h/$Mpc but only at the $3\%$ at $k=10~h/$Mpc when running pure CDM simulations \cite{2015arXiv150305920S}; on the other hand, and more importantly, there are sources of systematics, such as neglected baryonic effects, that are expected to significantly affect the shape and amplitude of the matter power spectrum on small (fully non-linear) scales \cite{2011MNRAS.415.3649V} to a larger extent than the observational error bars.

A related question is what infrastructural resources will be needed to properly model these additional effects.
We will report on the discussion of all these different aspects in what follows.

\paragraph{Small-scale physics: baryonic effects, star formation, and radiative processes.}

Hydrodynamics simulations are more challenging than pure CDM simulations because there are astrophysical processes that take place on scales not resolved by the simulations, such as star formation, supernova and AGN feedback, and so on. In order to model these processes, subgrid models are commonly used \cite{2003MNRAS.339..289S, 2014MNRAS.444.1518V, 2015MNRAS.450.1937C}. Unfortunately, different groups employ different models, and the results may change dramatically. Moreover, subgrid models usually contain free parameters whose values are tuned to reproduce some low-redshift observations, rather than taking values specified {\it a priori} from theoretical arguments.

It is thus not clear how future simulations will be performed to reach the $1\%$ level at $k=10~h/$Mpc whilst including all baryonic effects. After discussing these issues we concluded that in the coming years, as more and more sophisticated hydrodynamic simulations will be run, it will be possible to parameterize the effects of baryons into a given functional shape with some free parameters. The value of the cosmological parameters can then be found by marginalizing over those baryonic parameters. This approach is dangerous, however: one could miss important effects of a different origin, which would be spuriously attributed to baryons.

\paragraph{Large-scale physics: relativistic corrections to N-body simulations and the effect of inhomogeneities.}

The vast majority of N-body simulations work in the Newtonian limit. 
While there is evidence that this approach is in principle accurate even on very large scales \cite{Chisari_2011},
these effects have never really been measured properly. The question of exactly how much systematic error
is introduced by this approximation remains open.

A recent study of small-scale relativistic effects \cite{Bruni:2013mua}, based on the post-Newtonian formalism, 
shows that the systematic errors are fairly small, contributing less than one part in $10^{-4}$ at $k=10~h/$Mpc.
On the other side of the spectrum, the magnitude of horizon-scale corrections, obtained through various approximation
schemes, are usually regarded as insignificant. It is however to be noted that even mildly non-perturbative evidence is
rather scarce on the largest scales. Should these effects prove to be relevant, there may be a simple way of
incorporating them in current codes via a time-dependent FLRW background.

The general feeling is that large-scale effects are going to be much less relevant, and much easier to handle, than small-scale ones.

\paragraph{Code correctness: validation \& verification.}
Over the last few years, a great effort has been made to test and compare N-body codes, and codes that extract various observables from the output of the simulations. For instance, there have been detailed comparisons of standard GR N-body codes \cite{2012MNRAS.423.1726S, 2015arXiv150305920S}, modified gravity N-body codes \cite{2015arXiv150606384W} and also of codes that identify dark matter halos \cite{2011MNRAS.415.2293K}, voids \cite{2008MNRAS.387..933C}, halo substructure \cite{2012MNRAS.423.1200O, 2013MNRAS.429.2739O, 2014MNRAS.438.3205P, 2014MNRAS.442.1197H}, galaxies \cite{2013MNRAS.428.2039K}, tidal debris \cite{2013MNRAS.433.1537E}, merger trees \cite{2013MNRAS.436..150S}, halo mock generation \cite{2014arXiv1412.7729C}, and galaxy mass reconstruction \cite{2014MNRAS.441.1513O}, to mention just a few (see Ref.~\cite{2013MNRAS.435.1618K} for a review on the current status of structure finding in N-body simulations). Comparison projects of these (often complex) numerical techniques are crucial to identify any worrying systematics in the theoretical predictions.

Code comparison remains the main avenue for code validation and verification in cosmology, but it is complicated by the fact that  
there is no universal agreement on some algorithmic aspects such as halo finding; different groups simply use 
different approaches. More work in this area is required to reach a general consensus.

\paragraph{Computational resources.}
Simulations needed for planned large-scale structure surveys (for example, to compute covariance matrices) need to cover big volumes and at the same time resolve sufficiently small scales. This requires a lot of computational resources, which in practice implies writing somewhat lengthy proposals to supercomputer facilities, with very low success rates. 
It seems that getting time to investigate new exotic scenarios is much easier than for important in-depth modeling, parameter studies, and double-checking of \lcdm{}. It is not clear however that this power can be leveraged in an optimal way; to get there, we would need to invest significant human resources, which is harder to get than computing power. The easier strategy is to put up with a little inefficiency and slightly longer runtimes.

\subsection{New particles and structure formation}
\noindent{\it Plenary speaker: C. Boehm*}
\label{sec:newpart}

Despite some unsolved problems on small scales (see Sect.~\ref{sec:smallscale}), the cold dark matter hypothesis has in general been able to successfully describe the cosmological observations. This is mainly due to the fact that such a component is only necessary to explain purely gravitational phenomena, i.e., there is no need for the missing component to have any other types of interactions, with itself or with baryons, in order to explain the cosmological observations. It is not too difficult to come up with theoretical models of candidate dark matter particles that provide the required gravitational effects, which is why a long list of dark matter models exists. This is not where we would like to stop, however; our cosmological model is not complete unless we know what the nature of the invisible matter is. That is, we need to know the particle physics properties of dark matter.

Fortunately, most dark matter candidates do have non-gravitational implications that provide various possibilities for detecting or constraining them through processes that are not gravitational; we will discuss some of these possibilities in the next section. Here we ask a different question: in addition to particle physics processes that are sought for through direct, indirect, and collider experiments (see Sect.~\ref{sec:dmdetection}), do dark matter candidates provide any effects that can be probed through cosmological observations, e.g. from their impact on structure formation? In other words, are all CDM or beyond CDM scenarios degenerate in terms of the resulting properties of the large-scale structure, or do they make different predictions for some cosmological observables?

As will be discussed in Sect.~\ref{sec:dmdetection}, weakly interacting massive particles (WIMPs) are still the most popular candidates for dark matter. There are numerous types of WIMPs proposed based on different particle physics theories, and various experiments are ongoing to test either the entire WIMP hypothesis or individual WIMP theories through different particle physics processes (see e.g. Refs.~\cite{Scott:2011yv,Akrami:2011vh}). This is possible because WIMPs do in fact interact non-gravitationally, with each other and with standard model particles. A ``magic'' property of WIMPs is that one does not need to specify their particle physics nature in order to explain cosmological observations. Even generic properties of WIMPs, i.e. their masses and cross-sections, do not need to be known for cosmological purposes. All existing observations of the cosmological structure appear to be consistent with the WIMP hypothesis, as WIMPs are too heavy to free-stream significantly, and too weakly interacting to damp the cosmological fluctuations. There are effects that are expected to be detectable on very small scales for some classes of WIMPs, however.

\paragraph{Damping effects on the power spectrum.} One needs some guiding principle in order to study possible effects of dark matter particles on the formation of structure, and to distinguish different models through the measurements of the large-scale structure. Such a principle is often provided by {\it damping} effects on the cosmological matter power spectrum, on small scales. Such effects are generated by the so-called {\it free-streaming} of the dark matter particles and/or their {\it collisional damping}. A rough estimate of the free-streaming scale shows that it depends only on the time when dark matter particles become non-relativistic, and therefore only on the mass of the particles; the heavier (i.e. colder) the dark matter particles, the smaller the free-streaming (damping) scale~\cite{Boehm:2000gq,Boehm:2004th}. This implies that warm or hot dark matter particles leave an observable fingerprint on the power spectrum of the matter distribution, as the small-scale structure is damped more strongly in those cases~\cite{Viel:2013apy,Boyarsky:2008xj}. This means that more of the very small clumps are generated in the primordial Universe for cold dark matter compared to warm dark matter~\cite{Boehm:2014vja}. Two interesting warm dark matter candidates are keV sterile neutrinos~\cite{2012PDU.....1..136B} and gravitinos. In particular, gravitino, as a next-to-lightest supersymmetric particle (NLSP), can explain a reduction in power on dwarf galaxy scales, and is compatible with the LHC and BBN/CMB constraints~\cite{Allahverdi:2014bva}. Both sterile neutrinos and gravitinos have extremely weak interactions. 

A more realistic calculation of the free-streaming scale shows that it depends on three moments in cosmic history: $t_{\rm dec}$, when the particles decouple; $t_{\rm nr}$, when the particles become non-relativistic; and $t_{\rm eq}$, the time of matter-radiation equality. $t_{\rm dec}$ and $t_{\rm nr}$ are determined by the interaction strength and mass of the dark matter particles respectively. $t_{\rm eq}$, on the other hand, depends on the history of the Universe. As $t_{\rm eq}$ is assumed to be the same for all dark matter candidates, they are usually classified based on two quantities: their interaction rate $\Gamma$, and their mass $m_{\rm DM}$~\cite{Boehm:2000gq,Boehm:2004th}. In a more realistic framework, one should include the so-called collisional (or interaction) damping in addition to free-streaming (or self-damping). Both effects are there, but collisional damping happens first, before the time of decoupling when dark matter particles start to free-stream. Cosmological fluctuations are therefore damped first by collisions. In terms of the implications for cosmic structure, interacting dark matter affects the matter power spectrum by some additional damping on very small scales, generating very small clumps in the primordial Universe. Such small-scale effects are in agreement with present observations of the cosmological structure, and are only at the limit of our ability to measure the matter power spectrum on the relevant scales. Currently, one can only exclude some values of the elastic scattering cross-sections~\cite{Boehm:2006mi}.

We should add here that some dark matter candidates affect small scales through an extra damping, which comes from strong self-interactions of the particles. Self-interacting dark matter (SIDM) may provide a natural solution to the core/cusp problem (see Sect.~\ref{sec:smallscale}), and this has in fact been the initial motivation for introducing SIDM models~\cite{Elbert:2014bma}. Although the extra damping (through self-interaction~\cite{Buckley:2014hja} or collisions with standard particles such as radiation~\cite{Schewtschenko:2014fca}) is a feature of various dark matter models, the existing large-scale structure constraints so far favor no observable deviations from the standard cold dark matter picture. Such deviations on very small scales may however be observable in the CMB~\cite{Wilkinson:2013kia}, and can be looked for using future small-scale CMB experiments.

In addition to damping, dark matter particles can in principle also decay through interactions with other particles or self-interactions. Such decaying dark matter (DDM) particles may affect the properties of cosmic structure both primordially and at late times. Such models are strongly constrained, as if the decay is too fast one would expect to see large emissions of electromagnetic radiation from the decay (unless two types of dark matter particle are introduced). In addition, simulations of the small-scale structures in a Milky Way mass halo show that the abundance and structure of subhalos are significantly altered for DDM compared to CDM, even if the DDM decay time is as large as (e.g.) $40~{\rm Gyr}$~\cite{Wang:2014ina}. 

\paragraph{WIMPs under pressure?} Although most models of dark matter proposed so far have been based on the WIMP hypothesis, they seem to be under pressure to survive. Normal and heavy WIMPs, although still able to explain the gravitational effects we expect from dark matter particles, have cross-sections that are too small to affect the large-scale structure significantly. It is therefore very hard to detect their particle properties through measurements of structure formation. In addition, if they are heavy, no signal is expected to be detected in particle experiments, such as the LHC. Because they have almost no observable effect on the LSS, the majority of efforts for testing and constraining WIMP models are still on the particle physics side, through direct, indirect, and collider experiments. If not too heavy, WIMPs are arguably about to be excluded, as their parameter space should soon be fully explored though the particle and astroparticle experiments. If they are very heavy, on the other hand, they may be completely invisible, as no particle experiments will be able to detect them. If that happens, WIMPs will enter a situation where they may survive forever theoretically, while interest in them may be lost in about 5-10 years unless some breakthrough occurs.

The situation for lighter WIMPs is more promising, as their effects on large-scale structure should soon be observable. The problem with these dark matter candidates is more theoretical, however; it is much harder to construct viable theories for light WIMPs, and such theories are usually much more constrained. Let us point out here that there are also non-WIMP theories for dark matter that could be tested through their effects on structure formation. One interesting example is ultra-light axions \cite{2015arXiv151007633M}. Although such scenarios have very small (almost unobservable) effects on the CMB power spectrum, their effects on the matter power spectrum can be clearly distinguished from that of cold dark matter.

\paragraph{Is it particle dark matter at all?} We end this section by emphasizing the fact that although the existence of some kind of dark matter is strongly suggested by all cosmological observations, no dark matter particles have been found so far. As such, we are allowed to ask whether dark matter particles really exist. What would be the right thing to do if the situation continues to be the same in the coming years? We should not forget that as dark matter has been ``observed" so far only gravitationally, it is still legitimate to think that it might just be a purely gravitational effect, e.g. only an artifact of modifications of gravity on large scales. This route has not been sufficiently explored, perhaps mainly because existing modified gravity alternatives to dark matter have not been successful in explaining even the most basic cosmological observations, such as the peaks on the CMB power spectrum. It seems very difficult to come up with a theory of gravity that mimics dark matter on all scales, and explains all of the different cosmological observations.

To conclude, we would like to point out again that CDM has been incredibly successful in explaining various large-scale structure observations in the linear regime, for example the measured power spectra of the CMB and late-time matter distribution. Therefore, if one wants to test beyond CDM candidates for dark matter using cosmological observations, the right place to look seems to be the small scales, where the non-linear regime becomes important. As we discussed above, most non-CDM theories predict observable effects on such scales; the same is expected for modified gravity alternatives to dark matter.

\subsection{Dark matter direct detection: does it matter, and what if it never happens?}
\label{sec:dmdetection}
\noindent\textit{Discussion session chairs: J. H. Davis, S. Riemer-S{\o}rensen \& D. Spolyar}

Dark matter is now an observational fact. We see the effects of dark matter from rotation curves~\cite{1970ApJ...159..379R}, gravitational lensing~\cite{2013SSRv..177...75H}, the Bullet cluster~\cite{2004ApJ...606..819M,1998ApJ...496L...5T}, and even in the cosmic microwave background~\cite{2015arXiv150201589P}.  In fact, the CMB shows that nearly 85\% of the matter in the Universe is dark matter~\cite{2015arXiv150201589P}. 

The first evidence for dark matter dates to the mid-1920's, when Knut Lundmark first saw the effect of dark matter in rotation curves \cite{1925MNRAS..85..865L}. The luminosity of a galaxy falls off exponentially toward the edge of a galaxy. From Newton's Law of Gravitation, the velocity of stars should drop as well if the only matter present was visible, but in fact the rotation curves of spiral galaxies stay constant. Assuming the Law of Gravitation holds, extra ``dark" (invisible) matter is therefore necessary.  The dark matter hypothesis was not fully believed for nearly another 50 years until the ground breaking work of Rubin \& Ford~\cite{1970ApJ...159..379R}, who used more modern techniques to carefully measure the rotation curves of galaxies confirming the initial results of Lundmark. Subsequently, a consensus picture has evolved: dark matter is ``cold'',\footnote{The temperature refers to the thermal freeze-out of the particle. Heavier particles freeze out earlier and are thus colder than lower-mass particles.} and does not interact with itself or through electromagnetism.

On the theory side, there are many cold dark matter candidates, including WIMPs (see Sect.~\ref{sec:newpart}), axions, sterile neutrinos etc. Many of these models are naturally found in extensions of the Standard Model, such as with supersymmetry (SUSY) in the guise of the lightest supersymmetric particle for WIMP dark matter. 

\paragraph{Looking for dark matter.}
In many instances, dark matter could decay or annihilate into Standard Model particles, potentially allowing {\it indirect detection}. For instance, the Fermi satellite searches for dark matter annihilation into gamma rays from dwarf galaxies~\cite{2015arXiv150302641F, Scott:2009jn}. We can also look for a signal in neutrinos. There are limits on the spin-dependent dark matter scattering cross-section from capture in the Sun~\cite{2014PhRvD..89a3010L}, e.g. from IceCube and Super Kamiokande. In some cases, dark matter can scatter in the Sun and become trapped by its gravitational potential, leading to a build-up in the solar core. If dark matter self-annihilates into neutrinos, we can observe the neutrinos created due to the dark matter annihilation in detectors on Earth, allowing upper limits to be placed on the scattering cross-section.
 
Dark matter could also scatter off ordinary matter ({\it direct detection}). As the detectors become larger and larger, new backgrounds can become important.
We also discussed the future of dark matter direct detection experiments~\cite{2013arXiv1310.8327C,Akrami:2010cz} and the potential for upcoming tonne-scale experiments~\cite{Akrami:2010dn} to reach the so-called neutrino floor. This is where direct detection experiments become so large that the number of neutrinos from the Sun and cosmic ray decays in the atmosphere become of a similar size to the potential signal. Hence at this stage it becomes important to be able to separate the two conflicting signals effectively.

\paragraph{Beyond the neutralino.}

There are many different models of dark matter, but it is fair to say that most dark matter scenarios look at WIMPs, and are constructed in the context of SUSY (see e.g. Ref.~\cite{Akrami:2011vh}).
Without any detections of supersymmetry at the LHC yet, the standard SUSY neutralino dark matter candidate has lost part of its shine. As such, we discussed alternative candidates such as keV sterile neutrinos and axion-like particles. 

\begin{figure}[t]
\centering
\includegraphics[width=.6\hsize]{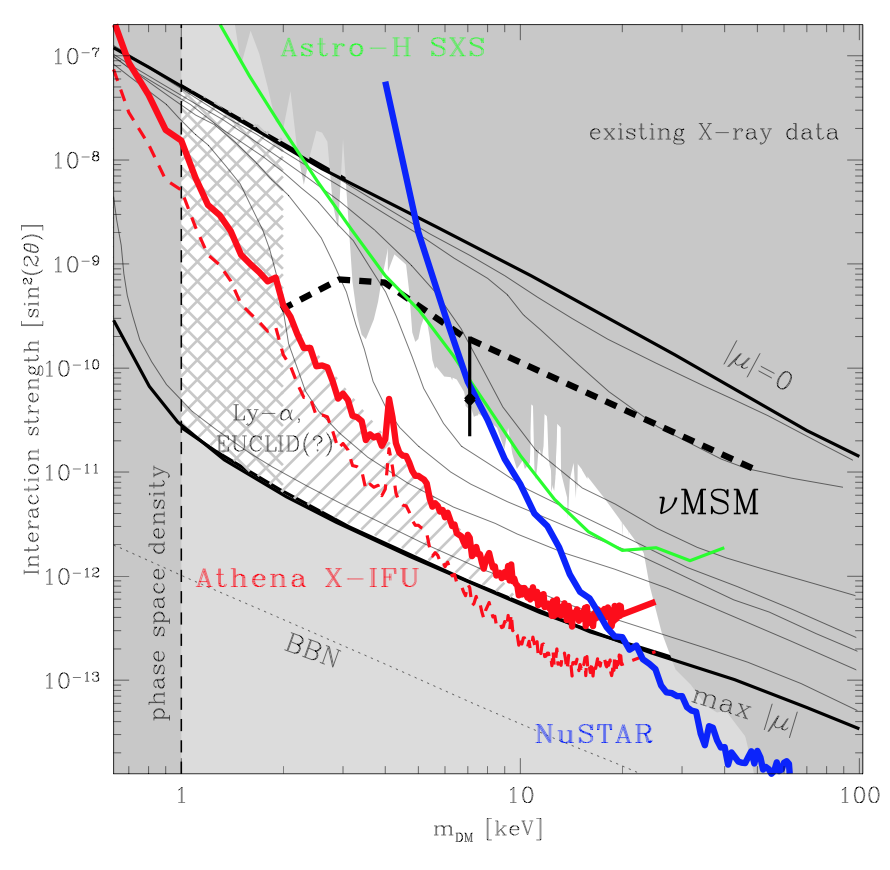}
\caption{Known constraints on the mass and mixing angle of sterile neutrino dark matter (grey shading). Black solid thin and thick curves show the theoretical predictions for various values of the primordial lepton asymmetry leading to the observed dark matter density today. The shaded grey area to the right is constrained by non-observations in X-ray spectra, while the 3.5 keV signal suggested by Ref.~\cite{2014ApJ...789...13B} is the data point with uncertainties. The hatched range shows the sensitivity reach of the future Lyman-$\alpha$ and weak-lensing probes. The red thick solid curve shows the sensitivity limit of Athena X-IFU, calculated assuming the minimal Segue 1 dSph signal, for a 1 Msec exposure. The dashed thin red curve is the sensitivity limit for the average mass estimate. The green curve shows the sensitivity of Astro-H / SXS, and the blue curve corresponds to the sensitivity of NuSTAR, also for 1 Msec long exposures. (Reproduced from Ref.~\cite{2015arXiv150902758N}.)
\label{fig:sterile}
}
\end{figure}

\begin{figure}
\centering
\includegraphics[width=.6\hsize]{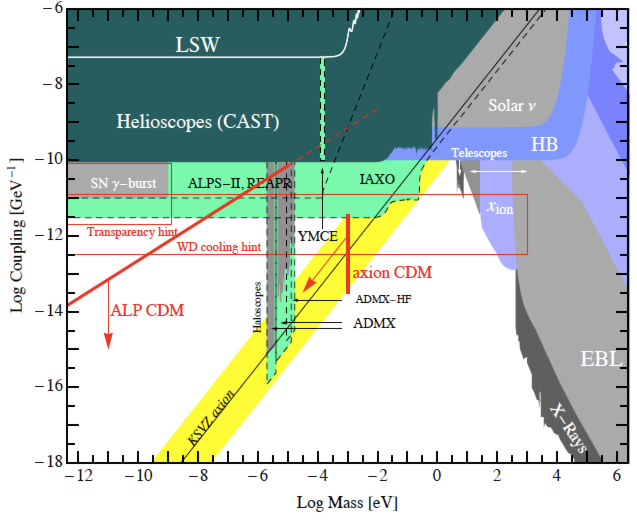}
\caption{The parameter space for axion-photon coupling versus mass of the axion-like particles. The standard QCD axion solving the strong CP problem is the yellow band. The width of the yellow band gives an indication of the model-dependence in this coupling, though the coupling can even be tuned to zero. (Reproduced from Ref.~\cite{2013PhRvD..88c5023G}.)
\label{fig:axions}
}
\end{figure}

The detection of excess line emission in stacked X-ray spectra of galaxy clusters~\cite{2014ApJ...789...13B} has significantly boosted the number of alternative candidates that could give rise to such signals, even though its origin remains debated (see Ref.~\cite{2015arXiv151000358I} and references therein for an overview of current detections and non-detections). 
The original paper suggested keV sterile neutrinos (e.g. Ref.~\cite{2012PDU.....1..136B} and references therein) as a possible explanation (see Fig.~\ref{fig:sterile}), but a wealth of alternatives have been suggested including minimal decaying dark matter, axion like particles (ALPs), non-thermal two-component dark matter, and milli-charged dark matter particles, as well as the decay of excited dark matter states, annihilating dark matter, and dark matter decaying into axion-like particles with further conversion to photons in the cluster magnetic field (see Ref.~\cite{2015arXiv151000358I} for an extensive list of references).
ALPs seem like a `natural' extension to the Standard Model, as they solve the strong CP problem, and there is enough flexibility within the models to provide alternative 
candidates. We discussed searches for axions with e.g. the ADMX experiment~\cite{2013PhRvD..88c5023G}, as summarized in Fig.~\ref{fig:axions}.

\paragraph{Astrophysics shines light on dark matter.}
Today, we look at dark matter through the lens of theory, which motivates the many searches for dark matter.  We could instead take a more agnostic approach however, and see what astrophysics can teach us about the particle physics properties of dark matter. Two open questions that still need to be understood are: Is dark matter really cold? And does dark matter interact with itself?

Dark matter is currently believed to be cold, which implies that clumps of dark matter exist on scales down to Earth mass and even smaller.  If dark matter was warm, its free-streaming length would prevent the appearance of halos with a virial mass below $10^9$ M$_\odot$. Cold dark matter simulations (without baryons) imply that the Milky Way should have many more dwarf galaxies than we observe (see Sect.~\ref{sec:smallscale}); one possible explanation of this ``missing satellite'' problem is warm dark matter~\cite{2015ApJ...803...28C}. In the warm dark matter scenario, the total number of halos that could host dwarf galaxies is dramatically reduced. 
  
\begin{figure}[htb]
  \centering
  \includegraphics[width=.6\hsize]{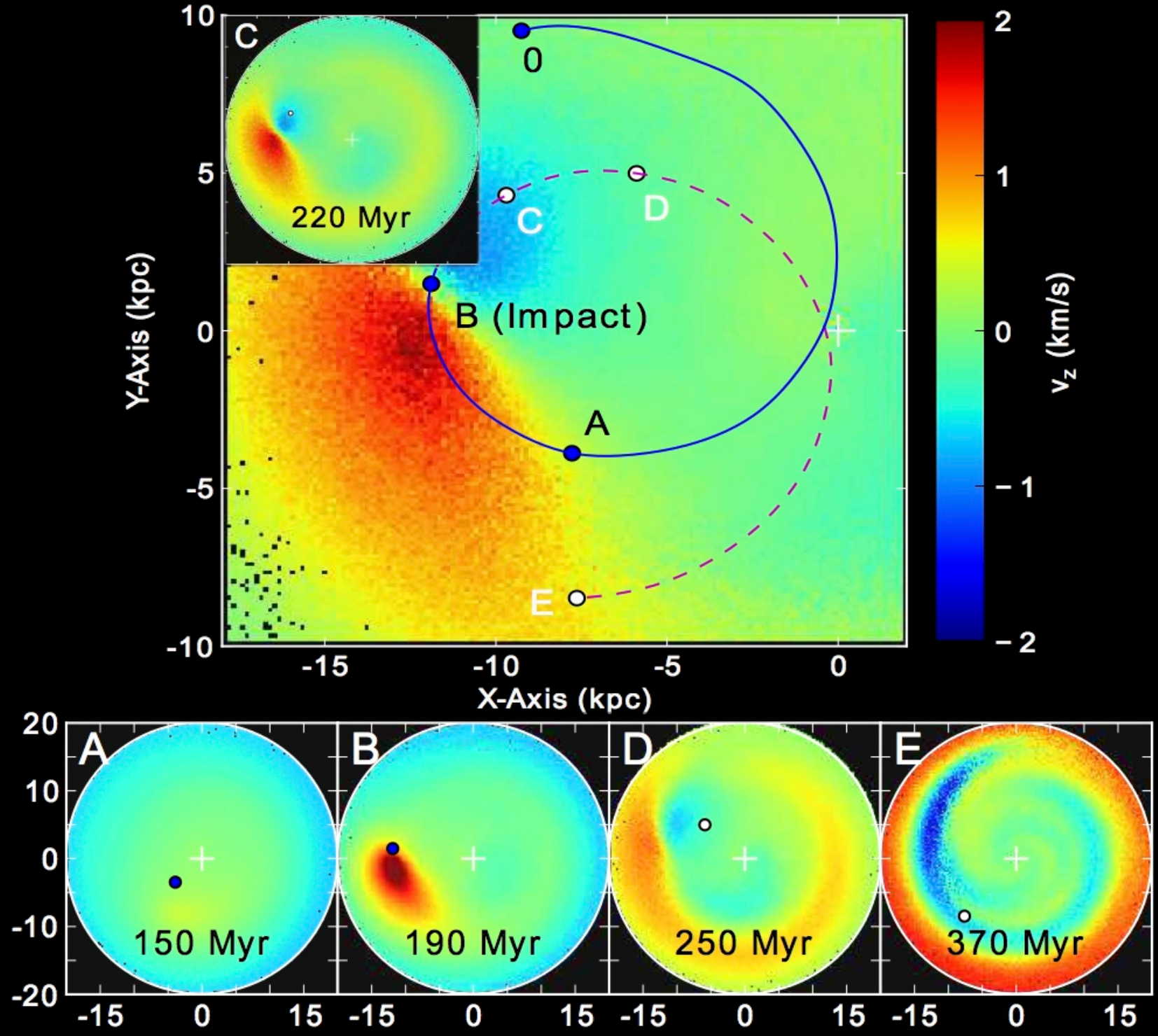}
\caption{The white cross represents the center of the galaxy. The white dot shows the position of the sub-halo above the disk. The blue dot shows the sub-halo below the disk.  Shown is the vertical velocity perturbation of the stars in the disk.  The time shown is the time elapsed since the start of the simulation. The sub-halo has a virial mass of $10^9$ M$_\odot$ and a tidal mass of ~$10^8$ M$_\odot$. The halo begins around 30 kpc above the disk and descends vertically down through the disk, and passing through it around 9 kpc from its center. (Reproduced from Ref.~\cite{2013arXiv1310.2243F}.)
\label{fig:dm:1}
}
\end{figure}

There are other ways of probing the existence of substructure. Ref.~\cite{2013arXiv1310.2243F} found that, as sub-halos pass through the galactic disk, the sub-halo can leave a tell-tale imprint in the motion of the stars in the disk, 
by drawing them up and pushing them down (see Fig.~\ref{fig:dm:1}).
Present missions such as Gaia and future follow up missions could detect the perturbations from halos with a virial mass far below $10^9$ M$_\odot$. The detection of a halo with a mass much less than $10^9$ M$_\odot$ would be a dramatic confirmation of the cold dark matter scenario.

Dark matter may also have a self-interaction, which could potentially be very large.  The present constraint on the interaction strength of dark matter comes from the shapes of clusters, and is fairly weak, $< 0.1$ cm$^2/$g~\cite{2013MNRAS.430..105P}.  
Interacting dark matter could explain why dwarfs have cores \cite{2000PhRvL..84.3760S} (where the dark matter density becomes constant towards the center of the halo).
The standard cold dark matter  scenario (without baryons) typically has a cusp, where 
the dark matter density diverges towards the halo center. This disagreement goes by the name of the cusp/core problem, as discussed in Sect.~\ref{sec:smallscale}.

Dwarf galaxies are dark matter-dominated. Stars can be used to trace-out the gravitational potential, which gives the dark matter profile of the galaxy. Unfortunately, the orbital structure of the stars is degenerate with the dark matter profile. Hence, depending upon what we assume about the orbital structure of the stars, one might think the galaxy has a core, when in fact the halo has a cusp. Most studies only use the 2D spatial information (density of stars) and the velocity of the stars along the line of sight. Without further information one can not determine the dark matter profile; see Refs.~\cite{2014MNRAS.440.1680R,2015IAUS..311...16M} for more details.

With additional velocity or spatial information, one can determine the orbital structure of the stars and infer the true dark matter density. Ref.~\cite{2014MNRAS.440.1680R} showed that inclusion of the 3D position of stars can break the degeneracy and determine the dark matter profile. They used RR Lyrae, which are variable stars.  The variability is related to the absolute luminosity of the stars, which allows an observer to determine the true distance to the star.  With additional velocity information, one can also break the degeneracy; see Ref.~\cite{2007ApJ...657L...1S}.

In summary, while the standard cold dark matter scenario has been very successful, many open questions still remain, particularly: what is the dark matter?  Along these lines,
model building purely with supersymmetry is a little pass\'e, and one should consider alternatives.  The determination of the properties of dark matter will
need to use many different channels, both from direct and indirect detection. From a more agnostic perspective, astrophysics still has much that it can teach us about the nature of dark matter, such as how cold it is, and whether it interacts with itself.

\section{Inflation and the early Universe}

The elegance and simplicity of the inflationary paradigm have led to its rapid and widespread acceptance, despite its far removal from the realm of direct testability. Doubts have begun to creep in in recent years however, as the decisiveness of indirect observational tests has been questioned, and concerns about its predictivity have multiplied. Is inflation really the compelling, and above all {\it true}, description of the early Universe that justifies its inclusion as one of the pillars of modern cosmology? Or is it an overly-flexible, under-predictive model that simply defers the initial conditions problem rather than solving it, as some have suggested \cite{2011SciAm.304d..36S}?

In this section we cast a critical eye over inflation, and ask whether it really succeeds at what it is ostensibly designed to do -- explain the initial state of the Universe. We also take a look at current and future observational tests, and ask the question of whether it is truly falsifiable, even in principle.



\subsection{The inflationary paradigm: more problems than solutions?}
\label{sec:inflationproblems}
\noindent{\it Discussion session chairs: M. Rinaldi \& J.-P. V\"aliviita}

\noindent The initial  question in our discussion is about definitions: how do we define, in modern terms, the ``inflationary paradigm?'' There is a general consensus on a minimal set of elements that the paradigm should have, which are that:
\begin{itemize}
\item Initial conditions for inflationary perturbations should be compatible with homogeneity and isotropy (see, however, Ref.~\cite{Borde:2001nh}, where it has been shown that (on average) no expanding spacetime is geodesically complete to the past, which strongly suggests that inflation cannot resolve any singularity problems). The Bunch-Davies vacuum seems to be the best set-up for primordial quantum fluctuations.
\item Inflationary expansion must be associated with some degree of freedom beyond gravity. Whatever this degree of freedom is (scalar, vector or gauge field, $f(R)$ extension of gravity, extra dimensions, etc.), the associated effective potential should be simple, with a minimal amount of fine-tuning.
\item Small patches are expected to inflate into large, flat, homogeneous, and isotropic volumes with almost Gaussian perturbations. (This condition is strictly linked to observations.)
\end{itemize}
The next question is whether (and how) we need to modify the inflationary paradigm in view of the recent results from Planck \cite{2015arXiv150202114P}.
The first observation is that the Planck results clearly point at inflationary potentials with a plateau or a Higgs-like shape. However, this is not enough to nail down the exact shape of the potential. In fact, as in the Standard Model of particle physics, the true potential might be extremely complicated, even if it looks simple locally. If we adopt this point of view, however, we must surrender to the idea that inflation is not really a predictive theory since it is impossible to determine the true shape of the potential outside the energy regime accessible by our probes. On the other hand, one can argue that the strength of inflation is instead that it does not depend at all on the details of the potentials, as it is a general phenomenon. This idea is particularly evident in the case of the $\alpha$-attractor models \cite{Galante:2014ifa} (see also Ref.~\cite{Linde:2014nna}).

Recently, some researchers have pointed out that plateau-like potentials may in fact seriously jeopardize the inflationary paradigm, in sharp contrast with the common belief that they are a desirable feature \cite{2013PhLB..723..261I}. They stress that the Planck data show that the energy density associated to the potential during inflation is about 12 orders of magnitude below the Planck density ($M_{\rm Pl}^{4}$). The typical inflationary paradigm, originally suggested by Linde~\cite{Linde:1983gd}, assumes equipartition of the energy density between kinetic and potential at the onset of inflation, according to the ``chaotic'' distribution $(\partial\phi)^{2} \sim V(\phi)\sim M_{\rm Pl}^{4}$. The findings of Planck sharply contrast with this assumption, since the potential energy density is several orders of magnitude smaller, rendering the equipartition configuration exponentially unlikely. As a result, gradients rapidly dominate at the onset of inflation, enhancing inhomogeneities rather than washing them out. In addition, plateau-like potentials require a much higher degree of fine-tuning than power-law ones, and this worsens the situation.

These controversial ideas have triggered a heated debate, and some authors (see e.g. Ref.~\cite{2014PhLB..733..112G}) have pointed out that the assumptions of Ref.~\cite{2013PhLB..723..261I} are too unrealistic, focusing in particular on their indirect assumption that the potential is the same all the way up to the Planck scale (which relates to our previous point on the true shape of the potential), and that the curvature term is exactly vanishing. Indeed, a small negative curvature is sufficient to restore the correct size of the homogeneous patch at the onset of inflation. In addition to this, multifield inflation can resolve this problem through multiple inflationary stages, along the lines proposed in Ref.~\cite{1988PhLB..202..194L}. More recently, by means of numerical simulations it has been shown that inflationary expansion can happen even in the presence of large inhomogeneities \cite{2015PhRvD..92f3519C} (see Ref.~\cite{Linde:2014nna} for additional discussions).

We believe that the question is not settled yet, and the debate will continue for some time -- at least until more precise information about the polarization signal from primordial fluctuations reveals the amount of relic gravitational waves (which is related to the initial potential energy density, so far only bounded from above by observations). 

Looking to the future, other possible observations could be made that can set the inflationary paradigm on a firmer (if not definitive) footing. For example, if a direct measurement of the tensor spectral index, $n_T$, and/or the non-Gaussianity parameter, $f_{\rm NL}$ (which is quite difficult right now) gives the theoretically-expected results, many alternative explanations of inflation would hardly stand. Conversely, there exist ``anti-smoking guns'' that could potentially kill the inflationary paradigm at once. The most powerful one would certainly be the discovery of a very large-scale asymmetry in the cosmic microwave background. A too-large value of $f_{\rm NL}$ would also be very hard to reconcile with single-field inflation.

Among the various alternative models of inflation, in recent years modified gravity theories have gained much popularity; see e.g. Ref.~\cite{2010LRR....13....3D} for a review. These are inspired by the proposal of Starobinsky \cite{1980PhLB...91...99S} back in 1980, based on the simple Lagrangian $R+aR^{2}$, and still in very good agreement with Planck data. The quadratic term in the Ricci scalar $R$ was inserted as a phenomenological manifestation of quantum corrections to the Einstein-Hilbert Lagrangian. In principle, however, any scalar obtained by combining the Ricci tensor and its contractions can be inserted into the Lagrangian, and particular combinations are suggested by more sophisticated quantum corrections to gravity, as shown in pioneering works such as Ref.~\cite{1977PhRvD..16..953S}. The exact form and true nature of these higher-derivative corrections are deeply rooted in quantum gravity, and more theoretical investigations are necessary to advance in this field. Inspired by these quantum gravity corrections, several models of inflation have been proposed in an $f(R)$ form, although the work of Stelle in Ref.~\cite{1977PhRvD..16..953S} and subsequent calculations in (e.g.) string theory support a very particular form of the quantum gravity corrections that are not of the $f(R)$ type. At the basic level, current observations cannot really distinguish between various versions of $f(R)$. This can be understood by remembering that, with a suitable conformal transformation, all $f(R)$ theories can be turned into scalar-tensor theories with a standard kinetic term and some kind of potential.\footnote{An exception exists for pure quadratic theories, as the conformal transformation may become singular \cite{2015arXiv150503386R, 2015PhRvD..91l3527R}.} In a way, the problem of the functional form of $f(R)$ goes back to the problem of finding the exact shape of the scalar potential. Thus, one possible way to distinguish between various $f(R)$ theories would be a careful measurement of non-Gaussianity. 

\subsection{Can we prove it was inflation? fundamental limits to observations}
\noindent\textit{Discussion session chairs: S. Clesse, F. Finelli \& D. Steer}
\label{sec:infobs}

Our discussion begins with a short reminder of the Planck results, which are in remarkable agreement with a flat \lcdm{} model with nearly scale-invariant, 
Gaussian and adiabatic primordial scalar perturbations, as predicted by the simplest single-field slow-roll models of inflation. Four important questions related to the status
of inflation and the perspectives given by future experiments will then be addressed:
\begin{enumerate}
 \item What can be learned about the shape of the potential in slow-roll inflation?
 \item What are the minimal predictions and consistency conditions for multi-field inflation?
 \item What is the status of topological defects and their observable predictions?
 \item Do viable alternatives to inflation exist?
\end{enumerate}

\paragraph{Inflation after Planck.}
The most recent Planck results \cite{Ade:2013sjv,Adam:2015rua} are in very good agreement with a nearly flat Universe, and a primordial power spectrum of nearly Gaussian adiabatic density perturbations with a red tilt, together with a relatively small amount of primordial gravitational waves. These properties are expected for some of the simplest models of inflation, that invoke a single, minimally-coupled, slowly-rolling scalar field~\cite{2014PDU.....5...75M, Planck:2013jfk,2015arXiv150202114P}. The constraints on the relevant inflationary parameters are reported in Table~\ref{tab:Planck2015}.   

\begin{figure}[h]  
\begin{center} 
\includegraphics[scale=0.65]{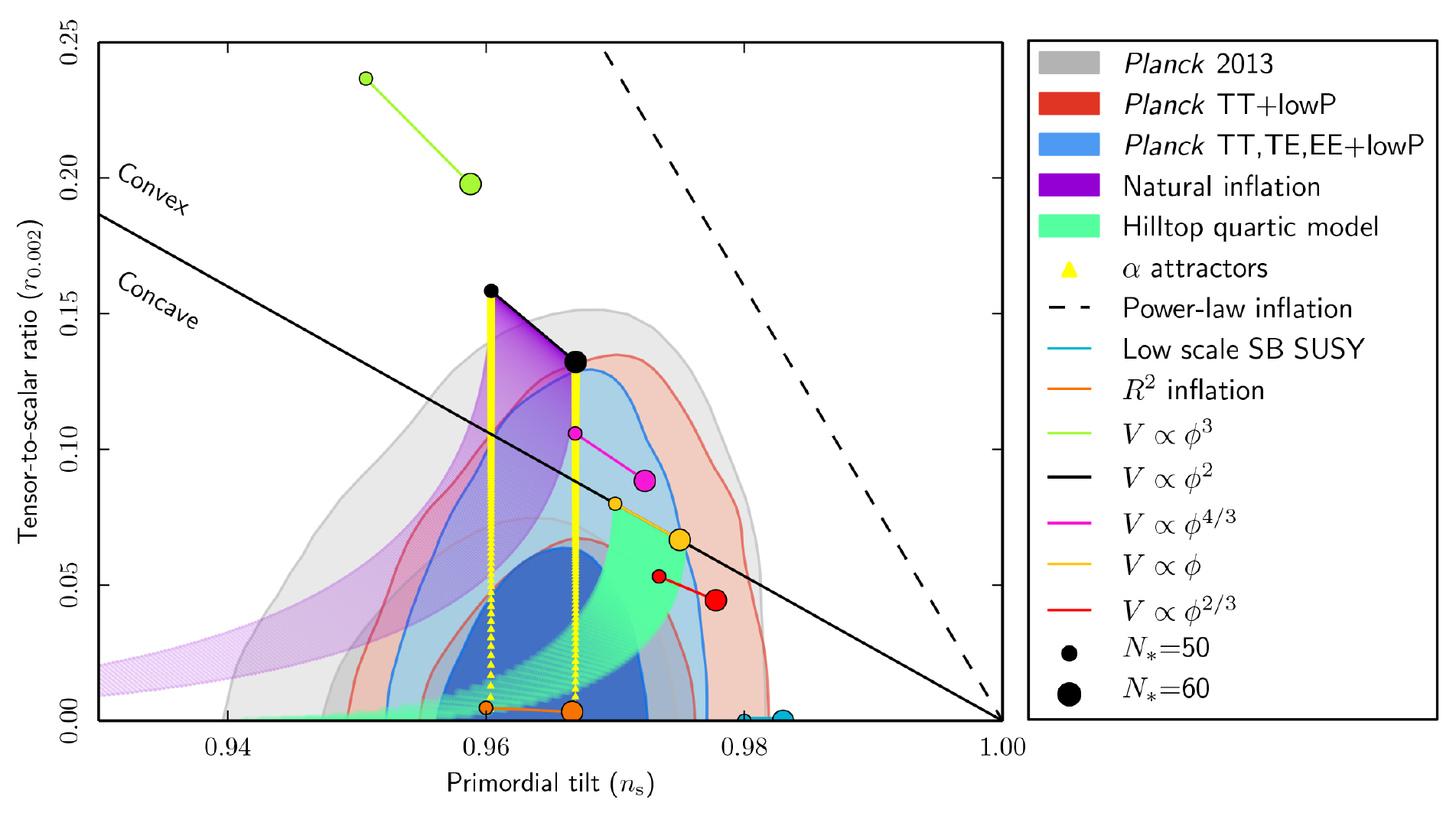}
\caption{Planck 2015 constraints 
in the $r-n_{\rm s}$ plane, and predictions for a selection of the simplest inflationary models including 
reheating uncertainties. Figure from Ref.~\cite{2015arXiv150202114P}, 
reproduced with permission from Astronomy \& Astrophysics, ESO; original
source ESA and the Planck Collaboration.  
\label{fig:Planck}}
\end{center} 
\end{figure}

As illustrated in Fig.~\ref{fig:Planck}, the Planck results also exclude with high 
significance (or strongly disfavor) some well-known inflationary models, such as the original hybrid inflation, polynomial potentials $V(\phi) \propto \phi^p$ with $p > 2$, 
intermediate inflation, power-law inflation, and the simplest supersymmetric F-term and D-term models. Furthermore, plateau-like concave potentials are preferred over convex ones.

The purpose of the remainder of this section is to discuss the capabilities of future experiments to further constrain the physics of inflation.

\begin{table}[h]
{\centering
\begin{tabular}{|c|c|c|} 
\hline
  & Parameter & Planck 2015 \\
  \hline
  Curvature  & $\Omega_{\rm K}$  & $-0.005^{+0.016}_{-0.017}$ \\
  Scalar spectral index & $n_{\rm s}$ & $0.968 \pm 0.006$ \\
  Running & $\mathrm{d} n_{\rm s} / \mathrm{d} \ln k$ & $-0.003 \pm 0.007$ \\
  Tensor to scalar ratio & $r_{0.002}$ (95\% CL) & $ < 0.11 $ \\
  Local non-Gaussianity (NG) & $f_{\rm{NL}}^{\rm{local}}$   & $2.5 \pm 5.7 $ \\
  Equilateral NG  & $f_{\rm{NL}}^{\rm{equil.}}$  & $-16 \pm 70 $  \\
  Orthogonal NG & $f_{\rm{NL}}^{\rm{ortho.}}$  & $-34 \pm 33 $  \\
  Isocurvature fluctuations & $\beta_{\rm{iso}}$ (uncorr. CDM, 95\% CL) & $< 0.039$   \\
  Cosmic strings (Nambu-Goto) & $G \mu$ (95\% CL) & $<1.8 \times 10^{-7}$ \\
\hline
\end{tabular}
\caption{Planck 2015 constraints on parameters relevant for inflation \cite{2015arXiv150201589P,2015arXiv150202114P,Ade:2015ava}.}
\label{tab:Planck2015}
}
\end{table}

\paragraph{What can we learn about the shape of the potential in slow-roll inflation?}

Here, we focus only on large-scale constraints on inflation and the primordial power spectrum; see e.g. Ref.~\cite{Bringmann:2011ut} for small-scale probes.

The Planck results are consistent with simple shapes of the inflaton field potential, as illustrated in Fig.~\ref{fig:Planck}. Current efforts to measure B-mode polarization  (ACTpol~\cite{2010SPIE.7741E..1SN}, SPTpol~\cite{2012SPIE.8452E..1EA}, PolarBear~\cite{2010arXiv1011.0763T}, BICEP3~\cite{2014SPIE.9153E..1NA}, SPIDER~\cite{2013JCAP...04..047F}, EBEX~\cite{2010SPIE.7741E..1CR}, ABS~\cite{EssingerHileman:2010hh}) and the next generation of CMB experiments (CLASS~\cite{Essinger-Hileman:2014pja}, PIPER~\cite{Lazear:2014bga}, LSPE \cite{2012arXiv1208.0281T}, LiteBird~\cite{2014JLTP..176..733M}, Core+~\cite{2011arXiv1102.2181T}) will be able to further constrain the tensor-to-scalar ratio, $r$, with a possibility to probe values of $r$ as small as $\sim 10^{-3}$ (subject to the complexity of the foreground contamination \cite{Ade:2015tva}). One can consider which inflationary models can be probed in the future, given the current measurements of the scalar spectral index $n_s$, by the standard approach of selecting a potential, or alternatively by using families of parameterizations for the slow-roll parameters \cite{Mukhanov:2013tua}. Even without a positive detection of primordial B-mode polarization, but with an upper limit of the order of $r \lesssim 10^{-3}$, the Lyth bound \cite{Lyth:1996im} could be verified, i.e. the excursion of the scalar field in the observable range would be of the order of the reduced Planck mass. The properties of slow-roll inflation can be further constrained with the use of future large-scale structure surveys. In particular, it would be possible to further squeeze the running of the scalar spectral index close to values of $\approx (n_s-1)^2 \approx 10^{-3}$ \cite{Basse:2014qqa}. 

21cm experiments will also play an important role in constraining inflation. By measuring the 21cm signal at high redshift, one can probe a much larger volume than with lower-redshift galaxy surveys, as well as observing several distinct epochs of cosmic history: the post-reionization and reionization epochs, and the end of the dark ages. Ultimately, with Earth-based~\cite{Tegmark:2008au,Clesse:2012th} and space-based ``omniscopes'' (which avoid atmospheric contamination at low frequencies), it is in principle possible to probe higher redshifts, up to $z\sim100$. Forecasts for optimal configurations of the SKA, in a very optimistic situation where the details of reionization are known, show that it is potentially feasible to improve the limits on the spectral index and the running by two or three order of magnitude~\cite{Mao:2008ug}. Even having in mind the limits of the semi-realistic assumptions of these predictions, the 21cm signal seems a promising avenue for probing the running of the scalar spectral index.

\paragraph{What are the minimal predictions and consistency conditions for multi-field inflation?}
Finding a non-zero amplitude for isocurvature fluctuations would be of key importance to probing multi-field dynamics during inflation. In the near future, the final measurement of the polarization pattern of CMB anisotropies from Planck will be the most important test for a mixture of adiabatic and isocurvature fluctuations motivated by multi-field inflationary models. However, one needs to keep in mind that, without convincing evidence for a non-zero isocurvature fraction, it is difficult to make definitive statements about multi-field dynamics during inflation. For instance, inflation might have been driven by many fields, but isocurvature perturbations could have been fully converted to adiabatic ones at the time of nucleosynthesis.

From the theoretical point of view, the consistency relation for local non-Gaussianity, $f_{\rm{NL}}^{\rm{local}} = 5 (n_s-1)^2/12$, which holds for single-field slow-roll canonical models \cite{Creminelli:2004yq} would be an ideal target to distinguish single-field from multi-field models. Achieving a sensitivity of $f_{\rm{NL}}^{\rm{local}} \sim 0.01$ would be required however, which seems beyond the scope of all planned and proposed experiments.

There exists at least one alternative to isocurvature fluctuations and non-Gaussianity in order to test multi-field scenarios. CMB spectral distortions probe the primordial power spectrum on a range of scales that is wider than for the primary CMB anisotropies. Since the large majority of single-field models in current agreement with CMB observations cannot induce modifications on the smaller scales probed by CMB spectral distortions \cite{Clesse:2014pna}, this is another promising avenue to test the possibility of multi-field dynamics during inflation~\cite{Jeong:2014gna,Chluba:2015bqa}.

\paragraph{Inflation and cosmic strings.}
A number of inflationary models end with the production of a network of cosmic strings~\cite{Turok:1985tt,Stebbins:1987cy}, which leave a further imprint on the CMB. Planck has established bounds on the parameters related to topological defect models~\cite{2015arXiv150201589P}, such as the string tension: $G \mu \lesssim \rm{few} \times 10^{-7}$. Future probes such as lensing at radio frequencies, CMB spectral distortions, and radio bursts should improve these limits to varying degrees. Indeed, under certain optimistic assumptions, future 21cm experiments might even be able to constrain $G\mu$ in the range of $10^{-10}-10^{-12}$~\cite{Khatri:2008zw}.

\paragraph{Alternative models to inflation.}
A number of alternative models to inflation have been proposed, such as pre Big-Bang scenrios \cite{Gasperini:1992em}, ekpyrosis \cite{Khoury:2001wf}, matter bounce \cite{Finelli:2001sr}, and string gas cosmology \cite{Brandenberger:2006xi} (see Ref.~\cite{Brandenberger:2009jq} for a review).

In the first three alternatives mentioned above, structure formation relies on the amplification of quantum fluctuations by geometry, whereas string gas cosmology is based on a thermal origin. All these models have different predictions for tensor modes from standard inflation: pre-Big-Bang and ekpyrotic scenarios predict a blue tilt; the matter bounce driven by a canonical scalar field predicts a small red tilt, but with a different consistency condition for the tensor sector; and string gas cosmology predicts a slightly blue tilt (also for Galileon inflation, when the null energy condition is violated \cite{Kobayashi:2010cm}). In addition to these different predictions for the tensor spectrum, there are other observables that are also expected to be different in inflation compared to its alternatives. These include expectations for the running of the scalar spectral index in ekpyrotic and matter bounce models~\cite{Lehners:2015mra}, and predictions for the bi- and tri-spectra in ekpyrotic models~\cite{Lehners:2013cka,Fertig:2015ola}. 

With the numerous on-going and planned experiments to measure B-mode polarization, we could envision testing some of these alternative predictions for tensors against inflation, particularly if non-zero primordial gravitational waves in the CMB polarization pattern are detected.

\section{Spatial symmetries and the foundations of cosmology}
\label{sec:symm}

A handful of assumptions about the symmetry and statistical properties of spacetime have been instrumental in enabling the construction of the standard cosmological model, to the point that they are now an indispensable part of its foundations. A growing list of theoretical concerns with \lcdm{} give cause to re-examine these assumptions however, in case they could be the source of some of the problems.

In this section, we first examine how such seemingly fundamental assumptions as statistical homogeneity and isotropy can be tested in practice.\footnote{See also the summary of recent developments in Ref.~\cite{2015arXiv151203313B}.} We then move on to the question of detecting anomalies, with particular reference to apparent anomalies in the CMB. How seriously should we be taking these possible hints of deviations from isotropy, and what more can be done to decide their significance?

\subsection{Testing the foundational assumptions of \lcdm{}}
\label{sec:maartens}
\noindent{\it Plenary speaker: R. Maartens}

The \lcdm{} model relies critically on various physical assumptions. Perhaps the most fundamental assumption is the Cosmological Principle (CP), i.e. that the Universe is isotropic and homogeneous on large scales. A second fundamental assumption is that general relativity is the correct theory of gravity to describe the Universe beyond the quantum regime (see Sect.~\ref{sec:gravcosmo}).  

\paragraph{The Cosmological Principle.}
The CP is critical to the \lcdm{} model, and indeed to all dark energy models: accelerated expansion implies  dark energy only if we impose the CP. Modified gravity models also rely on the CP. Consequently, all cosmological tests of GR are based on the CP.

How do we test the CP? We cannot directly test homogeneity, since we cannot observe the cosmic microwave background or large-scale structure on 3D spatial hypersurfaces. We observe on the past lightcone at effectively a single cosmological time, which means that we can only directly test isotropy about our worldline. Isotropy about all galaxy worldlines implies homogeneity. If our worldline is not special, i.e. if we adopt a Copernican Principle, then we can deduce the Cosmological Principle on the basis of isotropy.

What is the best basis that we have for isotropy? Everyone would likely give the answer as `the CMB'. But on its own, an isotropic CMB does {\em not} necessarily imply an isotropic spacetime~\cite{1978MNRAS.184..439E}. The fundamental result that delivers geometric isotropy from an isotropic CMB is based on a 1968 theorem by Ehlers, Geren and Sachs, subsequently strengthened by Ellis, Treciokas and Matravers in 1985. Suitably generalized to include cold dark matter and dark energy, the statement is as follows \cite{Clarkson:2010uz}:
\begin{quotation}
\noindent{\it If collisionless radiation has a geodesic and expanding 4-velocity, if it has vanishing dipole, quadrupole and octupole, and if matter is pressure-free and dark energy has no anisotropic stress -- then the spacetime is Friedmann-Lema\^itre-Robertson-Walker (FLRW). }
\end{quotation}

This result is based on analysis of the fully non-linear Einstein-Liouville equations in a general spacetime, and is currently our best motivation for the FLRW geometry that underpins \lcdm{}. However, the Universe is not FLRW -- at best, it is {\it statistically} isotropic and homogeneous on large enough scales. Bridging the gap between the exact result and the real Universe entails a set of unresolved complexities in cosmology \cite{1995ApJ...443....1S,2010JCAP...03..018R}. How do we perform covariant averages in GR \cite{2010arXiv1003.4020V,Andersson:2011za,Clarkson:2011zq}? How do we define statistical homogeneity in spacetimes that are not perturbed FLRW? How do we find the transition scale to homogeneity-on-average without assuming perturbed FLRW? Confronted by these daunting problems, we can be forgiven for adopting a pragmatic approach: we assume that the Universe has a perturbed FLRW geometry on large enough scales, and devise consistency tests of this assumption. If we find no violation, this strengthens the evidence for the CP. If we find a single statistically significant violation, this could rule out the CP. There are various types of consistency test (see Refs.~\cite{2011RSPTA.369.5115M,Clarkson:2012bg} for a review).

\begin{itemize}

\item {\it Geometric tests} look for violations of the tight relationship between distances, Hubble rate and curvature in an FLRW spacetime \cite{2008PhRvL.101a1301C,2009PhRvL.102o1302Q,2015PhRvL.115j1301R}.  They rely on supernova data and measurements of the baryon acoustic oscillation scale.

\item {\it Probing inside the lightcone via galaxies,} which carry a fossil record of their star formation history. We then test whether galaxies at the same lookback time (which is derived from the radial BAO scale) have the same fossil record \cite{2013ApJ...762L...9H}.

\item {\it Probing inside the lightcone via the Sunyaev-Zeldovich effect.} The thermal SZ effect can test for anisotropy at distant clusters \cite{1995PhRvD..52.1821G,2012PhRvL.109e1303C} while the kinetic SZ effect can detect anomalous radial cluster velocities \cite{2008JCAP...09..016G}. Future measurements of SZ polarization will probe the remote CMB quadrupole and the cluster transverse velocity \cite{1997PhRvD..56.4511K}.

\item {\it The matter dipole} will be measured by next-generation all-sky galaxy surveys, and deviations from the CMB dipole would signal a breakdown of the CP \cite{2015aska.confE..32S}.

\end{itemize}

\paragraph{General relativity and \lcdm{}.}

GR is routinely assumed to underpin the standard model of cosmology, and it is clearly essential for analysis of the CMB. For current galaxy surveys, which probe subhorizon (and in fact sub-equality, $k > k_{\rm eq}$) modes, a Newtonian approximation is adequate. However, the huge-volume galaxy surveys planned by Euclid, SKA, and LSST will probe horizon-scale modes where GR effects can become significant -- providing the opportunity to mine new information from large-scale structure and to develop new tests of gravity on horizon scales (again see Sect.~\ref{sec:gravcosmo}). 

The GR effects arise because the  fractional number overdensity of galaxies is observed on the past lightcone (see Refs.~\cite{2014CQGra..31w4002B,2014CQGra..31w4001Y} for reviews). In Newtonian gauge,
\begin{eqnarray}
\delta^{\rm obs}_g 
&=&  b\delta_{\rm cs}
  -\frac{(1+z)}{{H}} (n^i\partial_i)^2 V 
-\int_0^\chi d\tilde\chi\,(\chi-\tilde\chi){\tilde\chi \over\chi}\nabla_\perp^2 (\Phi+\Psi)
\nonumber\\&&{}
+ 3{H\over(1+z)} V
-\left[1+\frac{\dot{H}}{{H}^2}+2\frac{(1+z)}{\chi {H}}\right] n^i \partial_i V 
+\left[2+\frac{\dot{H}}{{H}^2}+2\frac{(1+z)}{\chi {H}}\right]\Phi -2\Psi  + {1\over {H}} \dot\Psi 
\nonumber\\&&{}
+ 2  \frac{1 }{\chi}\int_0^{\chi} {d\tilde \chi}\left(\Phi+ \Psi\right) 
+\left[1+\frac{\dot{H}}{{H}^2}+2\frac{(1+z)}{\chi {H}}\right] \int_0^{\chi} {d\tilde \chi} {\big(\dot\Phi+\dot\Psi \big) \over (1+z)},
\label{delgr}
\end{eqnarray}
where 
$\chi$ is the comoving distance, $\nabla_\perp$ is the transverse gradient, $n^i$ is the direction of the galaxy and its peculiar velocity is $v_i=\partial_i V$.
The comoving-synchronous matter overdensity is
$\delta_{\rm cs}=\delta -3aHV$ and the metric potentials are defined by
$ 
ds^2=-(1+2\Phi)dt^2+a^2(1-2\Psi)d {\bf x}^2
$. (For simplicity, we have neglected magnification bias and redshift evolution of number density.) The first line contains the standard overdensity and Kaiser terms, together with the magnification contribution to clustering. The following lines are the GR effects that are suppressed on subhorizon scales but dominate on horizon scales. Collectively they have the form $i\mu A(H/k)+B(H/k)^2$, where $\mu=n^ik_i/k$.

In principle, these GR effects give us access to the peculiar velocity field and the gravitational potentials. In practice, cosmic variance grows on horizon scales to smother the signal of these effects. Even the biggest future galaxy surveys will be unable to detect the GR effects on their own -- i.e., using a single tracer of the matter distribution \cite{2015arXiv150507596A}. However, the multi-tracer method, which combines two  tracers to beat down the cosmic variance, will allow us to detect the GR effects \cite{2012PhRvD..86f3514Y,2015PhRvD..92f3525A,2015ApJ...812L..22F}. This opens a new window onto the horizon-scale Universe in three dimensions. 

 If we do not assume GR, but only that gravity is a metric theory that obeys energy-momentum conservation, then we will also be able to extend tests of GR to horizon scales \cite{2015ApJ...811..116B}. Next-generation galaxy surveys will also provide a powerful probe of primordial non-Gaussianity, which introduces a term $\propto f_{NL}(H/k)^2$ in $\delta^{\rm obs}_g$ via the galaxy bias. The multi-tracer method will allow us to detect $f_{NL}$ below the cosmic variance limit of the CMB \cite{2015PhRvD..92f3525A,2015ApJ...812L..22F}. Since the GR effects partly mimic the $f_{NL}$ effect, it is essential to take them into account when measuring primordial non-Gaussianity \cite{2015MNRAS.451L..80C,2015arXiv150506179R}.
 
\subsection{Model-independent tests of \lcdm{}}
\label{sec:modelindep}
\noindent{\it Discussion session chairs: E. Di Dio, B. Hu \& I. Sawicki}

In this section we discuss how to test the assumptions behind \lcdm{} -- from basic 
statements about the Universe, to the particular models of gravity and initial conditions that the model uses. The assumptions can be roughly classified into two groups: (1) tests of the cosmological background geometry, assuming GR+$\Lambda$; and (2) tests of GR+$\Lambda$ assuming a geometry, i.e. a homogeneous and isotropic FLRW spacetime. In principle we should do both at the same time, but this is unrealistic considering
the observational accuracy of current and future surveys. Besides these, we also consider the modeling of bias in beyond-\lcdm{} scenarios, which needs to be treated more seriously if consistent tests of gravity are to be performed.

\paragraph{Tests of geometry, assuming GR+$\Lambda$.}
The standard model of cosmology is based on the assumptions of statistical spatial isotropy and homogeneity. Since most cosmological observations are restricted to lie along the past lightcone, spatial homogeneity cannot be probed directly, even with an ideal survey. Homogeneity is only inferred from spatial isotropy under the assumption of the Copernican Principle. It is possible to construct consistency relations to test homogeneity, however; see (e.g.) Ref.~\cite{2008PhRvL.101a1301C,2014PhRvD..90b3012S} for distance-based and Ref.~\cite{Uzan:2008aa} for time-drift-based consistency relations. Other tests are based on probes like the Sunyaev Zel'dovich effect \cite{1995PhRvD..52.1821G} or strong lensing \cite{2015PhRvL.115j1301R}, which involve light scattered from inside the past lightcone and therefore can break at least some geometric degeneracies.

To probe spatial isotropy, in principle one needs to observe isotropy of the matter distribution everywhere on the past lightcone. Four independent observables (angular diameter distances, number counts, bulk velocities and lensing) are enough to establish spatial isotropy \cite{Ellis1985315, 1994CQGra..11.2693M, Clarkson:2010uz}. While the isotropy of the CMB alone is not enough to show spatial isotropy by itself (although see Ref.~\cite{2012PhRvL.109e1303C}), in combination with the Copernican Principle it becomes a fundamental observable to probe the geometry of the Universe. Indeed, it has been shown that a vanishing CMB dipole, quadrupole and octupole imply exact spatial homogeneity if the Copernican Principle is assumed \cite{1983AnPhy.150..487E}.

We would like to emphasize that statistical and geometrical homogeneity and isotropy are not equivalent. Observations only allow us to probe statistical isotropy, and this can only be related to geometrical isotropy in some average sense. In general relativity, both the mechanism and interpretation of the required averaging procedure are poorly understood \citep{2010arXiv1003.4020V}. It is also not clear that observation of a nearly (or on-average) isotropic universe implies spatial homogeneity under the Copernican Principle (see Ref.~\cite{Rasanen:2009mg} and references therein).

Just as inhomogeneous models of the Universe containing only dust can mimic acceleration because we only observe on the lightcone \citep{2000A&A...353...63C}, so too can smaller inhomogeneities affect observed distances, even in a universe that is homogeneous. In fact, non-linear structures that are typical within \lcdm{} can bias the measurement of $w$ by several percent, introducing a potentially irremovable systematic error due to cosmic variance \cite{Marra:2012pj, Valkenburg:2013qwa}.

\paragraph{Tests of GR+$\Lambda$, assuming geometry.}
Assuming homogeneity and isotropy of the background spacetime breaks the lightcone degeneracy, and allows us to relate redshifts to a particular cosmic time. It also allows us to assume that the distribution of galaxies is statistically homogeneous and isotropic, without which one cannot use number count data (such as redshift-space distortions) to infer properties of the perturbations. We cannot strictly disentangle the composition of the dark sector even with this assumption though, due to the lack of independent knowledge about the bias between observed galaxies and the dark matter perturbation. All cosmological probes depend on observing tracers moving on geodesics (light emitted by free-falling galaxies and its deflection along the way), and are therefore sensitive only to geometry (gravitational forces) and not composition.

Breaking this ``composition degeneracy'' requires some sort of parameterization, even for the background. For example, without a parameterization for $w$, it is impossible to measure $\Omega_M$ from distances directly; they are only sensitive to the total expansion rate, $H(z)/H_0$, and a function $w(a)$ can always be chosen to mimic the same distances for a different dark matter fraction. This is called the \emph{dark degeneracy} \cite{Kunz:2007rk, Amendola:2012ky}. Thus it is only by measuring the properties of large-scale structure that this degeneracy can be broken and $\Omega_M$ determined \cite{Kunz:2015oqa}.

Given a distance measurement, it is possible to use the weak-lensing shape distortion of galaxies, and the measurements of peculiar velocities of galaxies through redshift-space distortions, to reconstruct the lensing potential ($\Phi+\Psi$) and the Newtonian potential ($\Psi$). In principle, one can therefore measure the gravitational slip $\eta$ (where $\eta \neq 1$ is an unambiguous signature of modified gravity \cite{Saltas:2014dha}) without making reference to galaxy bias \cite{Motta:2013cwa, Amendola:2013qna}.  Without assuming a value and form for the galaxy bias function, one cannot determine the underlying dark matter perturbation and so make a measurement of the effective Newton's constant $\mu$, however.

In the absence of a model for modified gravity or dark energy, $\eta$ and $\mu$ are free functions of space and time that must be parameterized. It is (so far) customary to do this in the quasi-static (QS) regime, where the form of the parameterization can be simple \cite{Bertschinger:2008zb}. The validity of the QS approximation imposes strong requirements on the dark energy theory however -- in particular, that its sound speed be close to $c$ \cite{2015arXiv150306831S}. An alternative approach, inspired by effective field theory, that exploits the diffeomorphism-invariant structure of GR \cite{Gubitosi:2012hu, Bloomfield:2012ff, Hu:2013twa, Bellini:2014fua} allows for a parameterization that does not depend on the assumption of quasi-staticity (although it does depend on modeling the dark energy sector as containing a single extra degree of freedom).

If a modified gravity theory contains an extra degree of freedom, that will have its own initial conditions and its own correlation statistics, which could influence observations in a non-trivial manner \cite{Burrage:2015lla}. The non-observation of such behavior could potentially be used to eliminate the possibility of there being an extra scalar degree of freedom in the gravity sector, i.e. the class of Brans-Dicke theories including $f(R)$ models.

\paragraph{Bias modeling.}
The predictions of dark energy models cannot be connected to observations without some sort of bias model. There are two relevant biases here: the density bias and velocity bias of galaxies. The latter is of key importance in redshift-space distortion observations; it is statistical in nature, since we sample the velocity field preferentially in locations where galaxies are \cite{Baldauf:2014fza}, but the galaxy velocity is the same as the dark matter velocity. The galaxy density bias contains this statistical element too, but there is also a systematic part that depends on the physics that makes a particular type of galaxy evolve in a given dark matter halo. One therefore expects the density bias to be much more sensitive to modifications of gravity than velocity bias.

\lcdm{}+GR is unique in that the dark matter density and the two Newtonian potentials are related by constraints, so it doesn't matter which of them the bias parameters are defined with respect to. Simulations of models that include massive neutrinos already begin to suggest that the bias should actually be related to the mass clustered in physical space \cite{Villaescusa-Navarro:2013pva}, however. In modified gravity theories, the bias could be much more complicated: growth is scale dependent, there are extra degrees of freedom that produce gravitational slip, and there are environment-dependent screening mechanisms that could affect galaxy formation. parameterizations of bias will inevitably have to be extended to allow for such physical processes, so that constraints on modified gravity parameters are not misinterpreted.

Observationally, there is already evidence of some tension between the vanilla \lcdm{} model and the data, such as the somewhat inconsistent measurements of $\sigma_8$ from Planck and CFHTLenS \cite{2015arXiv150201589P}. A serious consideration of the implication of this is only just beginning however \cite{Kunz:2015oqa, Lesgourgues:2015wza}. Notwithstanding these works, one may take the view that, without a compelling alternative to \lcdm{} that is based on well-motivated physics, and which can explain (and predict) the tension, the community is unlikely to take much notice of any deviation in such model-independent tests.

\subsection{Anomaly detection: a how-to guide}
\label{sec:peiris}
\noindent{\it Plenary speaker: H. Peiris*}

The experimental landscape of cosmology for the next decade appears thrilling. Various ground-based and balloon-borne CMB experiments, such as BICEP++, ACTpol, SPT3G, PolarBear, EBEX and SPIDER, are already taking data (and will continue doing so), or are planned. Several proposals for 4th-generation CMB satellite missions, such as CMBPol, EPIC, CoRE, and LiteBird, as well as spectroscopic space missions such as PIXIE and PRISM, have been put forward. In addition, several experiments will map the large-scale structure of the Universe with unprecedented precision; these include ground-based telescopes for photometry (such as DES, Pan-STARRS and LSST), and spectroscopy (such as HSC, HETDEX, and DESI), as well as space-based telescopes Euclid, WFIRST, and SPHEREx. Plans for new cosmological probes will also be brought to fruition, which will significantly add to our observational knowledge. These include $21 {\rm cm}$ experiments, such as the SKA and its pathfinders, and gravitational wave detectors, such as Advanced LIGO and NGO pathfinder. A shared science goal of these experiments is to tie our knowledge of the early and late Universe together. Cross-talk between all of the various probes and types of data is critical for the success of the overarching goal behind them.

This huge amount of data will not be useful unless we are able to interpret them; after all, the reason for collecting the data is to better understand the Universe, and to build a consistent and powerful model that is able to explain various cosmological phenomena in a unified framework. There are, in general, two types of model that one can build in order to describe a set of data:
\begin{enumerate}
 \item {\it Mechanistic (physical) models}: These are based on the laws of physics, which make forward modeling feasible. The types of analysis one can perform in this class include parameter estimation and model comparison. These models are used to test theoretical predictions.
 \item {\it Empirical (data-driven) models}: These models only characterize relationships in the data. They are not quantitatively based on physics, although they could be qualitatively motivated by physics; forward modeling is therefore infeasible. Such models may be used to postulate new theories, or to generate statistical predictions for new observables.
\end{enumerate}
Modeling in the next decade will however face four main challenges:
\begin{enumerate}
 \item {\it Era of Big Data}: We will be dealing with very large datasets; this means that we need to come up with clever ways of compressing, filtering, sampling, and making inferences from the data. 
 \item {\it Small signal-to-noise}: Frontier research inevitably involves finding new effects close to the noise level.
 \item {\it Large model space}: Not only will we have a large number of proposed models, many of the models will have large numbers of free parameters.
 \item {\it Cosmic variance}: This is a fundamental limitation to our ability of modeling the Universe, as we have only a single realization of an inherently random cosmological model.
\end{enumerate}
The ideal goal of collecting new data, and with higher precision, is to test our standard theoretical framework, falsify it if it is wrong or incomplete, and come up with new theories (or models) that describe the data better. Since the precision of the observations is already quite high, we do not expect to detect huge deviations from the predictions of the standard model after collecting new data. It describes existing observations highly accurately, and we are very much approaching the cosmic variance limit in some regimes. It is also becoming difficult to build experiments with very low levels of noise, and therefore we expect any signals of new physics to be close to the noise level, i.e. we will be dealing with situations where signal-to-noise is very small. That is exactly where ``anomalies" become important. Anomaly detection drives scientific discovery, as it is correlated with the cutting edge of the research frontier, and thus inevitably involves small signal-to-noise. Anomalies can be simply defined as {\it unusual data configurations}, i.e. deviations from expectations. These could be outliers, unusual concentrations of data points, or sudden changes of behavior. Such anomalies may arise from chance configurations due to random fluctuations, systematics (unmodeled astrophysics, instrument/detector artifacts, and data processing artifacts), or genuinely new discoveries. Clearly the latter is what matters for science, but in order to make sure that it is some new physics behind an anomaly, it is important to ensure that the other possibilities are excluded. 

In cosmology, anomalies are often discovered using {\it a posteriori} estimators, and this tends to enhance the detection significance. What we mean by a posteriori detections here is that one finds some features in the data that may look unlikely to happen, without expecting them based on any theoretical model. One may for example see a pattern in the CMB that looks unusual, although no theoretical framework predicts that pattern. Even though the pattern may really be a signature of new physics, one should be careful here: human eyes have evolved to see patterns in data, and seeing such a pattern does not necessarily mean that they are real and of any meaning. One will then prove, for example using simulations, that such a pattern cannot occur in the standard framework. Another thing that one should be careful about before claiming new physics is the so-called {\it look-elsewhere effect}, a statistical effect that influences the significance of observing a local excess of events when a signal is sought for in a range of a specific quantity without {\it a priori} knowledge of where in the range the signal should appear \cite{2014IAUS..306..124P}. The look-elsewhere effect is particularly severe if the detection significance is moderate. The significance calculation must therefore account for the fact that an excess could equally be considered as a signal anywhere in the possible range; this effect can be taken care of by taking into account the probability of observing a similar excess anywhere in the given range. Now the question is how to judge whether an anomaly represents new physics if an alternative theory is absent (which is often the case in cosmology).

In astronomy, the prevalence of systematics --both ``known unknowns'' and ``unknown unknowns''-- combined with increasingly large datasets, the prevalence of ad hoc estimators for anomaly detection, and the look-elsewhere effect, can lead to spurious false detections. This means that anomaly detection leading to discoveries of new (astro)physics needs a combination of physical understanding, careful experimental design to avoid confirmation bias, and self-consistent statistical methods. Here we suggest four different ways for investigating whether particular anomalies have to do with new physics; ideally all these procedures should be applied to the anomalies before claiming a discovery:

\paragraph{Assessing anomalies.} Here one employs various statistical techniques to account for the look-elsewhere effect by directly assessing the properties of particular anomalies within a broad range of possible similar anomalous features. This is important in order to ensure that the anomalies are not due to chance, and that some mechanism must be at work to explain them.

\paragraph{``Just-so'' models.} These are designer theories that stand in for ``best possible'' explanations of an anomaly. The proposal is as follows: 1) Find a {\it designer theory} (or {\it just-so model} as called in statistics) that maximizes the likelihood of the anomaly; 2) Determine the available {\it likelihood gain} for the just-so model with respect to the standard model; 3) Judge if this is {\it compelling} compared to the ``baroqueness'' of the model.

\paragraph{Data-driven models.} Here one builds a phenomenological model that captures the anomalous features, and then makes predictions for new data based on this model. A concrete example is the observed anomaly in the statistical isotropy of the CMB data, as we will discuss in the next sections. The anomaly has so far been found only in the temperature data. One can therefore build a theoretical model, driven by the CMB temperature data, to provide testable predictions for the statistics of the CMB polarization or large-scale structure data. These predictions will then go beyond a posteriori inferences, and if confirmed, would serve as a smoking gun for the anomaly to be a signature of new physical phenomena.

\paragraph{Blind analysis.} Here one tries to come up with an experimental design that minimizes false detections due to the experimenter's biases. One of the main challenges in data analysis is to have a thorough understanding of the data and systematics for convincing detections. This is particularly hard in cosmology as, for example, one has to deal with complex sky masks, inhomogeneous noise, and various foregrounds in the case of the CMB data; and seeing effects, sky brightness, stellar contamination, dust obscuration, spatially-varying selection functions, Poisson noise, and photo-z errors in the case of large-scale structure data. Although the ``known unknowns'' can play very important roles in our inferences, they are easier to take care of by employing robust Bayesian techniques. Dealing with unknown unknowns is more difficult. These can be mitigated by using ``blind analysis" algorithms. Blind analysis is based on the fact that the {\it value} of a measurement does not contain any information about its {\it correctness}, therefore knowing the value of the measurement is of no use in performing the analysis itself. In blind analysis, the final result, as well as the individual data on which it is based, are kept hidden from the analyst until the analysis is essentially complete. Blind analysis is important as it helps to avoid the experimenter's (subconscious) bias, as data collection, data analysis, and inference all involve a human stage, which represents unquantifiable systematic uncertainties. The experimenter's bias could be introduced at any of the following stages:
\begin{enumerate}
 \item Looking for bugs when a result does not conform to expectations (and not looking for them when it does).
 \item Looking for additional sources of systematic uncertainty when a result does not conform.
 \item Deciding whether to publish a result, or to wait for more data.
 \item Choosing cuts while looking at the data.
 \item Preferentially keeping or dropping outlier data.
\end{enumerate}
In summary, although various persistent anomalies have been detected in cosmological tests of the standard $\Lambda$CDM model (to be reviewed briefly below), assessing whether such inconsistencies represent new physics requires overcoming pitfalls associated with multiple testing and experimenter's subconscious bias. Case studies illustrate practical strategies, such as just-so models, data-driven models, and blind analysis. We should also add that although it is critical to assess the concordance of a standard model using combined probes \cite[e.g.][]{Raveri:2015maa}, anomalous violations of the concordance do not necessarily tell us that we should go beyond the model unless we ensure that systematics are understood and under control; this is the key in searches for new physics.

\subsection{An uncooperative universe: large-scale anomalies in the CMB}
\label{sec:starkman}
\noindent{\it Plenary speaker: G. D. Starkman}

The cosmic microwave background radiation is our most important source of information about the early Universe. Many of its features are in good agreement with the predictions of the so-called standard model of cosmology -- the Lambda Cold Dark Matter Inflationary Big Bang Theory. Those features are usually described in terms of the statistics of the $a_{\ell m}$, the coefficients of a spherical harmonic expansion of the temperature (or of the E or B mode of polarization). Why? Because our canonical theory, \lcdm{} tells us that those should be statistically-independent Gaussian random variables. Moreover, the Universe is meant to be statistically isotropic, 
i.e. $\left \langle a_{\ell m} a^\star_{\ell'm'} \right \rangle = \delta_{\ell\ell'}\delta_{mm} C_\ell$.
The observed angular power spectrum, $C_\ell$, can be used to fit model parameters with surprising precision.

It has long been noticed that the  large-angle fluctuations of the microwave background temperature are uncooperative with ``the program'' --  they exhibit several statistically significant anomalies that have persisted from  WMAP to Planck~\cite{2004ApJ...605...14E,2004MNRAS.354..641H,Jaffe:2005pw,Hoftuft:2009rq,Ade:2013nlj, 2014ApJ...784L..42A,Ade:2015hxq,Copi:2003kt,Schwarz:2004gk, Copi:2005ff,Copi:2006tu,Copi:2008hw,Copi:2010na,Copi:2013cya,Copi:2013jna} (see also Ref.~\cite{2015arXiv151007929S} for a recent critical review).

The first of these is that if we look at the whole sky, the lowest multipoles seem to be correlated  both with each other and with the geometry  of the Solar System. A useful tool for identifying and studying those alignments are the Maxwell multipole vectors, which replace the spherical harmonic representation of each $\ell$ by the product of $\ell$ dipoles, aka multipole vectors. Those dipoles (or more accurately, their cross products) are surprisingly aligned with one another across $\ell=2,3$, and with various foregrounds.  Effectively, the quadrupole and octupole form a ring of extrema perpendicular to the ecliptic plane and pointed at the cosmological dipole.

On the other hand, when we look just at the part of the sky that we most trust -- 
the part outside the galactic plane -- there is a dramatic  lack of large-angle correlations. So much so that it challenges basic predictions of the standard model, for it is difficult to explain within the context of statistically independent $C_\ell$. It requires instead that there be non-vanishing co-variance amongst the $C_\ell$.

There are no models that offer robust predictions of both anomalies. The simplest explanation is that they are statistical flukes within \lcdm{}. There is some potential for testing this ``fluke hypothesis'' by examining how predictions for new observables (e.g. the temperature-lensing potential cross correlation) will be altered if we live in a rare realization of \lcdm{}. There is also potential for testing phenomenological ``models'' even as we search for fundamental models. For example, if the absence of angular correlation in temperature reflects a physical vanishing of 3D spatial correlations, then one might expect that polarization correlation functions would also exhibit a lack of large-angle correlations.

To summarize, there are signs that the Universe is not statistically isotropic, and signs that large-angle correlations are not as predicted, but there is no good explanation (yet).

\subsection{The reality of large-angle CMB anomalies}
\label{sec:cmbanomalies}
\noindent{\it Discussion session chairs: T. S. Pereira \& A. Ricciardone}

The cosmic microwave background data released by the Planck 
Satellite~\cite{Ade:2013nlj,Ade:2015hxq} confirms the existence of several large-scale statistical ``anomalies'' in the temperature maps, most of which were also observed in the WMAP data. Amongst these, the most robust seem to be: the quadrupole-octupole alignment~\cite{deOliveira-Costa:2003utu,Ralston:2003pf,Schwarz:2004gk,Copi:2005ff,Copi:2006tu,Copi:2010na,Copi:2013jna}; the hemispherical asymmetry in power between the ``northern'' and ``southern'' hemispheres~\cite{2004ApJ...605...14E,2004MNRAS.354..641H,Land:2005ad,Jaffe:2005pw,Hoftuft:2009rq,Ade:2013nlj, 2014ApJ...784L..42A,Ade:2015hxq}; the existence of a Cold Spot~\cite{Cruz:2004ce,Cruz:2006sv,Cruz:2006fy,Ade:2015hxq}; the lack of angular correlation between separations of $60$ degrees and larger~\cite{Copi:2008hw,Copi:2010na,Copi:2013cya,Ade:2013nlj}; and the point-parity asymmetry~\cite{Land:2005jq,Kim:2010gf,Kim:2010gd,Gruppuso:2010nd,Aluri:2011wv,Ade:2015hxq}. If not due to statistical flukes~\cite{Bennett:2010jb}, these anomalies represent both the possibility of new physics at high energy scales and/or a hint of unknown systematics in the map-cleaning procedure (see e.g. Ref.~\cite{2010PhRvD..81j3008P}). Therefore, testing the significance of these large-scale anomalies is crucial both for validating the standard cosmological model and for pushing forward the \lcdm{} model by better understanding the systematics. Our discussion is based on five different aspects of the anomalies, which we summarize below.

\paragraph{Beyond the statistical concordance model.}
From the statistical point of view, the \lcdm{} model implies that our Universe should be one realization of a Gaussian and statistically isotropic random field. When combined, Gaussianity and statistical isotropy (SI) imply that different CMB multipoles should be \emph{statistically independent}, which translates into a diagonal covariance matrix of the CMB multipolar coefficients~\cite{Abramo:2010gk}. The lack of Gaussianity, or SI, or both, leads to correlations between different CMB scales, which could be seen as anomalous features in the final map. Most of the attempts to address these anomalies are based on models which respect Gaussianity, but break SI. It is well known, however, that (untilted) anisotropic models of the Universe respect parity~\cite{Pitrou:2008gk,Pereira:2015pxa,Sundell:2015gra}, and thus cannot produce correlations between even and odd multipoles, which are hinted at by the quadrupole-octupole alignment. 

On the other hand, in the presence of non-Gaussianity (but respecting SI), statistical independence of CMB scales is lost, even at the Gaussian level. In fact, any non-Gaussianity in the form of a non-zero bispectrum will produce a correlation between a pair of modes~\cite{Lewis:2011au, Bartolo:2012fk}, which implies a breaking of (statistical) translational invariance~\cite{Scoccimarro:2011pz}, and consequently of parity. This idea has been explored in~\cite{Schmidt:2012ky}, where it was shown that a sufficiently divergent bispectrum in the squeezed limit will couple smaller CMB modes, producing a power asymmetry in the CMB. This result leads to an interesting possibility: if the anomalies are a genuine feature of non-Gaussianity, it can be used to learn about inflation, regardless of the bounds from statistics based on the measurements of $f_{\rm NL}$. Moreover, given that the physics behind non-Gaussianity (e.g. non-standard inflation, non-linearities, etc.) is completely different to the physics behind the statistical anisotropy (e.g. anisotropic spacetimes~\cite{Bartolo:2013uq}, higher-spin (than scalar) fields \cite{Bartolo:2012fk}, primordial domain walls~\cite{Jazayeri:2014nya}, anisotropic dark energy and so on), an in-depth understanding of the physics behind the anomalies is mandatory.

\paragraph{Are the anomalies correlated with large-scale structure?}
Stronger support for the cosmological relevance of CMB anomalies would result from their correlation with the large-scale structure of the Universe. In the standard lore, the primordial density fluctuations generated during inflation are transferred by linear physics into the present distribution of dark matter traced by galaxies. One can then expect that a primordial mechanism responsible for one (or more) of the anomalies should also be imprinted on the late-time Universe at large scales. Using this idea, Hirata~\cite{Hirata:2009ar} showed that the CMB dipolar anomaly produced by the curvaton mechanism is observationally ruled out, since it produces a dipole larger than the one actually observed in the distribution of quasars measured by the Sloan Digital Sky Survey. However, more effort is needed to unveil possible correlations between other directional anomalies, like the quadrupole-octupole alignment, and the spatial distribution of supernovae and/or quasars. More recently, the WISE-2MASS all sky infrared catalog~\cite{Kovacs:2013cga} has shown the presence of a super-void in the direction of the CMB Cold Spot. A first attempt to understand a possible correlation between these observations was made in Ref.~\cite{Finelli:2014yha}, where it was found that a Lema\^{i}tre-Tolman-Bondi (LTB) void could fit both the Cold Spot and the super-void. 

A closing remark on this issue is that the scales probed by current LSS experiments are far below the scales probed by the CMB, and, as such, deeper/wider surveys will be required to further investigate such correlations, it they exist.

\paragraph{The role of a posteriori statistics and the look-elsewhere effect.}
There is something of a consensus amongst cosmologists that the statistical significance of CMB anomalies is an important open issue, and that the role of a posteriori statistics and the look-elsewhere effect (see Sect.~\ref{sec:peiris} for definitions of the two) has to be treated carefully during data analysis. There is always the possibility that interesting features in the data will bias the choice of statistical methods in their analysis, which would make the interpretation of their statistical significance even harder. One should note, however, that features at low $\ell$ (large scales) are produced by very different physics than those at high $\ell$ (small scales). Therefore, a strong detection of an anomaly at any $\ell$ should be taken seriously in itself. 

A posteriori/look-elsewhere effects would be weakened if the CMB anomalies are found to also affect polarization maps~\cite{Yoho:2015bla}. Moreover, the measurements in temperature at low $\ell$ are drastically limited by the cosmic variance, so an analysis in polarization would be a powerful cross-validation of isotropy violation and/or non-Gaussianity in the early Universe. This possibility, however, depends on future experiments with better sensitivity to the polarization signal.

\paragraph{Can the anomalies be due to unknown systematics?}
In the detection of CMB anomalies, a non-negligible role could be played by systematic effects and the masking procedure. The agreement in the detection of the anomalies with both the WMAP and Planck satellites, using independent data analysis and techniques, strongly suggests that the origin of the anomalies cannot be attributed to residual systematics, however.

\paragraph{How much work should we put into search for an explanation?}
Unfortunately, there is always the possibility that the anomalies result from pure statistical fluctuations. In fact, the WMAP team has argued against their physical interpretation~\cite{Bennett:2010jb}, suggesting that their significance is mostly due to human-biased selection. If we include the fact that low-$\ell$ anomalies are limited by cosmic variance, it is inevitable to question the importance that we should ascribe to them. Regarding this point, there was a broad consensus in the discussion session that CMB anomalies are robust against systematics, not to mention the fact that they look like the typical signatures one expects from simple extensions of the \lcdm{} model. Thus, CMB anomalies are likely to offer an interesting window to the largest scales of the Universe, which will otherwise only be accessible through improved CMB and LSS probes. This leaves a window open to the possibility of new physics.

\section{Towards a new standard model?}

\subsection{Radical solutions (in the spirit of Hoyle)}
\label{sec:magueijo}
\noindent{\it Plenary speaker: J. Magueijo*}

The biggest problems affecting \lcdm{} have so far failed to yield against two decades of broad and concerted effort by the cosmological community. So is it perhaps time to consider more radical solutions -- to do away with some of the standard model's foundational sacred cows and replace them with a wildly different approach?

\paragraph{Radicalism: more harm than good?}
The problem is that radical approaches are often much worse than the models they seek to replace. Cosmology has converged on the \lcdm{} model over many decades of painstaking theoretical and observational work from hundreds of different angles, and has its basis in the consistency of many independent results that tie in closely with other successful areas of physics (e.g. atomic, nuclear, and particle physics). It is not clear that one could construct a substantially different theory of cosmology while also preserving the successes of this synthesis. As exciting as the thought of a complete quantum mechanics-style revolution may be, one is much more likely to wreck the whole structure when pulling its foundations apart.

Some concepts that are currently held to be indispensable may nevertheless turn out to be false -- unnecessary `red herrings' that have been pointing us in the wrong direction. Possible examples include ghosts, renormalizability, and naturalness -- are we really so sure that these are necessary features of a good theory?

Without a successful alternative theory to guide the way, there is considerable subjectivity in which ideas it is considered acceptable to discard. Magueijo has highlighted the possibility of discarding foliation diffeomorphism invariance and local Lorentz invariance (in some agreement with the discussion in Sect.~\ref{sec:lcdmalt}); the constants and laws of nature may evolve in time, and space/time could be emergent properties of some underlying fundamental theory. These possibilities have been considered by a number of physicists in the past (Dirac being one notable example), and interesting theories can indeed be constructed with some of these properties.

\paragraph{Variation of fundamental constants.}
Let us take a closer look at the `radical' idea that the laws of physics may not be constant -- in particular, the case where the fundamental constants that characterize them are not actually constant. There have been a number of well-known attempts to construct such theories, such as the Brans-Dicke theory \cite{Brans:1961sx} of a varying gravitational constant, $G$, or the varying fine-structure constant theories \cite{Bekenstein:1982eu} (for which there is claimed, but disputed, observational evidence \cite{1999PhRvL..82..884W}; see Ref.~\cite{2011LRR....14....2U} for a review).

Varying speed of light (VSL) theories have also been developed (see Ref.~\cite{2003RPPh...66.2025M} for a review). There are many ways of constructing VSL theories -- Lorentz-violating and covariant/Lorentz-invariant \cite{2000PhRvD..62j3521M}, bimetric theories\footnote{These theories should not be confused with the bimetric theories discussed in Sect.~\ref{sec:gravcosmo}.}~\cite{2003IJMPD..12..281M}, and ones with deformed dispersion relations \cite{1998Natur.393..763A, 2004CQGra..21.1725M}, for example. Some are able to solve key problems in cosmology \cite{Albrecht:1998ir}, including explaining cosmic acceleration \cite{2000PhLB..477..269C} and producing an (almost) scale-invariant primordial power spectrum \cite{2003JCAP...07..004C} and other results that inflationary theory is normally invoked to explain \cite{1993IJMPD...2..351M}. In other words, VSL theories can reproduce multiple key features of \lcdm{} while invoking a very different physical explanation. Importantly, however, they do have testable differences that can be used to distinguish them from \lcdm{}, including (e.g.) a lack of primordial gravitational waves and specific consistency relations for the CMB bispectrum \cite{2010PhRvD..82d3521M}.

While this is just one example, it does show that radical departures from the standard picture can bear fruit -- and are not restricted to providing explanations of individual phenomena while ignoring everything else. A corollary is that the observational success of the overarching \lcdm{} framework is not, in itself, enough to prove that its individual components are necessary features of the true cosmological theory -- fundamental components {\it can} be replaced with radically different ideas without destroying the whole synthesis.

\paragraph{Intersection of fundamental physics and cosmology.}
While many cosmologists look towards fundamental physicists for a theory of quantum gravity (QG) that can explain the cosmological constant problem (e.g. see Sect.~\ref{sec:ccdisaster}) and other issues, there is a growing disconnect between the directions of the two fields. Some accuse the fundamental physics community of having forsaken empirical validation, replacing it with considerations based purely on mathematical elegance and `explanatory power' \cite{2014Natur.516..321E}. Similar allegations have been leveled at inflation \cite{2011SciAm.304d..36S}. Cosmological processes are so far from the everyday experience of human beings that perhaps it is unreasonable to expect us to be able to test the more extreme reaches of fundamental physics; nevertheless, one would still like {\it some} kind of guard against self-delusion. After all, plenty of (arguably) more beautiful and compelling ideas than string theory have fallen by the wayside in the face of empirical evidence.

Others are concerned that some (empirically-driven) fundamental theorists are too concerned with matching their results to \lcdm{}, rather than coming up with independent ideas that could perhaps upset the standard cosmological model. Should quantum gravity theories necessarily predict an inflaton field, for example? One needs to establish {\it some} way to connect a prospective fundamental theory with established physics however, although opportunities can surely be missed by insisting too strongly on reproducing certain features of lower-energy theories that may not be necessary.

More radically, cosmologists could be barking up the wrong tree altogether. Perhaps a theory of quantum gravity can explain the observational features of \lcdm{}, but without needing to admit a formal ``\lcdm{} limit'' (i.e. without the theory effectively reducing to \lcdm{} in a particular low-energy limit). As-yet unknown dynamics in a QG theory could naturally set the initial conditions of the Universe, for example, without requiring anything that looks like inflation.

\paragraph{Perspective.}
No-one yet knows how the theoretical maladies of cosmology will be solved, if they can be solved, or even if they need to be solved. As more `conventional' attempts to find solutions have failed to make headway, however, it becomes tempting to try more radical ideas. As evidenced by past `paradigm shifts' in physics, radical ideas are often necessary for progress, and we, as a community, must be open to their exploration. Certainly, there is no point in being dogmatic about \lcdm{} when there is consensus that it cannot be the full picture.

Still, it should be a {\it principled} radicalism that we insist upon. Smashing the foundations of the standard cosmological model is all well and good, but the end result cannot be considered successful unless it is a truly predictive theory -- one that not only fits the bulk of current and future data, but explains it as a non-trivial consequence of its deeper structure \cite{2015arXiv150609143G}. Simply introducing additional unconstrained degrees of freedom to fit-out deviations will not do. An alternative theory should ideally strengthen the connections between cosmology and the rest of physics too, as \lcdm{} has done so ably; theories with special constructions that disconnect the causes of cosmological phenomena from their possible consequences elsewhere look feeble.

But even if evolution, rather than revolution, is needed to fix up \lcdm{}, there may still be something to recommend a more radical stance -- perhaps a shake-up of our perspective, rather than our theory, is what has been needed all along?

\subsection{Novel observables in cosmology}
\noindent\textit{Discussion session chairs: E. Bellini \& M. Quartin}

In this section, we discuss the importance of pursuing new and complementary tests of the \lcdm{} model. We choose to define these ``new observables'' as any cosmological tests beyond the four most established ones: Type Ia supernovae luminosity distances, angular and radial baryon acoustic oscillations, weak gravitational lensing (based on shear measurements), and traditional CMB analysis (which focus mostly on linear perturbation quantities).

Such a broad definition does not allow for a comprehensive coverage of possible new observables. Instead, we focus first on particular topics of interest among the participants of the discussion session, and later on a general approach to the role of such novel observables. We enumerate below some promising candidates that were highlighted.
\begin{itemize}
 \item One of the limitations of traditional weak-lensing analysis is the lack of knowledge of the intrinsic orientation of the lensed galaxy, which thus becomes a nuisance parameter to be marginalized over. It was however recently shown in Ref.~\cite{Brown:2010rr} that since the polarization of galactic light is unaffected by lensing, this would allow for an independent measurement of the original orientation, thus improving the lens reconstruction. This technique should therefore be further investigated and applied to future surveys.

 \item The observation of both $E$ and (non-primordial) $B$ modes of polarization of the CMB is now an established field, and measurements in the former case have already been done at very high significance levels. Both modes are nevertheless modes of linear polarization. CMB circular polarization, on the other hand, is usually ignored (see Ref.~\cite{Mainini:2013mja} for one of the few attempts) because it is not expected to be present. A confirmation of this expectation is nevertheless an important cross-check to be made in the future.

 \item Relativistic (non-cosmological) effects that contribute to the observed redshift of sources were also agreed to be of crucial importance. Among these, priority should be put on separating the gravitational~\cite{Croft:2013taa}, kinematic, and lensing redshift contributions. The first should be especially important for clusters, which create deep gravitation wells; the second, for low-redshift objects~\cite{Hui:2005nm,Davis:2010jq}; and the last, for high-redshift objects such as supernovae~\cite{Takahashi:2011qd,Quartin:2013moa}.

 \item Real-time cosmology observations (based on changes in the sky on a time-scale of a few years~\cite{Quercellini:2010zr}) were also regarded as promising in the long-term. With the advent of the E-ELT, it is possible to measure the redshift of individual quasars. Alternatively, the SKA could be capable of measuring the average redshift drift of large atomic hydrogen (HI) regions~\cite{Klockner:2015rqa}.

 \item Many other interesting topics were raised (21cm observations~\cite{McQuinn:2005hk}, including probing ultra-large scales; strong-lensing effects including arcs, multiple images and time delays; the possibility of future neutrino microwave background observations; gravitational waves; etc.).
\end{itemize}
Particular topics aside, the main take-home message is that doing a large amount of independent cross-checks on \lcdm{} should be a priority in the near future. Traditionally, such tests are performed as an afterthought for most experiments. The consensus among the two dozen participants was instead that our community should re-double their efforts and even design experiments (or at least pipelines for current experiments) not only to look for alternative measures of known quantities, but also to measure expected ``null-signals.'' The latter, which includes examples such as the CMB circular polarization discussed above, probably poses a bigger challenge, as funding agencies may be reluctant to finance expensive projects unless they {\it expect} to discover something new.

In the end, this is the classical duality between precision and accuracy in science. The former is easy to establish (it is directly related to the experimental variance of an observable), whereas inferring the latter is much less obvious since it is related to the chosen underlying model. One way to probe accuracy is by looking for tensions between different data, so cross-checks with as many observables as possible is one way to highlight the limitations of a model. This shows the importance of cross-correlating all observables to the fullest extent possible. Possible tensions arising from these cross-checks would provide important hints on how to move beyond \lcdm{}.

\subsection{Summary and directions: what next?}
\noindent{\it Plenary speaker: A. Heavens}

As Tom Shanks once said, there are only two things wrong with \lcdm{}: $\Lambda$; and CDM.  The trouble is that, from most cosmologists' perspective, there is nothing wrong with it at all -- there is no strong evidence anywhere for a failure.  Moving away from \lcdm{} is also not trivial -- the direction to take is not clear at all.  \lcdm{} is like Hotel California; it is very hard to leave, and most, if not all, efforts to do so have ended with some insurmountable obstacle.  From a Bayesian perspective, the prospects for falsifying \lcdm{} on large scales are not good without new observables, as constraints from existing observables are so close to \lcdm{} that models with more parameters are likely to be disfavored by Bayesian evidence calculations.

\begin{figure}
\begin{center}
\includegraphics[width=0.8\hsize]{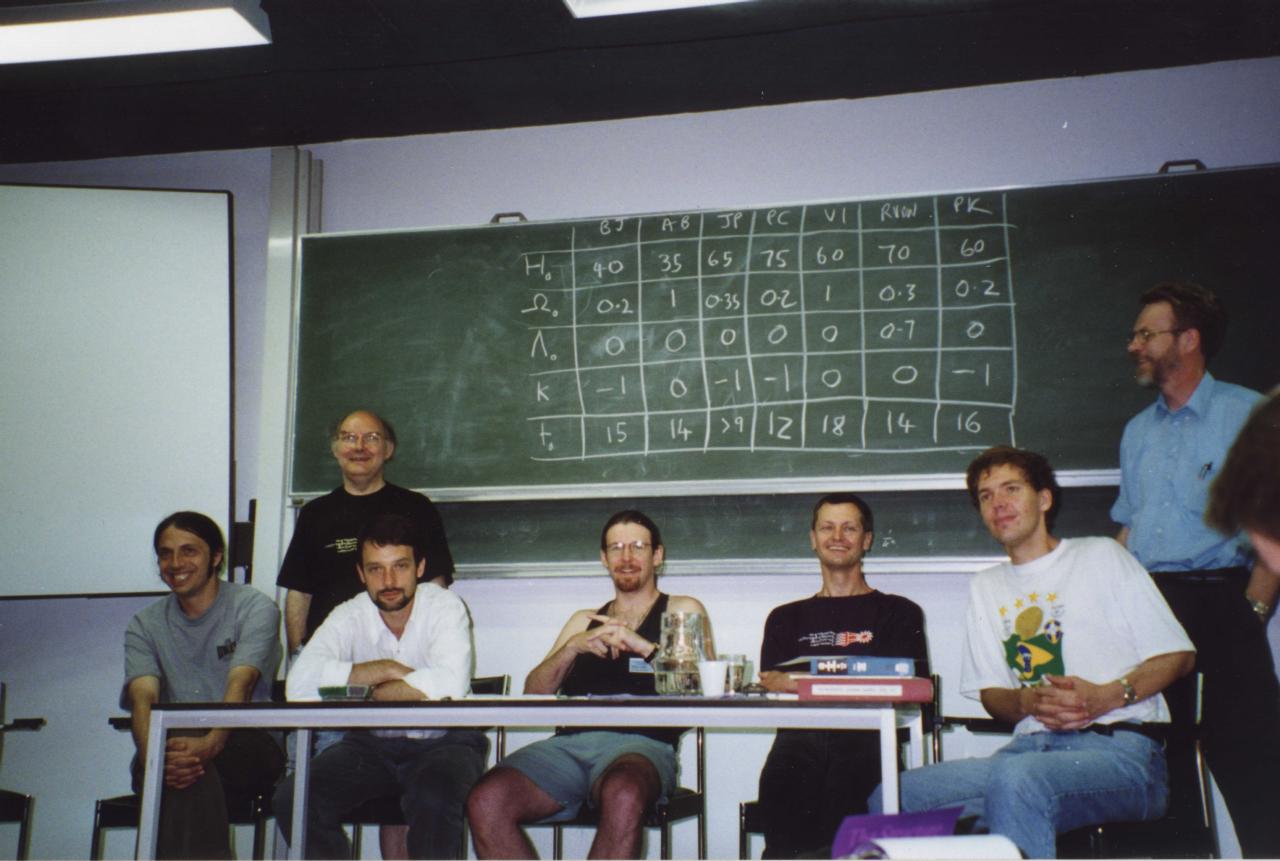}
\includegraphics[width=0.8\hsize]{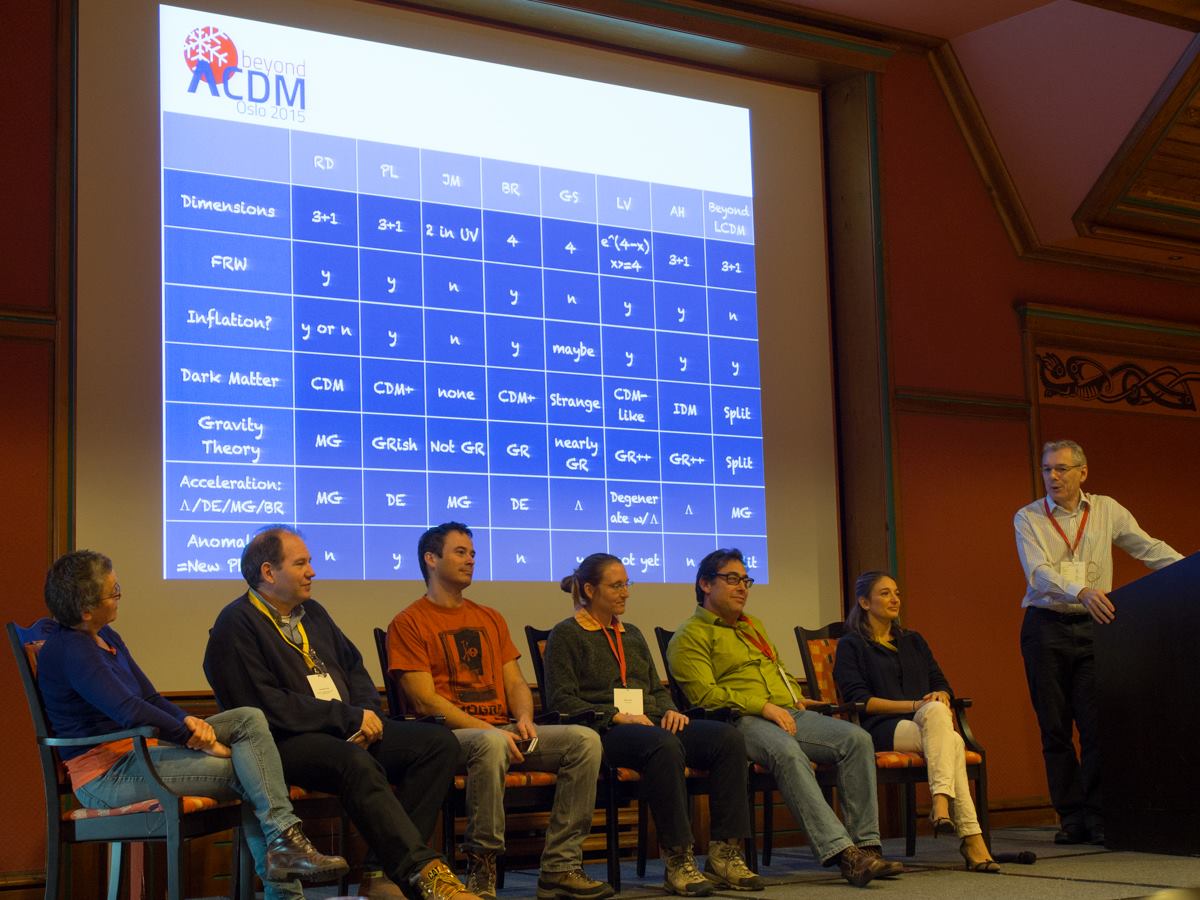}
\end{center}
\caption{{\it (Top)} The panel at Leiden, 1995. Bernard Jones, Alain Blanchard, John Peacock, Peter Coles, Vincent Icke, Rien van de Weygaert, and Peter Katgert (photo courtesy of Peter Coles). {\it (Bottom)} The panel at Beyond \lcdm{}, Oslo, 2015. Ruth Durrer, Per Lilje, Jo\~ ao Magueijo, Beth Reid, Glenn Starkman, Licia Verde. Chair Alan Heavens.}
\label{P1995}
\label{PLCDM}
\end{figure}

In 1995, various famous cosmologists were asked what they thought the cosmological parameters were (see Fig.~\ref{P1995}). Rien van de Weygaert chose \lcdm{} and so got it right. The final discussion panel for Beyond \lcdm{} had a different selection of questions, and none quite voted for vanilla \lcdm{} (Fig.~\ref{PLCDM} and Table~\ref{tab:vote}). Some advocated something very close to \lcdm{} (Beth Reid, Licia Verde, Alan Heavens), but others were much more radical, most notably Jo\~{a}o Magueijo and Glenn Starkman.  The last column is the consensus of the rest of the participants. Interesting reading.

\begin{table}[h]
{\centering \footnotesize
\begin{tabular}{|p{2.7cm}ccccccc|c|}
\hline
 & RD & PL & JM & BR & GS & LV & AH & B\lcdm{} \\
\hline
Spacetime dimens. & 3+1 & 3+1 & 2 in UV & 4 & 4 & $e^{4-x}, x \ge 4$ & 3+1 & 3+1 \\
FLRW? & Y & Y & N & Y & N & Y & Y & N \\
Inflation? & Y or N & Y & N & Y & Maybe & Y & Y & Y \\
Dark matter? & CDM & CDM+ & None & CDM+ & Strange & CDM-like & IDM & (split) \\
Gravity theory? & MG & GRish & Not GR & GR & Nearly GR & GR++ & GR++ & (split) \\
Acceleration? & MG & DE & MG & DE & $\Lambda$ & Degen. $w$/$\Lambda$ & $\Lambda$ & MG \\
Anomalies are new physics? & N & Y & Y & N & Y & Not yet & N & (split) \\
\hline
\end{tabular}}
\caption{Responses to questions about directions beyond \lcdm{} given by the participants of the panel discussion (R. Durrer, P. Lilje, J. Magueijo, B. Reid, G. Starkman, L. Verde, A. Heavens, and the audience, ``B\lcdm{}'').
\label{tab:vote}}
\end{table}

\subsection{What have we learned?}
\noindent {\it Editorial perspective: Y. Akrami \& P. Bull}

We are living in an exciting and frustrating time for cosmology. Excellent progress has clearly been made in our understanding of the physics of the Universe, with \lcdm{} providing a useful and compelling picture of the cosmos. The quality, and volume, of observational data is impressive, and improving remarkably rapidly. A number of difficult puzzles have been highlighted in recent years however, pointing to multiple large gaps in our understanding. These are a grave concern -- the variety and complexity of the topics discussed in this paper are testament to that.

They are also testament to the ingenuity and hard work that is being applied to try and patch up the cosmological theory. The extent of the literature concerning possible solutions to the cosmological constant problem alone is breathtaking. And yet, many feel that theoretical progress is not being made (or if it is, not nearly fast enough). There is excitement about the progress that will no doubt occur with forthcoming observations, but has our quest for a deep understanding of the fundamental picture stalled?

We are optimistic. While many of the problems appear to be resolutely failing to yield, we took away a sense of slow, steady, but definite theoretical progress from the discussions above. There are many leads to follow. The cosmological constant problem is now better posed, for example, and we now know of many solutions that do not work, and why they can't work. This landscape of discarded theories is a valuable source of new insights. The same can be said for the gravitational sector, where there is now a much clearer picture of what properties are necessary for a viable gravity theory. This work has also established the relative merits of a variety of observables, setting important targets for what should be measured in the coming decade to test \lcdm{} most stringently, and most strongly enhance the chances of a `breakthrough'.

By way of a conclusion, we highlight some of the points from the discussion above that we found most intriguing. If the community agrees, we will perhaps see some of them emerging as important research directions over the coming years.
\begin{itemize}

 \item Understanding the quantum dynamics of the ground state of the Universe may hold the key to the CC problem; mechanisms to cancel various vacuum contributions require further study (Sect.~\ref{sec:ccdisaster}). It is also important to realize that the vacuum has a history -- it changes as the Universe passes through various symmetry-breaking events, confusing whether the CC problem is an issue with the UV or IR (Sect.~\ref{sec:justlambda}).
 
 \item It is acceptable for new theories to break fundamental principles like locality, Lorentz invariance, and the action principle, but it's not OK for them to have ghosts (Sects.~\ref{sec:lcdmalt} and \ref{sec:magueijo}). Much future model building will probably focus on the effects of breaking these principles.
 
 \item Attempts to modify gravity by adding new fields appear to have reached their logical conclusion; more radical options that embrace non-locality and `emergence' are being adopted as new promising directions (Sect.~\ref{sec:mgdev}). The modified gravity community does follow `fashion trends' in model-building, but this is a sign of its strength and vibrancy, rather than a weakness (Sect.~\ref{sec:mgdev}).
 
 \item As datasets grow, statistical subtleties related to the increasing number of degrees of freedom will require model testers to take additional care not to unfairly rule-out models (Sect.~\ref{sec:modelsvsparams}). Cosmologists may also have to follow the lead of particle physicists and gravitational wave astronomers, and begin the routine use of blind analysis methods to guard against confirmation bias (Sect.~\ref{sec:stresstest}).
 
 \item Ultra-large scales are a good place to look for modifications to GR that are motivated as solutions to the dark energy problem (Sect.~\ref{sec:mgde}). The justification is that if a modified theory needs to introduce a new physical scale to explain cosmic acceleration, it is most natural to put this at the Hubble scale, $\mathcal{H}_0/c \sim 2\times 10^{-4}$ Mpc$^{-1}$. The anisotropic stress is another key modified gravity observable, and can be accessed by weak-lensing surveys (Sect.~\ref{sec:mgde}).
 
 \item There are serious degeneracies between the effects of baryons, massive neutrinos, and modifications to gravity on small scales, which need to be better understood before small-scale clustering can be used to test any of these things (Sect.~\ref{sec:mgde}). The definition of bias in beyond-\lcdm{} theories is subtle (as hinted at by simulations that include massive neutrinos), so understanding how it differs from its \lcdm{} behavior will be important for consistently testing GR with large-scale structure observables (Sect.~\ref{sec:modelindep}).
 
 \item Imperfect modeling of baryonic processes is a major obstacle to achieving sufficiently accurate simulations to support forthcoming large-scale structure surveys (Sects.~\ref{sec:sim} and \ref{sec:smallscale}). A complete understanding of baryonic physics still seems a distant prospect, so there is a good case for the formulation of a ``good enough'' phenomenological baryonic model as a stop-gap (Sect.~\ref{sec:cosmosims}). The parameters of this model could then be marginalized over in large-scale structure analyses, mitigating the systematic biases that would otherwise cause serious problems (Sects.~\ref{sec:huntingde} -- \ref{sec:lahav}).

 \item Computer time for simulations that investigate new and exotic scenarios appears to be easier to obtain than for important in-depth modeling and parameter studies, and the testing/confirmation of \lcdm{} results (Sect.~\ref{sec:cosmosims}). If too strong a bias is present, it could hinder important work that is needed to understand and clarify the ``standard'' results.
 
 \item Data seem to favor plateau-like inflationary potentials, but there is some debate over whether these can generate sufficiently homogeneous initial conditions in a natural way (Sect.~\ref{sec:inflationproblems}). Inflation models based on modifying GR are gaining in popularity, and may have a justification from results in quantum gravity (Sect.~\ref{sec:inflationproblems}); string theory calculations suggest a particular (non-$f(R)$) form for quantum gravity corrections to the action. For the foreseeable future, the detection of B-mode polarization in the CMB will likely remain the most powerful test of inflation (and alternatives) available to us (Sect.~\ref{sec:infobs}).
 
 \item Large-angle CMB anomalies persist in multiple datasets, making instrumental/data analysis systematics look increasingly unlikely as an explanation (Sect.~\ref{sec:cmbanomalies}). There is a lack of compelling models (particularly ones that can explain more than one anomaly at a time) however, and the possibility that they are just a statistical fluctuation remains (Sect.~\ref{sec:starkman}).
 
 \item While many cosmologists are looking to quantum gravity to solve the CC problem (Sect.~\ref{sec:surv}), there is a sociological danger that the two fields will diverge too far from one another to make a useful connection possible (Sect.~\ref{sec:magueijo}). Additionally, ``quantum cosmologists'' may consider relaxing the requirement to reproduce \lcdm{} from their theories -- QG processes may explain certain features of cosmology independently, without a formal \lcdm{} limit (Sect.~\ref{sec:magueijo}).
 
 \item Particle physics is the great hope for solving the CC problem, and a primordial B-mode detection is the great hope for understanding inflation (Sect.~\ref{sec:surv:inflation}). CMB polarization, spectroscopic galaxy surveys, weak-lensing surveys, and 21cm intensity mapping in the Epoch of Reionization are the observables that most will be watching over the coming decade. Conversely, CMB temperature anisotropies, local Hubble rate measurements, and laboratory tests of GR were thought to be less interesting (Sect.~\ref{sec:surv:obs}).
 
\end{itemize}

\newpage
\section{Cosmological opinions: a survey}
\label{sec:surv}
\vspace{-0.5em}\noindent {\it Plenary speaker: R. Durrer*}

Questions of promising directions, the significance of anomalies, and how compelling a given theory is are intrinsically subjective, but it is still useful to be able to gauge the opinions of other researchers on such topics -- if only to check how far from the mainstream one's own opinions are. To that end, we conducted an informal poll of the participants of the Beyond \lcdm{} conference, as well as others in the community, asking for opinions on topics such as the seriousness of the cosmological problem and the likely impact of various observational probes. We present the results of the poll in this section.

It is important to take this survey in the spirit that it was intended: an informal (and slightly tongue-in-cheek) poll of the attitudes and opinions of active researchers in cosmology. It is not in any way scientific, is surely subject to all manner of biases, and lacks a sufficient sample size to be considered truly representative. Regardless, it gives some impression of what working cosmologists are thinking, and the general mood with respect to \lcdm{} and its alternatives.

All of the questions were multiple choice, so the numbers in the tables that follow correspond to the number/percentage of survey participants that chose a given option -- the percentages do not add up to 100\%. The specified timeframe, where relevant, was ``the next decade''. \\

\vspace{-0.5em}
\subsection*{\lcdm{}: problems}

The first question asked, in general terms, what the most serious problems are with \lcdm{}. Unsurprisingly, the cosmological constant problem came out on top. A more indeterminate concern, that the model would continue to fit observations well enough without being better understood from a physical perspective, was the second most popular response, pointing towards a general worry that \lcdm{} may be too good at explaining data without being able to provide deeper insights into important problems. A number of other responses were also popular, but more notable is one of the least popular -- apparently very few people think that there is {\it nothing} wrong with \lcdm{}. \\

\rowcolors{1}{}{lightgray}
{\footnotesize\centering\begin{tabular}{p{0.2cm}p{13.5cm}rr}
\hline
\multicolumn{2}{p{30em}}{\bf 1. What are the biggest deficits and challenges of the \lcdm{} paradigm?} & {\bf ~~~~No.} & {\bf ~~~\%~} \\
\hline
& The cosmological constant problem & 40 & 66 \\
& \lcdm{} will remain the best-fit model to the data while not being understood theoretically & 29 & 48 \\
& There are no compelling alternatives & 19 & 31 \\
& It can't explain small-scale structure (e.g. dwarf galaxies) & 14 & 23 \\
& Cold dark matter & 13 & 21 \\
& The model is fine-tuned & 13 & 21 \\
& Inflation is a general idea with no clear implementation in particle physics & 13 & 21 \\
& Baryonic effects are too difficult to model & 11 & 18 \\
& Confirmation bias & 9 & 15 \\
& The coincidence problem & 8 & 13 \\
& Effects of inhomogeneities and anisotropies & 8 & 13 \\
& Inflation isn't predictive enough & 7 & 12 \\
& The Big Bang singularity and high energy description of gravity & 6 & 10 \\
& GR cannot be relied upon on large length scales & 4 & 7 \\
& (Nothing is wrong with \lcdm{}) & 3 & 5 \\
& There is no dark matter & 2 & 3 \\
& Cosmic variance on ultra-large scales & 1 & 2 \\
& Other & 1 & 2 \\
\hline
\end{tabular}} \\ 

\subsection*{Observations and the future}
\label{sec:surv:obs}

Next, we asked what role future observations are expected to play in the development of \lcdm{}. Most respondents were confident that large experiments would be the dominant source of important advances, but that systematics and a lack of theoretical progress would slow them down. The question of what the experiments are expected to discover was met with a relatively conservative response; most expect that \lcdm{} will either be confirmed to higher precision, or that new things will be discovered that can simply be added into the existing framework. Still, a considerable number of respondents hoped for a result that would result in a fundamental change.

The source of the next big observational results? Respondents were most positive about four key observables -- CMB polarization, spectroscopic galaxy redshift surveys, weak-lensing surveys, and 21cm intensity mapping in the epoch of reionization. Direct gravitational wave detection and late-time 21cm intensity mapping were also identified as exciting prospects by a significant number. \\

{\footnotesize\centering\begin{tabular}{p{0.2cm}p{13.6cm}rr}
\hline
\multicolumn{2}{l}{\bf 2. The most important results will come from...} & {\bf ~~~~No.} & {\bf ~~~\%~} \\
\hline
& large experiments & 44 & 72 \\
& theorists & 24 & 39 \\
& simulations & 20 & 33 \\
& small experiments & 18 & 30 \\
& particle physicists & 11 & 18 \\
& other & 2 & 3 \\
\hline
\end{tabular}} \\

{\footnotesize\centering\begin{tabular}{p{0.2cm}p{13.6cm}rr}
\hline
\multicolumn{2}{l}{\bf 3. Observational progress...} & {\bf ~~~~No.} & {\bf ~~~\%~} \\
\hline
& will slow due to difficult systematics & 30 & 49 \\
& will slow due to lack of theoretical progress & 27 & 44 \\
& will speed up! & 12 & 20 \\
& will slow due to computational challenges & 9 & 15 \\
& will slow due to reduced funding & 5 & 8 \\
& other & 4 & 7 \\
\hline
\end{tabular}} \\

{\footnotesize\centering\begin{tabular}{p{0.2cm}p{13.6cm}rr}
\hline
\multicolumn{2}{l}{\bf 4. Next-generation experiments will...} & {\bf ~~~~No.} & {\bf ~~~\%~} \\
\hline
& confirm \lcdm{} to higher precision & 24 & 39 \\
& discover new things that can just be added to \lcdm{} & 24 & 39 \\
& discover new things that fundamentally change \lcdm{} & 19 & 31 \\
& discover something that completely overturns \lcdm{} & 6 & 10 \\
& other & 2 & 3 \\
\hline
\end{tabular}} \\

{\footnotesize\centering\begin{tabular}{p{0.2cm}p{13.6cm}rr}
\hline
\multicolumn{2}{p{14.2cm}}{\bf 5. For cosmology, the most valuable/exciting observables over the next decade will probably be:
} & {\bf ~~~~No.} & {\bf ~~~\%~} \\
\hline
& CMB polarization & 33 & 54 \\
& Galaxy redshifts (spectroscopic) & 31 & 51 \\
& Weak lensing (shear/convergence) & 30 & 49 \\
& 21cm intensity mapping (EoR) & 29 & 48 \\
& Gravitational waves & 25 & 41 \\
& 21cm intensity mapping (late times) & 22 & 36 \\
& Galaxy redshifts (photometric) & 18 & 30 \\
& CMB lensing & 16 & 26 \\
& Peculiar velocities (Kinetic SZ) & 14 & 23 \\
& Dark matter direct detection & 14 & 23 \\
& High-redshift galaxies & 12 & 20 \\
& Matter distribution on ultra-large scales & 12 & 20 \\
& Neutrinos & 12 & 20 \\
& CMB scattering (Thermal SZ/Rayleigh/spectral distortions) & 11 & 18 \\
& Strong lensing & 10 & 16 \\
& Various cross-correlations & 10 & 16 \\
& Other large-scale structure/matter distribution observables & 8 & 13 \\
& Local tests of gravity & 8 & 13 \\
& Supernovae & 7 & 12 \\
& Cosmic rays & 6 & 10 \\
& Particle collisions (LHC) & 5 & 8 \\
& Local Hubble rate measurements & 3 & 5 \\
& Transients & 3 & 5 \\
& Proper motions of stars & 3 & 5 \\
& Variation of fundamental constants & 3 & 5 \\
& Laboratory tests of gravity & 2 & 3 \\
& CMB temperature & 1 & 2 \\
& Other & 1 & 2 \\
\hline
\end{tabular}} \\

\subsection*{The cosmological constant problem and dark energy}
\label{sec:surv:cc}

We asked several questions about the cosmological constant problem and its relation to particle physics and alternative theories. Most thought that a better understanding of particle physics would lead to a solution of the cosmological constant problem, although modified gravity was also a popular option. The majority expected dark energy to be indistinguishable from a cosmological constant, although a significant number are holding out for a more radical solution (``something completely different'').

The relatively undeveloped state of alternative theories with respect to \lcdm{} was mostly put down to the increased difficulty in working with them, although some degree of self-consciousness was also blamed -- alternatives can be seen as `wacky'.

As far as particle physics is concerned, many expect cosmology to provide a fundamental discovery about neutrinos in the near future -- presumably a measurement of the sum of the neutrino masses, or the discovery of a sterile neutrino. Hopes are also relatively high for a particle dark matter direct detection. The Higgs, on the other hand, is seen by most as unimportant for most work in cosmology, despite being the only scalar field currently known in nature. \\

{\footnotesize\centering\begin{tabular}{p{0.2cm}p{13.6cm}rr}
\hline
\multicolumn{2}{l}{\bf 6. The cosmological constant problem will be solved by...} & {\bf ~~~~No.} & {\bf ~~~\%~} \\
\hline
& better understanding of particle physics & 28 & 46 \\
& a modified gravity theory & 17 & 28 \\
& realising it's not a problem & 12 & 20 \\
& other & 12 & 20 \\
& a dark energy theory & 11 & 18 \\
\hline
\end{tabular}} \\

{\footnotesize\centering\begin{tabular}{p{0.2cm}p{13.6cm}rr}
\hline
\multicolumn{2}{l}{\bf 7. Dark energy will turn out to be...} & {\bf ~~~~No.} & {\bf ~~~\%~} \\
\hline
& indistinguishable from a cosmological constant & 29 & 48 \\
& something completely different & 20 & 33 \\
& a modification to GR & 16 & 26 \\
& a new scalar field & 9 & 15 \\
& related to dark matter & 7 & 12 \\
& a cosmological constant & 6 & 10 \\
& other & 2 & 3 \\
\hline
\end{tabular}} \\

{\footnotesize\centering\begin{tabular}{p{0.2cm}p{13.6cm}rr}
\hline
\multicolumn{2}{l}{\bf 8. Alternatives to \lcdm{} are much less developed because...} & {\bf ~~~~No.} & {\bf ~~~\%~} \\
\hline
& they are more difficult to work with than \lcdm{} & 39 & 64 \\
& you're considered a bit wacky if you consider alternatives & 23 & 38 \\
& supervisors don't offer projects on alternative theories & 11 & 18 \\
& they're not sexy & 8 & 13 \\
& other & 8 & 13 \\
& we can never distinguish them anyway & 7 & 12 \\
& there's no need to work on other models & 1 & 2 \\
\hline
\end{tabular}} \\

{\footnotesize\centering\begin{tabular}{p{0.2cm}p{13.6cm}rr}
\hline
\multicolumn{2}{l}{\bf 9. Particle physics and cosmology:} & {\bf ~~~~No.} & {\bf ~~~\%~} \\
\hline
& Cosmologists will discover something fundamental about neutrinos & 29 & 48 \\
& Particle dark matter will be discovered experimentally & 22 & 36 \\
& Particle physicists will discover something fundamental about neutrinos & 18 & 30 \\
& Completely new particles will be discovered with important implications for cosmology & 9 & 15 \\
& Particle dark matter will be found not to work & 8 & 13 \\
& Particle physicists will explain the nature of inflaton & 6 & 10 \\
& Other & 5 & 8 \\
& Neutrinos will remain mysterious & 1 & 2 \\
\hline
\end{tabular}}

{\footnotesize\centering\begin{tabular}{p{0.2cm}p{13.6cm}rr}
\hline
\multicolumn{2}{l}{\bf 10. The Higgs will turn out to be important in understanding...} & {\bf ~~~~No.} & {\bf ~~~\%~} \\
\hline
& none of these options & 38 & 62 \\
& inflation & 13 & 21 \\
& dark matter & 9 & 15 \\
& dark energy & 5 & 8 \\
& other & 1 & 2 \\
\hline
\end{tabular}} \\

\subsection*{Inflation and anomalies}
\label{sec:surv:inflation}

Hopes for improving our understanding of inflation are mostly pinned on the detection of a primordial B-mode signal, although many respondents were more pessimistic in that they expect the situation to remain foggy. A variety of other potential probes of inflation were met with a lackluster response, with considerably fewer people identifying primordial non-Gaussianity and large-scale anomalies as likely sources of progress. Most expected current anomalies (of all types) to remain however, with the general feeling being that they must be addressed. Most were also confident that new anomalies would also be found, and would also need to be addressed. \\

{\footnotesize\centering\begin{tabular}{p{0.2cm}p{13.6cm}rr}
\hline
\multicolumn{2}{l}{\bf 11. Our understanding of inflation will...} & {\bf ~~~~No.} & {\bf ~~~\%~} \\
\hline
& improve due to a primordial B-mode detection & 27 & 44 \\
& remain foggy & 23 & 38 \\
& improve due to a non-Gaussianity detection & 13 & 21 \\
& improve due to new/existing CMB/large-scale anomalies & 9 & 15 \\
& improve due to detection of features in the primordial power spectrum & 6 & 10 \\
& other & 5 & 8 \\
& improve because it will be ruled out & 4 & 7 \\
& improve due to a detection of non-zero spatial curvature & 3 & 5 \\
& improve due to philosophical developments & 2 & 3 \\
& improve due to string theorists & 2 & 3 \\
& get worse! & 1 & 2 \\
\hline
\end{tabular}} \\

{\footnotesize\centering\begin{tabular}{p{0.2cm}p{13.6cm}rr}
\hline
\multicolumn{2}{l}{\bf 12. Current anomalies will...} & {\bf ~~~~No.} & {\bf ~~~\%~} \\
\hline
& remain and must be addressed & 41 & 67 \\
& go away & 12 & 20 \\
& remain, but can safely be ignored & 8 & 13 \\
& overturn \lcdm{} & 4 & 7 \\
& other & 3 & 5 \\
\hline
\end{tabular}}

{\footnotesize\centering\begin{tabular}{p{0.2cm}p{13.6cm}rr}
\hline
\multicolumn{2}{l}{\bf 13. New anomalies will be found...} & {\bf ~~~~No.} & {\bf ~~~\%~} \\
\hline
& and must be addressed & 51 & 84 \\
& but can safely be ignored & 6 & 10 \\
& and will overturn \lcdm{} & 6 & 10 \\
& other & 3 & 5 \\
\hline
\end{tabular}}

\subsection*{Dark matter and small scales}

The majority of respondents expected particle dark matter to be discovered in the future, and for it to explain astrophysical dark matter observations. Other options, including modified gravity theories, were significantly less popular. The most likely solution to the small-scale problems was subject to a greater diversity of opinion, although baryonic effects still emerged as a clear leader. Many were also enthused about the applications of modifications of gravity here (although less so for MOND), with non-cold dark matter (interacting or warm) following close behind. Finally, baryonic physics was seen by most as a difficult issue, with an important role as a cause of systematic errors. Many were hopeful that they can be understood well enough through simulations to allow the correct interpretation of observations, however. \\

{\footnotesize\centering\begin{tabular}{p{0.2cm}p{13.6cm}rr}
\hline
\multicolumn{2}{l}{\bf 14. (Dark matter): In the future, we will...} & {\bf ~~~~No.} & {\bf ~~~\%~} \\
\hline
& discover new particle(s) with the right properties to be dark matter & 34 & 56 \\
& show that dark matter is modified gravity & 11 & 18 \\
& show that dark matter is not a particle & 8 & 13 \\
& never know what dark matter is & 6 & 10 \\
& other & 6 & 10 \\
\hline
\end{tabular}}

{\footnotesize\centering\begin{tabular}{p{0.2cm}p{13.6cm}rr}
\hline
\multicolumn{2}{p{14.2cm}}{\bf 15. If you were to work on one solution to small-scale problems, which one would you choose?} & {\bf ~~~~No.} & {\bf ~~~\%~} \\
\hline
& Baryons & 22 & 36 \\
& Modified gravity & 14 & 23 \\
& Interacting dark matter & 10 & 16 \\
& Warm dark matter & 8 & 13 \\
& Self interacting dark matter & 8 & 13 \\
& Observational systematics & 8 & 13 \\
& MOND & 7 & 12 \\
& Other & 1 & 2 \\
\hline
\end{tabular}}

{\footnotesize\centering\begin{tabular}{p{0.2cm}p{13.6cm}rr}
\hline
\multicolumn{2}{l}{\bf 16. Baryonic physics...} & {\bf ~~~~No.} & {\bf ~~~\%~} \\
\hline
& will remain difficult to simulate, and is an important systematic effect & 34 & 56 \\
& will be understood well enough through simulations to interpret observations correctly & 20 & 33 \\
& will remain difficult to simulate, but is a minor systematic effect & 4 & 7 \\
& will remain difficult to simulate, and is disastrous for observations & 4 & 7 \\
& will be completely understood through simulations & 2 & 3 \\
& other & 1 & 2 \\
\hline
\end{tabular}}

\section*{Acknowledgements}

We are grateful to all of the organizers and participants of the Beyond \lcdm{} workshop, held in Oslo, Norway from the 14th--17th January 2015. We would also like to thank A.~Linde and M.~Maggiore for useful comments.

PB is supported by an appointment to the NASA Postdoctoral Program at the Jet Propulsion Laboratory, California Institute of Technology, administered by Oak Ridge Associated Universities through a contract with NASA. PB also acknowledges support from European Research Council grant StG2010-257080. YA acknowledges support from DFG through the project TRR33 ``The Dark Universe.'' TB is supported by All Souls College, Oxford. EB (Bentivegna) is supported by the project ``Digitizing the universe: precision modeling for precision cosmology," funded by the Italian Ministry of Education, University and Research (MIUR). The work of SC (Clesse) is supported by the Return Grant program of the Belgian Science Policy (BELSPO). FF would like to thank the Kavli Institute for Theoretical Physics China in Beijing, where this contribution was partially written, for hospitality. CL acknowledges support from the Research Council of Norway through grant 216756, and from STFC consolidated grant ST/L00075X/1. RP is supported by the Swedish Research Council, contract number 621-2013-428. The contribution of MSP to this publication was made possible through the support of a grant from the John Templeton Foundation. IDS has been supported by the Funda\c c\~ao para a Ci\^encia e Tecnologia (FCT) through the Investigador research grant SFRH/BPD/95204/2013, as well as UID/FIS/04434/2013. IT acknowledges support from the FCT through the Investigador contract No. IF/01518/2014 and strategic project UID/FIS/04434/2013. FVN is supported by the ERC Starting Grant ``cosmoIGM'' and partially supported by INFN IS PD51 ``INDARK''. The research of MvS leading to these results has received funding from the ERC under the European Community's Seventh Framework Programme (FP7/2007-2013 Grant no. 307934). Part of the research described in this paper was carried out at the Jet Propulsion Laboratory, California Institute of Technology, under a contract with the National Aeronautics and Space Administration.

\newpage
{\footnotesize
\bibliography{blcdm}}

\end{document}